\documentclass[11pt]{article}
\usepackage{amssymb}
\usepackage{amsmath}
\usepackage{graphicx}
\usepackage{epsfig}

\newlength{\bredde}
\def\slash#1{\settowidth{\bredde}{$#1$}\ifmmode\,\raisebox{.15ex}{/}
\hspace*{-\bredde} #1\else$\,\raisebox{.15ex}{/}\hspace*{-\bredde} #1$\fi}
\newcommand{\sect}[1]{\setcounter{equation}{0}\section{#1}}

\newcommand{\beq}{\begin{equation}}
\newcommand{\eeq}{\end{equation}}
\newcommand{\nn}{\nonumber}

\newcommand{\Dirac}{\rlap{\hspace{-1.0mm} \slash} D}
\newcommand{\id}{1\!|}
\newcommand{\qq}{{\langle \bar{q}q\rangle}}

\newcommand{\Tr}{\mbox{Tr}}

\def\beqn{\begin{eqnarray}}
\def\eeqn{\end{eqnarray}}

\def\hchi{\hat{\chi}}
\def\hH{\hat{H}}
\def\hI{\hat{{\cal I}}_\nu}

\def\al{\alpha}
\def\eps{\epsilon}
\def\vareps{\varepsilon}
\def\re{{\Re\mbox{e}}}
\def\im{{\Im\mbox{m}}}

\begin{document}
\topmargin -1.4cm
\oddsidemargin -0.8cm
\evensidemargin -0.8cm

\title{\Large{\bf 
Matrix Models and QCD with Chemical Potential
}}

\vspace{2cm}

\author{\\{\sc G. Akemann}\\~\\~\\
Department of Mathematical Sciences \& BURSt Research Centre\\
Brunel University West London\\ 
Uxbridge UB8 3PH\\ United Kingdom
}

\date{}

\maketitle
\vfill
\begin{abstract}
The Random Matrix Model approach to Quantum Chromodynamics (QCD) 
with non-vanishing chemical
potential is reviewed. The general concept using global symmetries is 
introduced, as well as its relation to field theory, 
the so-called epsilon regime of chiral Perturbation
Theory ($\eps\chi$PT). 
Two types of Matrix Model results
are distinguished: phenomenological applications 
leading to phase diagrams, and an exact limit of the QCD Dirac operator 
spectrum matching with $\eps\chi$PT. All known analytic results for 
the spectrum of complex 
and symplectic Matrix Models with chemical potential
are summarised for the symmetry classes of ordinary and adjoint QCD,
respectively. These include correlation functions of Dirac operator 
eigenvalues in the complex plane for real chemical potential, 
and in the real plane for imaginary isospin chemical potential. 
Comparisons of these predictions 
to recent Lattice simulations are also discussed.  
\end{abstract}
\vfill

\newpage

\tableofcontents

\newpage

\sect{
Introduction}
\label{intro}

In the phase where chiral symmetry is broken and quarks and gluons are
confined Quantum Chromodynamics (QCD) is strongly coupled and perturbation
theory breaks down. Several methods alow to study this situation from first
principles. In an effective field theory approach only the lowest degrees of
freedom are taken into account, such as the pseudo Goldstone bosons in chiral
Pertubation Theory ($\chi$PT). Second, in Lattice Gauge Theory (LGT) QCD is
studied on a Euclidean space-time lattice. At zero baryon chemical potential
$\mu=0$ LGT has been very successful, e.g. in describing the particle spectrum
or the chiral phase transition. Turning on $\mu$ in the phase diagram, LGT
faces the so-called sign problem: the action becomes complex in general.
This invalidates the usual method to weight configurations by their Boltzmann
factor, as it ceases to be a real positive quantity.

In this article we will study a simple model called chiral Random Matrix Model
(MM) in order to understand some aspects of this problem. This includes 
insights into chiral symmetry breaking at non-zero chemical potential,  
and quantitative aspects of the Dirac operator spectrum close to zero 
momentum, illuminating the interplay between LGT and $\chi$PT. 
MMs can be mapped to field theory in
a particular limit called the epsilon regime or $\eps\chi$PT. 
Clearly QCD is not only a MM, and we will cover aspects  
dictated by global symmetries alone. 
Also our MM does not solve the sign problem on
the Lattice, but it provides a model that can be exactly solved despite this
problem, and that offers quantitative predictions for the Lattice. 
The three types of real, complex and symplectic chiral MMs match with the
classification of gauge theories in general (including QCD) into 3 classes, 
displaying different ways or the absence of the sign problem.

To start let us briefly summarise what has been achieved for MMs of QCD at 
zero chemical potential since their introduction in \cite{SV93}. 
An excellent review describing its concepts and 
success can be found in \cite{VW}. 
The introduction of MMs as a tool to study complicated physical systems by
means of symmetry predates QCD substantially, going back to Dyson and Wigner
in the late 1950 and early 1960. Initially developed for
heavy nuclei MMs have today found applications in most fields of physics (and
other sciences), as reviewed in \cite{GMW}. 
The first MM for the Dirac spectrum of 
QCD was written down and solved in \cite{SV93}, including 
all quenched and massless 
eigenvalue correlation functions for the MM Dirac operator eigenvalues.
It became apparent that the 3 fold 
classification of Dirac operators for different
field theories according to charge and complex conjugation symmetry
\cite{Jac3fold} 
exactly coincides with the 3 possible chiral symmetry breaking patterns
\cite{Peskin}. The corresponding 3 MMs are called chiral ensembles, having
real, complex and quaternion real matrix elements\footnote{Non-chiral MMs
as those of Dyson and Wigner can be used to study 
flavour symmetry breaking in 3 dimensions, see  
\cite{VZ3d,Szabo} and references therein.}. 
All eigenvalue correlation functions for these 3 chiral MMs 
ensembles were consecutively
computed, including an arbitrary number of massive quark flavours $N_f$ 
\cite{SV93,JacSU2,DNmass,WGW,AK,NN,APV},
and individual eigenvalue distributions \cite{DNW,WGW}.

The equivalence of such MMs to $\eps\chi$PT
was first realised on the level of partition functions \cite{SV93,HV}, 
leading to identical sum-rules for the Dirac operator \cite{LS,SV94}. 
As explained by Gasser and Leutwyler \cite{GL}
the $\eps\chi$PT regime also called microscopic domain 
is reached, when the zero-momentum Goldstone modes completely dominate the
partition function, decoupling from Gaussian propagating modes.

The MM density \cite{DOTV}
and density-density \cite{TV} correlations were shown to coincide with those
derived from $\eps\chi$PT alone, and consequently the individual eigenvalue
distributions to leading order \cite{AD03}.
Note that so far no higher density correlations, nor density correlations
in symmetry classes other than QCD have been computed from $\eps\chi$PT.

Furthermore it was understood precisely when the MM-$\eps\chi$PT
correspondence breaks down, at the so-called Thouless energy 
\cite{OV}, making the MM
approximation transparent. Even when going beyond the MM regime its techniques
have been useful to compute expectation values of vector and axial vector
currents, see \cite{DHJLL} and references therein. 

Below the Thouless energy MMs give precise analytic predictions 
for the low
Dirac operator eigenvalues lying inside the $\eps$-regime, as a function of 
quark masses and gauge field topology, for each of the 3 symmetry classes. 
Such quantitative predictions have been extensively checked in all 3 classes,
in quenched and unquenched simulations. First tests were done using staggered
fermions, 
giving confirmation in all 3 symmetry classes \cite{RMTlatt1} in the sector of 
topological charge $\nu=0$.
For staggered
fermions care has to be taken 
compared to the global symmetry in the continuum in 2 of the 3 classes. 

The predictive power of MMs has gained even more relevance in the important
development of chiral fermions, that possess an exact Lattice chiral
symmetry and well defined topology away from the continuum (see \cite{Ha04}
for a review). 
Here MMs have been used to test 
such algorithms at fixed gauge field 
topology, being confirmed for all 3 classes by various groups
\cite{RMTlatt2}. 
Subsequently staggered simulations have been improved, also showing the 
correct
topology dependence as overlap fermions 
\cite{RMTlatt3}.

In this article we will try to go through the same steps for MMs with
non-zero chemical potential, and compare the 
analytical prediction available for 2 of the 3 classes to 
Lattice data whenever possible. 
Very recently a review on the same topic appeared \cite{Kim}, 
focusing more on the relation to $\eps\chi$PT than the MM concepts, 
its solution and comparison to LGT covered here.

In order to emphasise the limitations of our approach let us recall some 
important questions of LGT at finite density and temperature $T$. 
One of the purposes underlying Lattice simulations is to better understand
the restoration of chiral symmetry, going hand in hand with deconfinement for
QCD. Major experiments such as RHIC as well as the transition in the early
universe happen at very small $\mu$ close to the temperature axis. Here
several techniques have been applied to avoid the sign or complex action 
problem, where we refer to \cite{atsushi}
for a review and \cite{Schmidt:2006us} for recent developments. 
Physical questions as for the free energy, pressure, equation of state,
screening mass or string tension
have been addressed, and the dependence on the number of
flavours and quark masses is being tested (see \cite{Karsch:2006sf} for a
review and references). 
However, none of these techniques actually solves
the sign problem. For that reason the phase diagram far away from the $T$-axis
is still current subject of debate \cite{dFP}.

Our simple MM cannot give answers to these physical questions, despite
providing a qualitative (flavour independent) phase diagram for QCD. 
The domain of our MM solution 
is at small or zero $T$ and small to moderate $\mu$. 
Its analytical predictions there can serve as a testing ground, 
both in the presence or absence of the sign problem. 
It also can tell us in which region simulations are
doable at finite volume and when the sign problem hits in, as reviewed in 
\cite{Kim}. 
Because of this situation so far we can only provide evidence for 
MM predictions by comparing to quenched Lattice QCD, 
and to unquenched LGTs not suffering from
the sign problem. 

Several topics using MMs at finite $\mu$
are not covered in this review. In section \ref{phase} we review
only aspects of 
the phase diagram for QCD. Similar results exist for two colours
\cite{KTV2} as well, and 
for a discussion of the chiral Lagrangian with chemical potential
for two colours and the adjoint representation we refer to \cite{Kogut}. 
The phase diagram of QCD can also be analysed using the complex zeros of the
partition function, where we refer to recent \cite{Stephanov:2006dn} 
as well as previous work \cite{Halasz:1996jg}.
Recently the phase of the fermion determinant has been computed
\cite{phaseletter}, which can be tested in Lattice simulations. 
In addition to chemical potential and temperature also colour degrees of
freedom can be taken into account in MMs 
\cite{Vanderheyden:2000ti} in order to model single-gluon exchanges, 
or the symmetries of the staggered operator can be added
\cite{Halasz:2000yn}. However, these additions typically 
destroy the exact analytical solvability.  
Last but not least a factorisation method has been advocated by the authors of
\cite{Ambjorn:2002pz}. It uses MMs as a toy model in order to tackle the 
sign problem in Lattice simulations.  
As mentioned earlier MMs enjoy many different applications in Physics. 
For applications of MMs with complex eigenvalues in other areas we refer to 
the review \cite{FS}.

This article is organised as follows. In the next section \ref{concept} we
explain the concept of our MM approach including a symmetry classification and
its relation to $\eps\chi$PT. After a brief view  on the resulting phase
diagrams in section \ref{phase}, 
we turn to the relation between Dirac operator eigenvalues
and chiral symmetry breaking in section \ref{exact}. 
Here we summarise what is know for complex
eigenvalue correlation 
functions in the class containing QCD, and confront with quenched lattice
results. Section \ref{aQCD} contains all results for symplectic MMs  
corresponding to field theories with fermions in the adjoint representation. 
Here we compare to unquenched Lattice results. 
In section \ref{Imu} 
we present the solution of a MM for the QCD class with imaginary isospin
chemical potential where unquenched simulations are straightforward. 
Conclusions and an outlook are given in section 
\ref{COP}.

\sect{Concept of Matrix Models at Finite $\mu$}
\label{concept}

\subsection{Global symmetries}
The general strategy 
in applying Matrix Models to Physics is to replace a
complicated operator by a random matrix with the same global symmetries,
trying to gain some information on its spectrum. In 1993 Shuryak and
Verbaarschot applied this idea for the first time to the Dirac operator 
$\Dirac$ in QCD \cite{SV93}, which was then extended to QCD with chemical
potential by Stephanov \cite{Steph}:
\beq
\Dirac(\mu)\equiv\Dirac+ \mu\gamma_0 \ \longrightarrow \left(
\begin{array}{cc}
0& \Phi+\id\,\mu\\
-\Phi^\dagger+\id\,\mu& 0\\
\end{array}
\right)\ .
\label{Dmudef}
\eeq
The matrix $\Phi$ has constant, that is space-time independent matrix
elements, typically
with identical Gaussian distributions and hence its name random. 
Here we have used already 2 global
symmetries: in the continuum chiral symmetry implies that 
$\{\Dirac(\mu),\gamma_5\}=0$, 
leading to an off-diagonal block structure. Second,
we have used that the Euclidean Dirac operator is anti-hermitian at zero
chemical potential, and that $\mu\gamma_0$ is hermitian.

The average over the gauge fields of the Yang-Mills action
is replaced by a random matrix average,
\beq
\exp[-S_{gauge}]\ \longrightarrow \ \exp[-\sigma^2 \Tr\ \Phi^\dagger\Phi] \ ,
\label{vev}
\eeq
when computing observables. The inverse 
variance $\sigma^2$ in the Matrix Model is to
be fixed later.
Using a lattice regularisation in a finite volume $V$ 
the Dirac matrix becomes finite dimensional, 
and we thus can choose the random matrix $\Phi$ of dimension
$N\times(N+\nu)$. 
Here we have furthermore restricted ourselves to a sector
of fixed topology $\nu$, or through the index theorem to a fixed number of zero
modes of $\Dirac$. Its eigenvalues are defined as 
\beq
\Dirac(\mu) \varphi_k \ =\ i z_k \varphi_k\ ,\ \ k=1,\ldots,V\ \
(\mu=0\Rightarrow z_k\in\mathbb{R}) \ .
\label{defDev}
\eeq
For zero $\mu$ they lie on the imaginary axis, whereas for nonzero $\mu$ 
the operator 
$\Dirac(\mu)$ has complex eigenvalues in general,
being a complex non-Hermitian operator.
The large-$N$ limit in the matrix size thus
corresponds to the thermodynamic limit $\lim 2N\sim V \to \infty$, 
keeping the size of the gauge group $N_C$ fixed. 
\begin{table}[h]
\begin{tabular}{c|c|c}
{$SU({2})$ fund.} &
{$SU({N_C\geq 3}),\ U(1)$ fund.} &
{$SU({N_C})$ adjoint}\\[2ex]\hline\\[1ex] 

{$\Dirac(\mu)=
\left(\begin{array}{cc}
0& \Phi+\id\mu\\
-\Phi^T+\id\mu \\
\end{array}\right)$ }  
&
{$
\left(\begin{array}{cc}
0& \Phi+\id\mu\\
-\Phi^\dagger+\id\mu& 0\\
\end{array}\right)$ }
&
{$
\left(\begin{array}{cc}
0& \Phi+\id\mu\\
-\Phi^\dagger+\id\mu& 0\\
\end{array}\right)$ }            
\\[4ex] 

  \fbox{$\Phi_{ij}$$\in\ \mathbb{R}$,\  $\beta_D=1$}
& \fbox{$\Phi_{ij}$$\in\ \mathbb{C}$ ,\ $\beta_D=2$}
& \fbox{$\Phi_{ij}$$\in$ real $\mathbb{H}$ ,\ $\beta_D=4$} \\[2ex]

{as $[C\sigma_2 K,\Dirac]=0$ } 
& & {as $[CK,\Dirac]=0$} \\[2ex]

$\exists$ finite fraction $ z_k \in \mathbb{R}$ 
& &$\forall k:$ $ z_k \notin \mathbb{R}, i\mathbb{R}$\\[2ex]\hline\\[1ex]

{$SU({N_C})$ adjoint} staggered 
& $SU({N_C\geq 3}),\ U(1)$ fund. 
& {$SU({2})$ fund.} staggered\\

& staggered & \\[2ex]

\end{tabular}
\caption{Table of global symmetry classes.}
\label{Dsym}
\end{table}

What are the matrix elements of $\Phi$? It turns
out that they can be further restricted by symmetry and fall into 3 different
classes.
According to the choice of gauge group 
and its representation it was shown for gauge groups $SU(N_C)$   
at $\mu=0$ \cite{Jac3fold} and at $\mu\neq0$ \cite{Kogut}  
that for either two colours $N_C=2$ or the adjoint
representation an anti-unitary symmetry allows to choose the matrix elements
$\Phi_{ij}$ to be real or quaternion real\footnote{For quaternions $\id$
  denotes the quaternion unity matrix.}, respectively.
For the class with $SU(N_C\geq3)$ in the fundamental
representation containing QCD 
no such symmetry exists, and the matrix elements are complex. 
These are the global symmetries that 
the random matrices and the exact Dirac operators
share, as shown in the table
\ref{Dsym}. This classification can of course be extended to any gauge groups
having a real, complex or pseudo-real representations\footnote{I would like to
  thank P. Damgaard for this comment.}, and we have added $U(1)$ to the table.

The anti-unitary 
symmetry in the real and pseudo-real case 
is provided by a product of the charge $C$ and complex
conjugation matrix $K$, and a Pauli matrix $\sigma_2$
for $SU(2)$ \cite{Jac3fold,Kogut}. We have introduced the Dyson index
$\beta_D$ to label the number of independent degrees of freedom per matrix
element of $\Phi$. 
In the class $\beta_D=1$ the non-symmetric, real matrix $\Dirac(\mu\neq0)$ 
has a real secular equation. Hence it 
can have real eigenvalues\footnote{In a different MM their expectation 
value growns
  $\sim\sqrt{N}$ \cite{Efetov}.} in addition to complex eigenvalues occuring in
conjugated pairs. 
For that reason this class has a real $\det[\Dirac(\mu)]$ which
is not necessarily positive. This sign problem is thus reduced from a phase
to a fluctuating sign. 
For an even number of two-fold degenerate flavours 
the product of all determinants remains positive and has no sign problem. 
For $\beta_D=4$ the determinant $\det[\Dirac(\mu)]$ is alway positive, as 
the eigenvalues of a real quaternion matrix 
occur in complex conjugated pairs \cite{Mehta2}.

Notice that for discretised Dirac operators on a Lattice their 
global symmetries
may change: 
when using staggered fermions the
symmetry classes $\beta_D=1$ and 4 have to be interchanged
\cite{Halasz:1995vd}. On the other hand Wilson type Dirac operators are in the
same symmetry class as the continuum. So far the transition from the discrete
to the continuum symmetry class has not been observed for staggered fermions
in the Dirac spectrum.

\begin{table}[h]
\begin{tabular}{c|c|rl}
& {$\mu=m=0$} &  {$\mu\neq0,\ m\geq0$}& \\[1ex]\hline\\[1ex]

$\beta_D=1$ & $SU(2N_f) \to Sp(2N_f)$ 
&  $SU(N_f)_L\times SU(N_f)_R \to$ &$ Sp(N_f)_L\times Sp(N_f)_R$\\ 
&&$\to$&$Sp(N_f)_V$ for  $m\neq0$\\[1ex]
$\beta_D=2 $& $SU(N_f)_L\times SU(N_f)_R \to SU(N_f)_V $ & $SU(N_f)_L\times
  SU(N_f)_R \to $&$ SU(N_f)_V$\\[4ex] 
$\beta_D=4 $& $SU(2N_f) \to SO(2N_f) $ 
&  $SU(N_f)_L\times SU(N_f)_R \to$&$ SO(N_f)_L\times SO(N_f)_R$\\
&&$\to$& $SO(N_f)_V$ for $m\neq0$\\

\end{tabular}
\caption{Table of spontaneous (left) and explicit (right)
chiral symmetry breaking patterns.}
\label{XSB}
\end{table}
\newpage
It turns out that the classification into 3 groups
exactly corresponds the 3 possible patterns of spontaneous 
chiral symmetry breaking
introduced by Peskin \cite{Peskin}, where a maximal breaking is assumed. 
Here we have dropped both the vector $U_V(1)$ and axial vector $U_A(1)$
factor as they are not affected by the spontaneous breaking. 
The case of symmetry breaking in 2 dimensions 
is special, and we refer 
to \cite{Leonid} for an exhaustive treatment of the Schwinger model and its
relation to MMs.

The mass term in $\chi$PT (see eq. (\ref{LchPT})) of course explicitly breaks
chiral symmetry. At $\mu=0$ the patterns are the same as for spontaneous
breaking in all 3 classes, see  table \ref{XSB}. 
A nonzero $\mu$-term also explicitly breaks chiral symmetry, and 
the breaking patters differs from the spontaneous one for $\beta_D=1,4$
as discussed in \cite{Kogut}, both for $m=0$ and $m\neq0$. 
There is a competition for the vacuum aligning to the
chemical potential or the mass term if $m\neq 0$, and we refer to 
\cite{Kogut} for details. 
Only in the $\beta_D=2$ class the explicit and spontaneous breaking patters 
remain the same at $\mu\neq0$.

This complication for $\beta_D=1,4$ at 
$\mu\neq0$ is to be contrasted with the global
symmetries in table \ref{Dsym}. Despite  breaking the anti-hermiticity
$\mu\neq0$ does not change these global symmetries. 
Therefore in the following we 
use table \ref{Dsym} as a guiding principle to build our MMs 
with $\mu\neq0$, following \cite{Steph} for $\beta_D=2$ and 
\cite{HOV} for $\beta_D=1,4$ where these MMs were first introduced. 
A second reason is that the effective field theories for $\beta_D=1$ and 4 
are much complicated due to the structure of the coset space for the Goldstone
manifold from table \ref{XSB}.  Only limited analytical results are 
available even at $\mu=0$, see 
\cite{SV94,Toublan:1999hi}.

\begin{figure}[-h]
\centerline{\epsfig{
figure=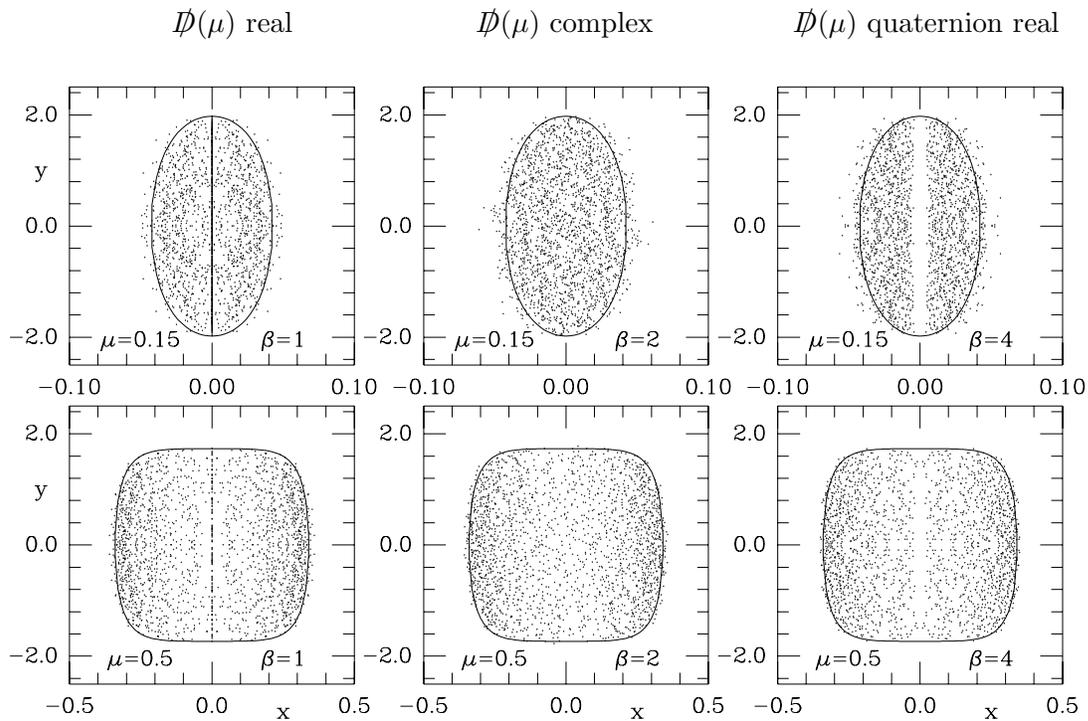
,width=20pc,angle=-90}
\put(-345,20){$\Dirac(\mu)$ real}
\put(-230,20){$\Dirac(\mu)$ complex}
\put(-110,20){$\Dirac(\mu)$ quaternion real}
}
\caption{Scatter plots from \cite{HOV}
of complex eigenvalues in the 3 different MM symmetry
  classes (columns) obtained numerically for a given $N\gg1$ and two different
  values of $\mu$. The larger $\mu$ (lower row) 
is just below the critical  $\mu_c \approx 0.527$
in these models.}
\label{HOVfig}
\end{figure} 
In \cite{HOV} the effect of chemical potential in the three different symmetry
classes  in table 
\ref{Dsym} was analysed using a MM, as shown in fig. \ref{HOVfig}.
It shows a scatter plot of complex MM eigenvalues obtained numerically 
for large random matrices. In the matrix model (as well as on the Lattice) 
the eigenvalues stay inside a bounded
domain. The shape of this boundary is a function of $\mu$ and was computed in a
saddle point calculation at $N\to\infty$ for a non-chiral MM 
\cite{Steph}
for $\sigma^2=N$: $(y^2+4\mu^2)(\mu^2-x^2)^2+x^2=4\mu^2(\mu^2-x^2)$ where
$z=x+iy$. 
This envelope is plotted as a full line in fig. \ref{HOVfig}. It splits into
two sets at $\mu=1$ 
a spurious transition which is beyond the chiral transition 
at $\mu_c\approx 0.527$ for $T=0$, see eq. (\ref{phasebound}). The
envelope generalises the Wigner semi-circle for the MM spectral density 
at $\mu=0$, which may also split
into two arcs due to the influence of 
an external field like temperature $T$ \cite{JV}.

Fig. \ref{HOVfig} illustrates the signature of 
the 3 symmetry classes: for $\beta_D=1$ there is always 
a finite fraction of eigenvalues remaining real, as for $\mu=0$ the
eigenvalues of $\Dirac$ lie along the imaginary $y$-axis. 
For $\beta_D=4$ a gap opens up on the $y$-axis, depleting the eigenvalues. If
we were to look along the real $x$-axis on the same scale - note the
difference of scale in $x$ and $y$ direction of the picture - we would see the
same depletion here (see e.g. fig. \ref{MMsig4} in section \ref{exact}).
Our aim is to compute all complex eigenvalue correlation functions 
analytically in the vicinity of the origin and compare them to Lattice
simulations where possible.
We have succeeded so far for the 2 classes  $\beta_D=2$ and 4.

The addition of the $\mu$-part in the MM in table \ref{Dsym} 
is not unique. Instead of
adding $\mu$ times unity $\id$, as was first introduced by Stephanov
\cite{Steph}, the symmetries also allow to add a {\it second} random matrix
$\Psi$ in the $\mu$-part, $\id\to\Psi$, of the same symmetry as $\Phi$ in the
corresponding class $\beta_D$ \cite{James}:
\beq
\Dirac(\mu) \ \longrightarrow \left(
\begin{array}{cc}
0& \Phi+\mu\Psi\\
-\Phi^\dagger+\mu\Psi^\dag& 0\\
\end{array}
\right)\ ,\ \ 
\Phi_{ij},\ \Psi_{ij}\ \in \mathbb{R} / \mathbb{C} / \mbox{real}\ \mathbb{H} 
\ \mbox{for}\
\ \beta_D=1 / 2 / 4 
\label{D2matrix}
\eeq
This choice of having 2 random matrices (two-MM) 
corresponds to a non-diagonal chemical
potential term in random matrix space. Introduced by Osborn 
\cite{James} for $\beta_D=2$ and the author \cite{A05} for $\beta_D=4$ 
this apparent
complication has the virtue to allow for a Schur decomposition of 
$\Dirac(\mu)$, leading to complex eigenvalue model. 
In the former model with $\Psi=\id$ \cite{Steph} 
and only 1 random matrix (one-MM) eq. (\ref{Dmudef}) this is not possible.
The two models agree for the partition function and density 
as we will see later,
illustrating the concept of MM universality. The same universality is expected
to hold also for a non-Gaussian distribution in eq. (\ref{vev}) but is much
harder to prove, for results on non-chiral MMs see \cite{A02}.

With all ingredients given so far we can write down our MM partition function 
with $N_f$ quark flavours of masses $m_f$ and chemical potentials $\mu_f$
in a fixed topological sector $\nu$ as follows
\beq
{\cal Z}^{(N_f,\nu)}= \int d\Phi_{N\times(N+\nu)} 
\prod_{f=1}^{N_f} \det\left(
\begin{array}{cc}
m_f& \Phi+
\mu_f\id
\\
-\Phi^\dag+
\mu_f\id
& m_f\\
\end{array}
\right)
\exp[-N
\Tr\ \Phi\Phi^\dag]\ .
\label{ZMM1}
\eeq
The fermions  integrated out in the field theory 
lead to the determinants of the
Dirac operator. 
Here we have chosen $\Psi=\id$, and
fixed the inverse variance to be $N$ in order to have a compact support in the
large-$N$ limit.  

These chiral complex ensembles introduced in \cite{Steph} for $\beta_D=2$ and 
\cite{HOV} for $\beta_D=1,4$ are to be contrasted with 
the 3 Ginibre ensembles for complex
eigenvalues introduced in \cite{Gin}. In  \cite{Gin} no determinants are 
present and 
the eigenvalues of the matrix $\Phi$ are considered, instead of those of 
the chiral matrix $\Dirac(\mu)$ in eqs. (\ref{Dmudef}) or (\ref{D2matrix}).


\subsection{Relation to field theory}\label{FTrel}

Apart from using global symmetries in our MM approach 
it appears so far that we have made the
approximation of static gauge fields with Gaussian fluctuations, in order to
arrive at our MM eq. (\ref{ZMM1}). 
In reality our approximation is that of static meson fields, and 
since the first introduction of MMs in 1993 
\cite{SV93} this approximation 
has been understood much more precisely. In the
large-$N$ limit MMs become  
equivalent to a theory of mesons in the limit where their 
zero-momentum modes completely dominate the partition function. This is the 
$\epsilon$-regime (or microscopic domain)
of $\chi$PT as explained by Gasser and
Leutwyler \cite{GL}, 
and we will point out this close relation now. Unfortunately, very
little is known about the corresponding group integrals over the Goldstone
manifold in the classes $\beta_D=1$ and 4: even at $\mu=0$ analytic results
only exist for degenerate masses for more than $N_f=2$ flavours \cite{SV94}.
Therefore we will restrict our discussion in this subsection to the 
$\beta_D=2$ symmetry class including QCD, where most explicit results are
known. 

The starting point is our Gaussian MM partition function eq. (\ref{ZMM1}). 
In a first step we rewrite the determinants in terms of Grassmann
vectors $\chi$, using 
$\det[A]=\int d\chi d\chi^\dag \exp[-\Tr \chi A  \chi^\dag]$. 
The random matrix $\Phi$ can now be integrated out, being quadratic in the
exponent. This leads to a fermionic integral with quartic terms in the
exponents. The introduction of an auxiliary complex matrix $Q$ of size 
$N_f\times N_f$, also called Hubbard-Stratonovic transformation, makes the
fermions $\chi$ quadratic again, and we can integrate them out to arrive at
\beqn 
{\cal Z}^{(N_f,\nu)}
&\sim& 
\int dQ dQ^\dagger  
\det[Q^\dagger]^\nu  
\det\left[Q^\dagger Q 
-Q^\dagger BQ^{\dagger\,-1}B\right]^N 
\mbox{e}^{-N \Tr \, (Q^\dagger-M)( Q-M)},
\label{Zsigma}
\eeqn
after shifting  $Q\to Q-M$ for real masses $m_f$.
Details can be found e.g. in \cite{AFV}.
Here we have abbreviated by $M=$ diag$(m_1,\ldots,m_{N_f})$ 
the diagonal mass matrix and by 
$B=$diag$(\mu_1,\ldots,\mu_{N_f})$ 
the Baryon charge matrix,  
the diagonal matrix of chemical
potentials. This representation also called sigma model is exact, so far no
approximation has been made. Notice the duality with eq. (\ref{ZMM1}), where
$N$ and $N_f$ have changed role: we now have a determinant to the power $N$
susceptible to a saddle point approximation. 
In particular after introducing also a temperature $T$ 
the discontinuities of the partition function leading to its
phase diagram can be analysed in this way when making further assumptions
about the saddle point, as will be discussed in the next section. 

The partition function 
can be computed without further assumptions, leading
to the chiral Lagrangian in the $\epsilon$-regime. 
If we choose\footnote{A different choice is possible, keeping $\mu_f$ fixed at
$N\to\infty$. This leads to a different result, see \cite{AFV} for details.} 
to scale the chemical potentials keeping  
$2N\mu_f^2$ to be fixed in the large-$N$ (volume) limit we obtain
\beqn 
{\cal Z}^{(N_f,\nu)}
&\sim& \mbox{e}^{-N\Tr \,M^2}  
\int dQ  
\det[Q^\dagger]^\nu \mbox{e}^{-N\Tr \, Q^\dagger Q}  
\nn\\
&&\times \exp\left[ N\Tr\left(M(Q+Q^{\dagger}) 
-Q^{-1}B Q^{\dagger\,-1}B\right)\ + \ N\Tr\ln(Q^\dagger Q) 
\ +\ {\cal  O}(1/N) \right],
\label{Z1qqQexp}
\eeqn
where we have expanded the logarithm.
Next we parametrise $Q=UR$ with $U\in U(N_f)$ unitary and $R$ Hermitian
positive definite matrices. In
the saddle point limit we obtain $R\sim\id_{N_f}
$ 
(after going to eigenvalues, see \cite{AFV}) so that the Gaussian term in
$QQ^\dag$ and the logarithm term disappear.   
Thus 
we can match with the following group integral   
of $\eps\chi$PT in the fixed sector of topological charge $\nu$, 
up to constant terms $\sim \exp[const\cdot\mu^2]$, 
\beq
Z_{\eps\chi PT}^{(N_f,\nu)}
= 
\int_{U \in U(N_f)} dU \ \det[U]^\nu\ \exp\left[\frac 12 \qq V {\Tr}M(U +
  U^{-1})
\ -\ \frac {V}{4}F_\pi^2
{\Tr} [U,B][U^{-1},B]\right]\ .
\label{ZchPT}
\eeq
It includes the pion decay constant $F_\pi$, and the chiral condensate $\qq$. 
The second, constant term in the commutator, ${\Tr} [U,B][U^{-1},B]=
2\Tr(UBU^{-1}B-B^2)$, can be included in eq. (\ref{Z1qqQexp}) 
by multiplying with an appropriate constant. 
This leads us to the following identification of MM and physical, 
dimensionful parameters:
\beqn
2N\mu^2_f &\longrightarrow& VF_\pi^2\mu^2_f \nn\\
2Nm_f  &\longrightarrow& V \qq m_f 
\label{scale}
\eeqn
In principle we could have introduced 
two parameters in the MM through its 
variance, $e^{-N\qq^2\Tr\Phi\Phi^\dag}$, 
and by rescaling $\mu_f\to \tilde{const}\cdot\mu_f$. 
At $\mu=0$ the MM predictions measured in units of $\qq$ are parameter
free. At $\mu\neq0$ the ratio of the two parameters has to be determined,
e.g. by a fit to lattice data. This is is a virtue and not a drawback, as such
a fit determines $F_\pi$ without having to look at finite-volume corrections
\cite{DHSS,DHSST}.

Let us explain how the group integral eq. (\ref{ZchPT}) 
is obtained in the $\epsilon$-regime of $\chi$PT, following
\cite{GL,LS} for $\mu=0$ and \cite{TV} for $\mu\neq0$. To leading
order the chiral Lagrangian is given by 
\beq
{\cal L}_{\chi PT}
=\frac{F_\pi^2}{4}\nabla_a \hat{U}(x) \nabla_a\hat{U}(x)^\dagger
-\frac12 \qq 
(M\mbox{e}^{i\frac{\Theta}{N_f}}\hat{U}(x)+
(M\mbox{e}^{i\frac{\Theta}{N_f}}\hat{U}(x))^\dag)
\label{LchPT}
\eeq
Here we have included a topological $\Theta$-term in the QCD gauge action
$S_{gauge}$: $+i\Theta\int
F\tilde{F}=i\Theta\nu$, related to the exact Dirac operator 
zero modes due to the index
theorem. The latter produces a factor 
$m_f^\nu$ from the Dirac determinant for each flavour, 
which
is why mass and $\Theta$-term always group as shown. The
space-time dependent field lives in the coset space $\hat{U}(x)\in SU(N_f)$.  
The covariant derivative $\nabla_a \hat{U}= \partial_a\hat{U}  
-\mu(B_a \hat{U}+\hat{U} B_a)$ introduces a
coupling to the chemical potentials via the 
Baryon charge matrix $B_a=(B,\mbox{\bf 0})$.
In the chiral limit
$m_f\to0$ the zero momentum modes lead to a divergence in the
propagator. It can be cured by splitting off
these zero-modes $U$ and treating them non-perturbatively: $\hat{U}(x)=
Ue^{i\varphi(x)}$, where $U\in SU(N_f)$ is now a constant matrix. 
The propagating modes  $\varphi(x)$ have to be expanded in the algebra, that
is in terms of the Pauli matrices for $N_f=2$ for example. 
In the $\eps$-regime of $\chi$PT in a finite volume box $V=L^4$
\cite{GL} the pion mass is counted as $m_\pi\sim1/L^2$  (in contrast to the
$p$-regime where $m_\pi\sim1/L$): 
\beq
m_\pi,\mu\ll1/L\ll\Lambda\ .
\eeq
The right inequality implies the validity of the chiral Lagrangian, with
$\Lambda$ being the scale of the lowest non-Goldstone modes. The left
inequalities imply that the path integral over $\hat{U}(x)$ 
factorises\footnote{The terms mixing propagating and zero modes are boundary
  contributions \cite{DHSST}.} into 
a constant $SU(N_f)$ group integral times a Gaussian integral over
the fluctuations of $\varphi(x)$. The latter gives only a constant that is
omitted. In a last step eq. (\ref{ZchPT}) is obtained by fixing the sector of
topological charge in the sum ${\cal Z}_{QCD}=\sum_{-\infty}^\infty
\mbox{e}^{i\Theta\nu}{\cal Z}^{(N_f,\nu)}$: 
the inverse Fourier transformation leads
to the prefactor $\det[U]^\nu$ and promotes the integral from $SU(N_f)$ to
$U(N_f)$. 
Let us emphasise that in the $\eps$-limit the leading order Lagrangian
eq. (\ref{LchPT}) becomes exact. 

What we also learn at this point is when the MM approximation breaks down: it
happens when the zero-momentum Goldstone modes cease to dominate and 
propagating modes start to contribute.
In analogy to condensed matter this scale was called Thouless energy, and
it is given by $E_c/(\Sigma V)\sim F_\pi^2\sqrt{V}$ in dimensionless units at
$\mu=0$ \cite{OV}. Clearly $\mu\neq0$ will influence the mixing and it was
found in \cite{OW} that the Thouless energy increases with $\mu$ on 
the lattice.

After pointing out the equivalence between MMs and the $\eps\chi$PT 
on the level
of partition functions - a result
which is new for more than 2 chemical potentials and was independently derived
in \cite{James06}
- we give some explicit 
results in which the partition function can be further calculated 
in terms of Bessel functions.
For $m+n=N_f$ quarks of opposite isospin chemical potential
$B=\mu$ diag$(\id_m,-\id_n)$ 
with rescaled masses $\{\eta_{j\pm}=V\qq m_{j\pm}\}$
respectively, 
we obtain for the group integral eq. (\ref{ZchPT}), as well as for the MMs
eqs. (\ref{ZMM1}) and (\ref{D2matrix})
\cite{SplitVerb1,AFV}:   
\beq
{\cal Z}_{\eps\chi PT}^{(N_f,\nu)}
\sim \frac{1}{\Delta_m(\{\eta_+^2\})\Delta_n(\{\eta_-^2\})}
\det_{1\leq f,g\leq n}\left[
\begin{array}{c}
\hI(\eta_{f+},\eta_{g-})\\
\eta_{g-}^{f-m-1}I_\nu^{(f-m-1)}(\eta_{g-})\\
\end{array}
\right].
\label{Zmn}
\eeq
Without loss of generality we have chosen $n\geq m$ here. With 
$\Delta_m(\{\eta^2\})=\prod_{k>l}(\eta_k^2-\eta_l^2)$ 
we abbreviate the Vandermonde determinant of squared arguments. 
Here we have introduced the notation
\beq
\hat{\cal I}_\nu(\eta_+,\eta_-)\ \equiv\ 
\int_0^1dt\ t\ e^{-2VF_\pi^2\mu^2t^2}
I_\nu(t\eta_+)I_\nu(t\eta_-)\ ,
\label{hIdef}
\eeq
constituting the first $m$ rows. The remaining rows contain 
$I_\nu^{(k)}(\eta)$ with increasing $k$, 
denoting the $k$th derivative of the modified $I$-Bessel
function with respect to its argument. 
Note that the partition function 
eq. (\ref{Zmn}) is real {\it only} 
if all masses are real (or purely imaginary).

When all quarks have the same chemical potential $\mu_f\equiv\mu$, 
we recover the familiar result \cite{LS}
\beq
{\cal Z}_{\eps\chi PT}^{(N_f,\nu)}
\sim \frac{1}{\Delta_{N_f}(\{\eta_-^2\})}
\det_{1\leq f,g\leq N_f}\left[
\eta_{g-}^{f-1}I_\nu^{(f-1)}(\eta_{g-})
\right]\ ,
\label{ZLS}
\eeq
which is totally independent of $\mu$. This comes as no surprise since pions
don't carry Baryon charge: $[U,B\sim\id\,]=0$. 
In particular this implies that the
Leutwyler-Smilga sum rules \cite{LS}
continue to hold, despite the eigenvalues being now complex.
The simplest example is given by  
\beq
\left\langle \sum_k{}^\prime \frac{1}{z_k^2}\right\rangle_{\eps \chi PT} \ =\
\frac{1}{4(\nu+N_f)}\ ,
\label{sumrule} 
\eeq
for $N_f$ massless flavours, for massive sum rules we refer to
\cite{poulmass}. Here $\Sigma{}^\prime$ denotes the sum over $z_k\neq0$ only.
For quarks including isospin partners sum rules
depend on $\mu$, as has been worked out explicitly for imaginary $\mu$
\cite{Luz}. Note that the group integral eq.  (\ref{ZchPT}) equally holds for 
imaginary chemical potential by the simple rotation $\mu\to i\mu$, flipping the
sign of $\mu^2$. On the level of the initial MM 
such a rotation is highly nontrivial, and for the corresponding MM and its
solution we refer to section \ref{Imu}. 

Let us make an important remark. The computation of the group integral 
eq. (\ref{ZchPT}) is relevant even for a theory of only pions: 
despite the fact that its partition function is $\mu$-independent, 
the generating functional
for its spectral density $\rho_{\Dirac}(z)$ in the replica approach 
\cite{Steph,SplitVerb1}
does depend on $\mu$ through 
a pair of conjugate quarks of masses $z$ and $z^*$. 
This statement directly relates to the failure of the quenched approximation
as explained in \cite{Steph}, and will become more
transparent in subsection \ref{XSBD} later. 

In addition to the above 
also partition functions with one or more bosonic quarks have to be
computed in the Toda lattice approach \cite{SplitVerb1}, giving rise to
inverse powers of determinants. In the simplest case
of 1 pair of conjugated bosonic quarks, denoted by negative $N_f=-2$, 
the partition functions is reading \cite{SplitVerb1}
\beq
Z_{\eps\chi PT}^{(-2,0)}
= 
\int \frac{dQ}{\det[Q]^2}\ e^{\frac 12 \qq V \Tr M(Q + Q^{-1})
- \frac {V}{4}F_\pi^2
\Tr \{Q,B\}\{Q^{-1},B\}}\ \sim\ 
e^{-\frac{V^2\qq^2(m^2+m^{*\,2})}{8\mu^2F_\pi^2V}} 
K_0\left(\frac{V^2\qq^2|m|^2}{4\mu^2F_\pi^2V}\right)
\label{ZchPTboson}
\eeq
for zero topology $\nu=0$. The integration goes over $2\times2$ positive
definite Hermitian matrices and has to be regularised. Only the leading
singular term is given here, which can be reproduced from MMs as well 
\cite{AOSV}. 
The analytic
behaviour and $\mu$-dependence of such bosonic integrals 
is strikingly different from fermionic
partition functions, and we refer to \cite{Splittorff:2006uu} for details. 

General partition functions with both fermions and bosons can be also used
directly as generating functionals for resolvents in the supersymmetric
approach. This has been used at $\mu=0$ in \cite{DOTV,TV}, 
in order to derive the spectral density of the Dirac operator directly from
$\eps\chi$PT, without using MMs.  
In the following sections we will use a different approach that does
not rely on bosonic or mixed partition functions.

\newpage

\sect{Phenomenological Application: Phase Diagrams of QCD}
\label{phase}

In this section we will derive a qualitative phase diagram for QCD using the 
MM partition function in terms of the sigma model from the last section. 
These results are called phenomenological as the link to the chiral Lagrangian 
is lost close to the phase transition, where $\chi$PT and in
particular the epsilon regime breaks down. 
Let us point out however, that for the two other classes, $SU(2)$ and adjoint
QCD, such an effective description can be used along the $\mu$-axis to
describe a phase transition due to diquark condensation. We refer e.g. to
\cite{Kogut} for further details.

\begin{figure}[-h]
\centerline{\epsfig{
figure=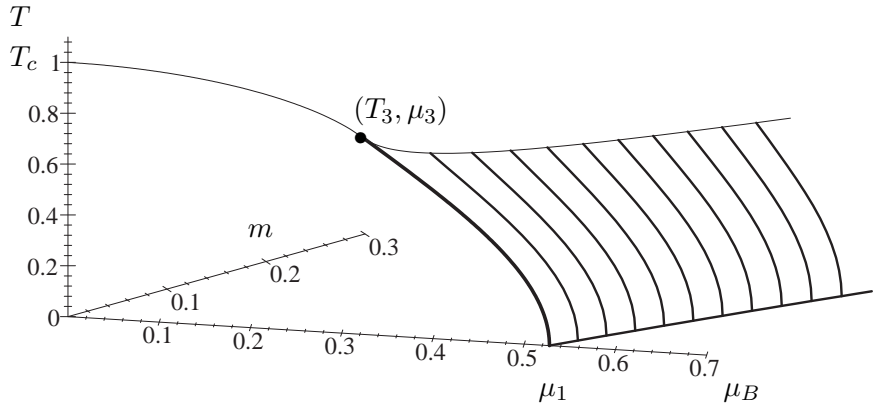
,width=30pc}
\put(-350,140){$T$}
\put(-350,125){$T_c$}
\put(-220,95){$\bullet$}
\put(-220,105){$(T_3,\mu_3)$}
\put(-260,60){$m$}
\put(-80,0){$\mu_B$}
\put(-150,0){$\mu_1$}
}
\caption{Phase diagram of QCD for $N_f=2$ light flavours, from \cite{HJSSV}.}
\label{QCDphase}
\end{figure}

Coming back to QCD let us write down the effective Lagrangian or potential,
following closely \cite{HJSSV}.
The effect of the lowest Matsubara frequency  $\pm \pi T$
or Temperature is introduced 
as in \cite{HJSSV}, 
\beq
\Dirac(\mu;T) \ \longrightarrow \left(
\begin{array}{cc}
0& \Phi+(\mu+iT\sigma_3)\id\\
-\Phi^\dagger+(\mu+iT\sigma_3)\id& 0\\
\end{array}
\right).
\label{DTmatrix}
\eeq
Because both the Euclidean time derivative and the $\mu$-term are
proportional to $\gamma_0$, $(\partial_\tau+\mu)\gamma_0$, 
they are replaced by 
a constant matrix times an off-diagonal unity matrix. In addition to the first
proposal \cite{JV} here half of the $\Dirac$ eigenvalues have positive and
half have negative frequencies, through the $\sigma_3$-term. 
Despite the dimensional reduction through the discretisation of the time 
direction the the global symmetries of $\Dirac$ are taken to be 
the same as in 4 dimensions. 
Note that the above $T$-term alone would preserve the anti-Hermiticity. There
are ways to include also higher frequencies by $T\to$ diag$(T_1,\ldots,T_N)$
as suggested 
in \cite{GSW}, but we will pursue
the simplest 
approximation \cite{HJSSV}. Also we only keep one single, real $\mu$ 
here for all quarks. 
All steps go through as in the previous section until 
eq. (\ref{Zsigma}). In terms of the auxiliary field $Q$  we obtain the
effective Lagrangian 
\beq
{\cal
  L}(Q,\{m\},\mu,T) \ =\ N\qq QQ^\dagger
-\frac{N}{2}\ln[(Q+M)(Q^\dagger+M)-(\mu+iT)][(Q+M)(Q^\dagger+M)-(\mu-iT)]
\ .
\eeq
The topological charge $\nu$ of order 1 can be neglected in the large-$N$
limit. Making the further assumption that the saddle point solution is
proportional to unity, $Q=\phi\,\id$,
leads to the following situation. 
Despite ${\cal L}$ not being polynomial the resulting saddle point equation
$\delta_\phi {\cal L}=0$ 
is a polynomial equation of 5$th$ order, just like following 
from a Landau-Ginsburg ansatz with a sextic potential. At zero mass one
obtains \cite{HJSSV}
\beq
\phi^4-2\left(\mu^2-T^2+\frac12\right)\phi^2 
+ (\mu^2+T^2)^2 + \mu^2-T^2\ =\ 0\ ,
\label{phasebound}
\eeq
leading to a second order line starting at $\mu=0$ and $T_c$ to meet a first 
order line starting from $T=0$ and $\mu_1$ in a tri-critical point 
$(T_3,\mu_3)=(\frac12\sqrt{\sqrt{2}+1},\frac12\sqrt{\sqrt{2}-1})$ 
\cite{HJSSV}. For equal nonzero mass $M=m\id$ the second order line 
becomes a cross over as shown in the full 3-dimensional
phase diagram fig. \ref{QCDphase}.
 
Using the input of $T_c=160$ MeV and $\mu_1=1200$ MeV physical scales can
be reintroduced, leading to a prediction for the location of a 
tricritical point.

Some remarks are in order. The same phase diagram has been obtained from
several other models as reviewed in \cite{Casalbuoni}, 
as well as partly from Lattice simulations along the $T$-axis and close to
$T_c$ following the transition (crossover) line
(see \cite{atsushi,dFP} for references). 
However, 
there is no flavour dependence in this MM prediction as the
dependence of the sigma model on $N_f$ is too weak:
the
phase diagram looks the same also for any number $N_f>2$, given there is no
other saddle point than $Q=\phi\,\id$.
On the other hand 
for $N_f=3$ flavours we know from Lattice studies that for zero $\mu$ 
the transition at $T_c$ depends on the masses, e.g. for equal mass it
becomes first order. We refer to \cite{Heller} for a recent review. 
As a further restriction 
$\qq$ is a fixed, $N_f$ independent input parameter in our MM. 
Because of the difficulties of performing Lattice simulations with a complex
action, the presence or absence of a tri-critical point is still a subject of
debate today (see \cite{dFP} for a most recent critical review).

\begin{figure}[-h]
\centerline{\epsfig{
figure=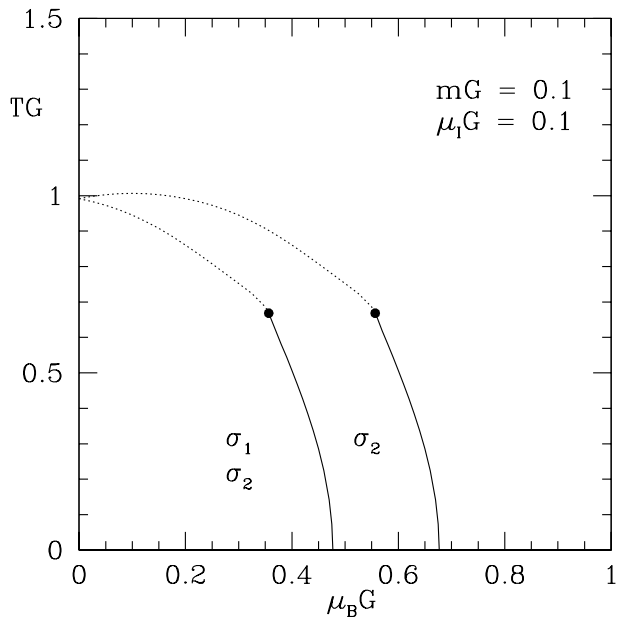
,width=20pc}
}
\caption{Phase diagram of QCD for $N_f=2$ light flavours in the $\mu_B$ plane
  for fixed $\mu_I$ and mass $m$, from \cite{KTV}. All quantities are in units
  of the inverse variance $G$.
}
\label{QCDmu12}
\end{figure}

The MM can be refined by including different chemical
potentials for the quarks, that is baryon $(\mu_B)$ and isospin ($\mu_I$) 
chemical potential. This
analysis has been carried out in \cite{KTV} for $N_f=2$ 
including a very detailed analysis of
the phase diagram. We only cite the most prominent feature, that is a doubling
of the phase transition lines  for fixed $\mu_I$ and mass $m$ as shown in 
fig. \ref{QCDmu12}. 
This can be easily understood as for $\mu_1\neq\mu_2$ the chiral restoration
for the up and down quark condensates $\langle \bar{u}u\rangle$
and $\langle \bar{d}d\rangle$ is different.
A similar effect can be expected if the masses are split, $m_u\neq m_d$, at
equal chemical potential $\mu_1=\mu_2$. The same split of one first order
transition into 
two was found analytical from MMs at $T=0$ \cite{AFV}, as can be
derived from 
eq. (\ref{Z1qqQexp}) of the previous section. It leads to the fact that 
the $N_f$-flavour partition function can be written as a
determinant of single flavour partition functions. In consequence the 
$N_f$-flavour partition function 
undergoes a transition whenever
one of its building blocks, a single flavour partition function,
has a phase transition (singular derivative).

\sect{Exact Limit: QCD Dirac Operator Spectrum}
\label{exact}

In this section we consider our MM as an exact limit of QCD, in the sense that
it coincides with the $\eps$-regime of $\chi$PT as has been
pointed out in subsection \ref{FTrel}. 
This assumes that $\chi$PT as an
effective theory itself can be consistently derived from QCD. 
In all the following we will consider temperature $T=0$ only.

\subsection{Chiral symmetry and the Dirac spectrum}
\label{XSBD}

\begin{figure}[-h]
\centerline{\epsfig{
figure=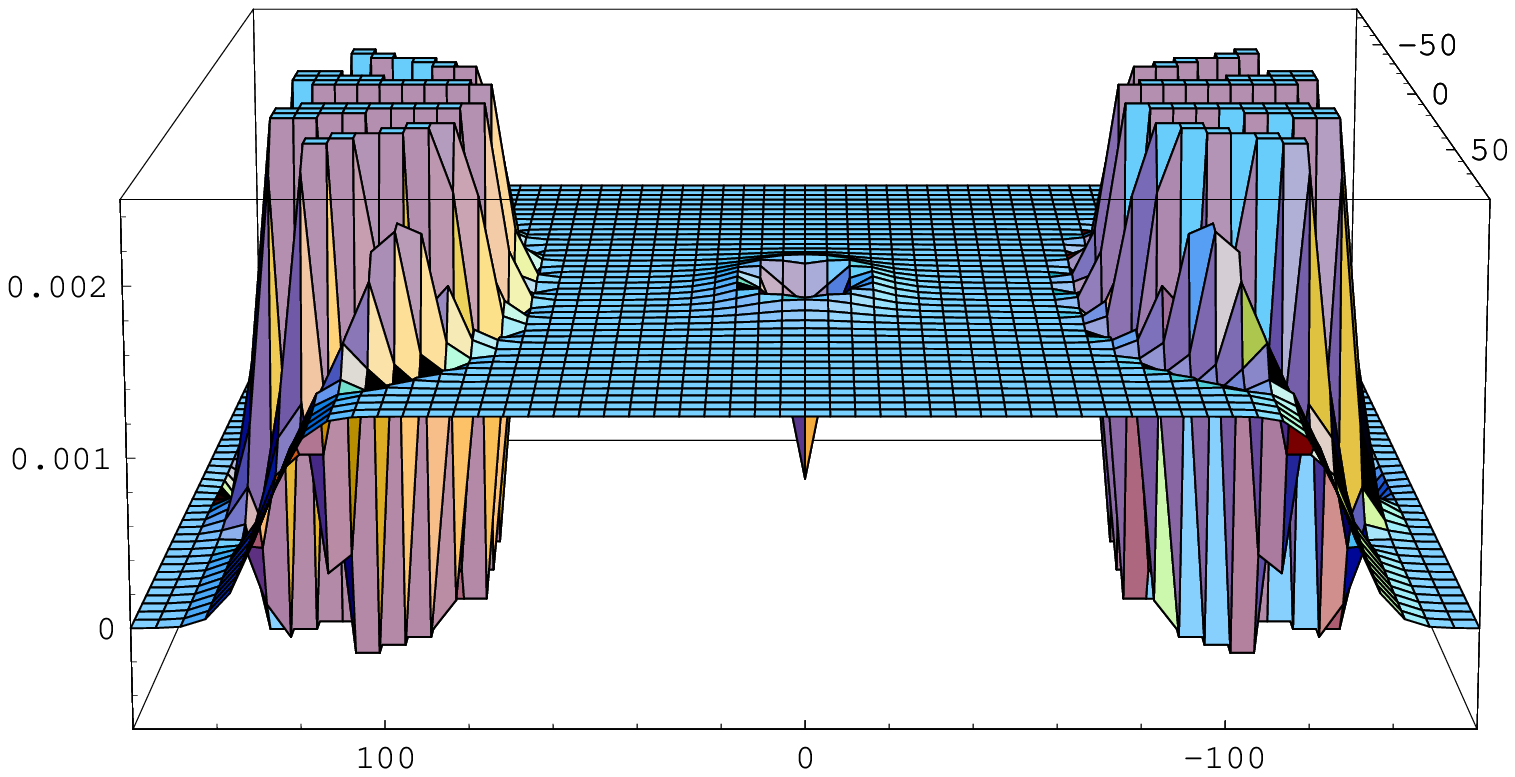
,width=20pc}\epsfig{
figure=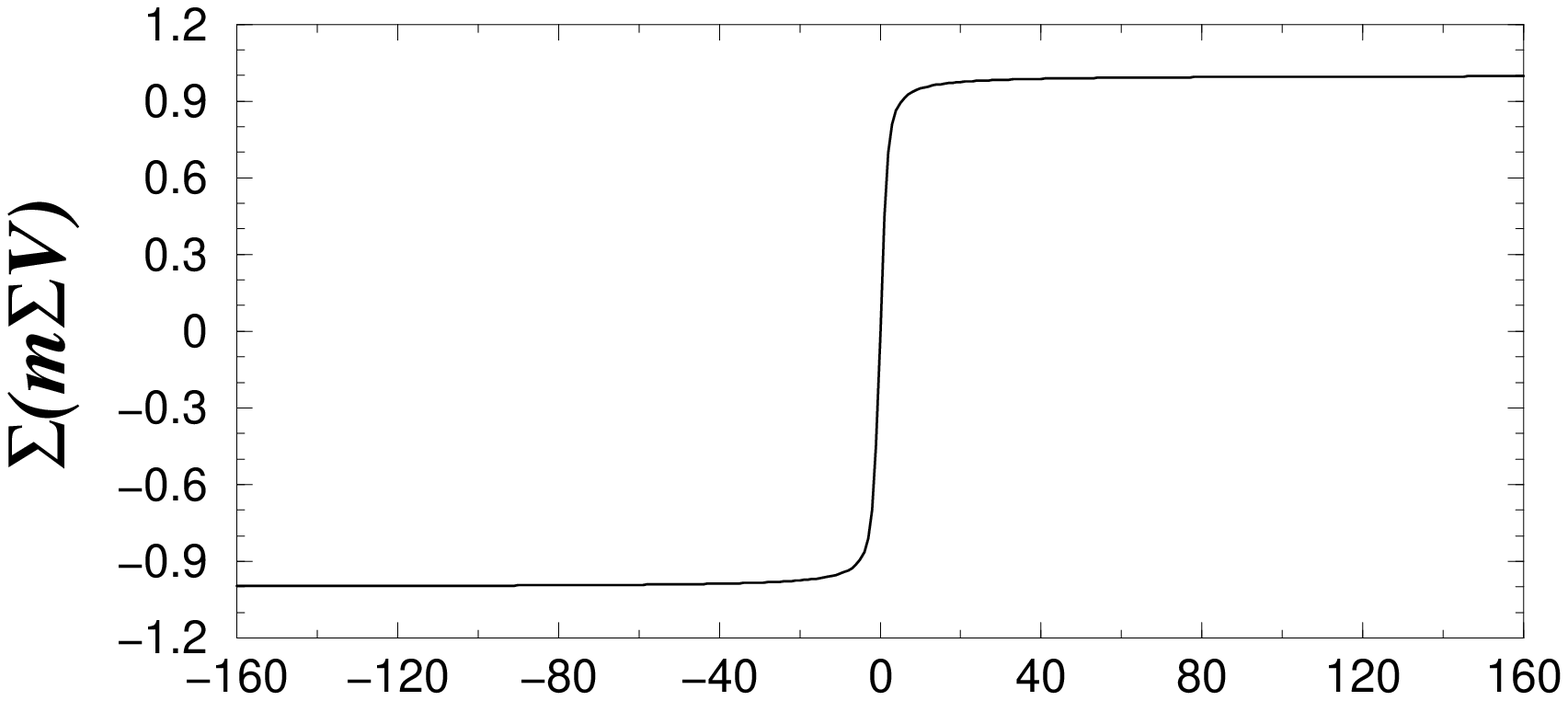
,width=20pc}
\put(-370,20){$\im\ \xi$}
\put(-50,45){$m\Sigma V$}
\put(-480,130){$\re(\rho_{S\ \beta_D=2}^{(N_f=1,0)}(\xi))$}
}
\centerline{\epsfig{
figure=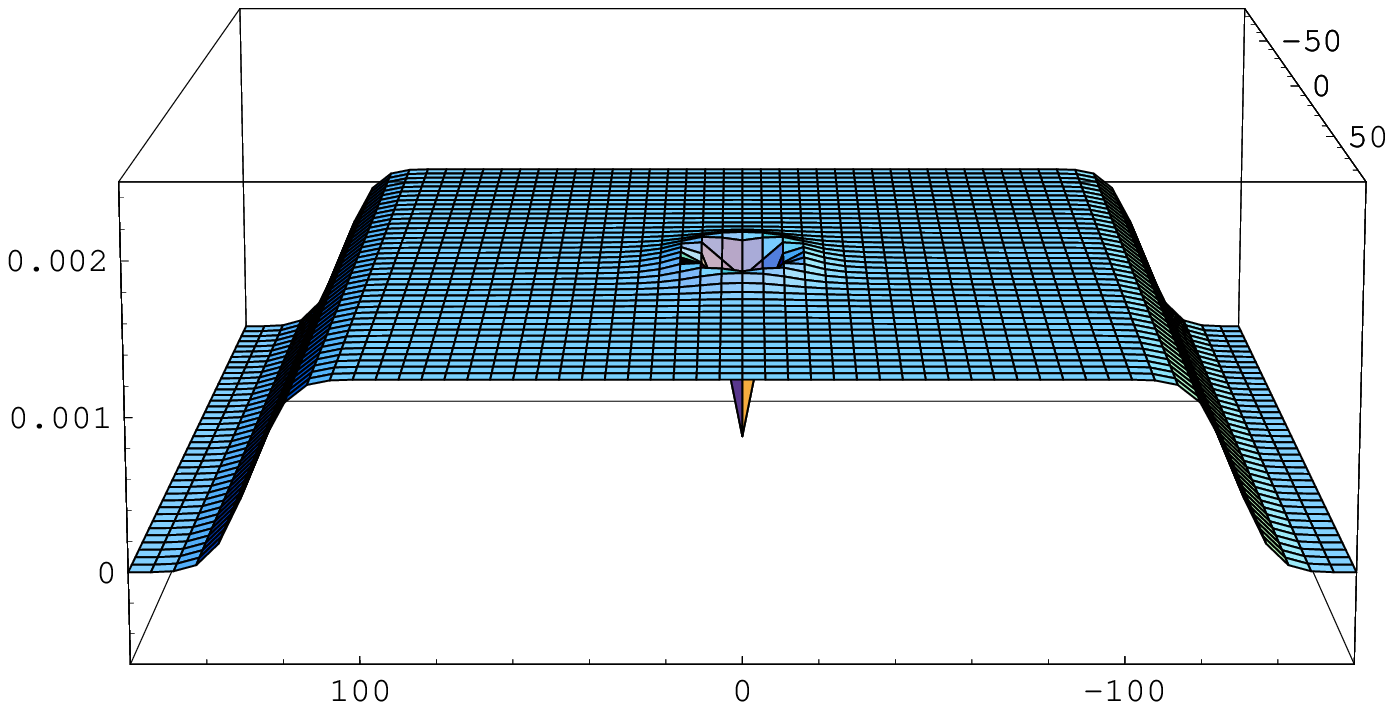
,width=20pc}
\epsfig{
figure=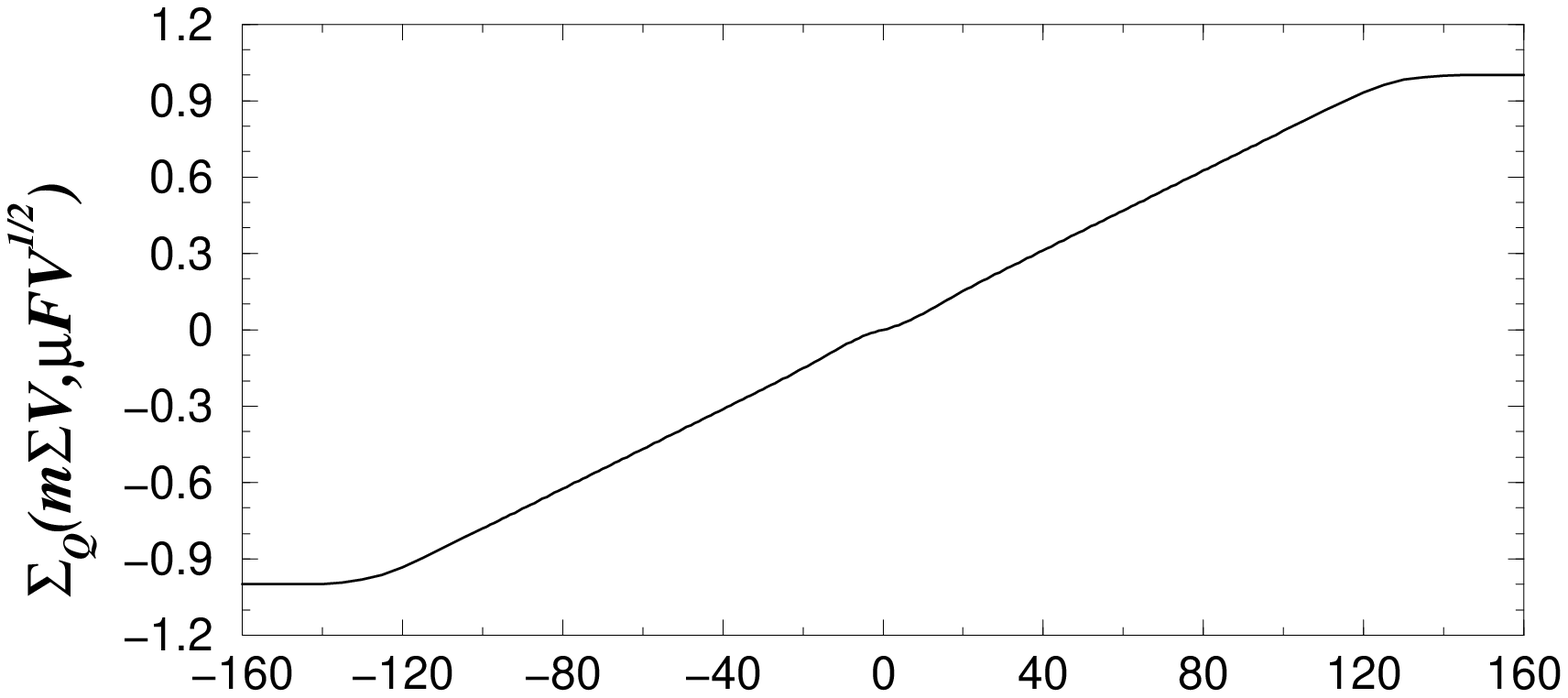
,width=20pc}
\put(-370,20){$\im\ \xi$}
\put(-50,45){$m\Sigma V$}
\put(-480,130){$\rho_{S\ \beta_D=2}^{(0,0)}(\xi)$}
}
\caption{The microscopic spectral density $\rho_S^{(N_f,v)}(\xi)$ 
at large volume, cut 
perpendicular to the real axis and rotated by $\pi/2$:
the real part of the complex valued, unquenched density from MMs
for $N_f=1$ flavour with mass $m\qq V=80$ (upper plots)
vs. the quenched density from $\eps\chi$PT (lower plots), both at rescaled 
$\mu F_\pi\sqrt{V} =8$. The  
resolvent $\Sigma(m\qq V)$ obtained numerically is shown for comparison.
All plots are from the second ref. of \cite{OSV}, with $\Sigma\equiv\qq$.
}
\label{MMsig}
\end{figure}

\begin{figure}[-h]
\centerline{\epsfig{
figure=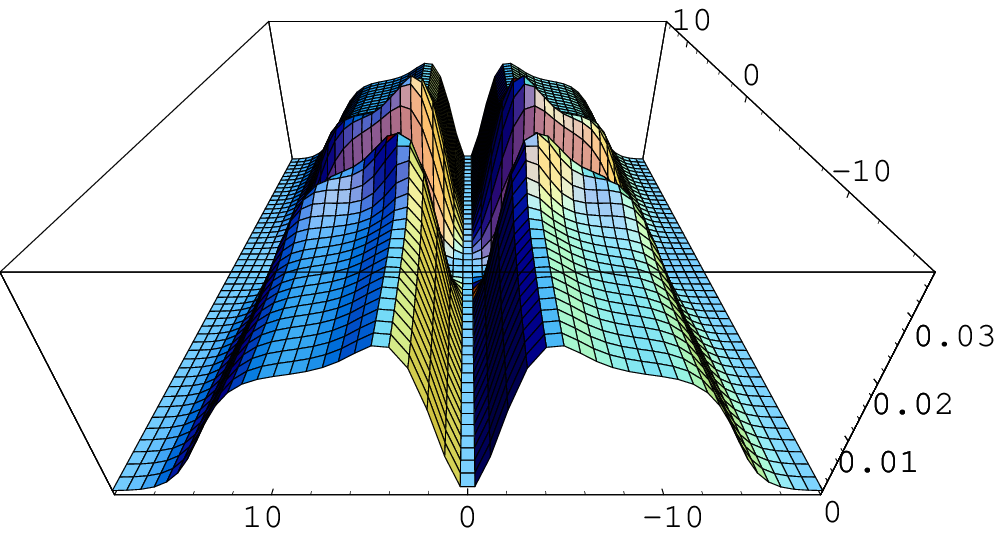
,width=20pc}
\epsfig{figure=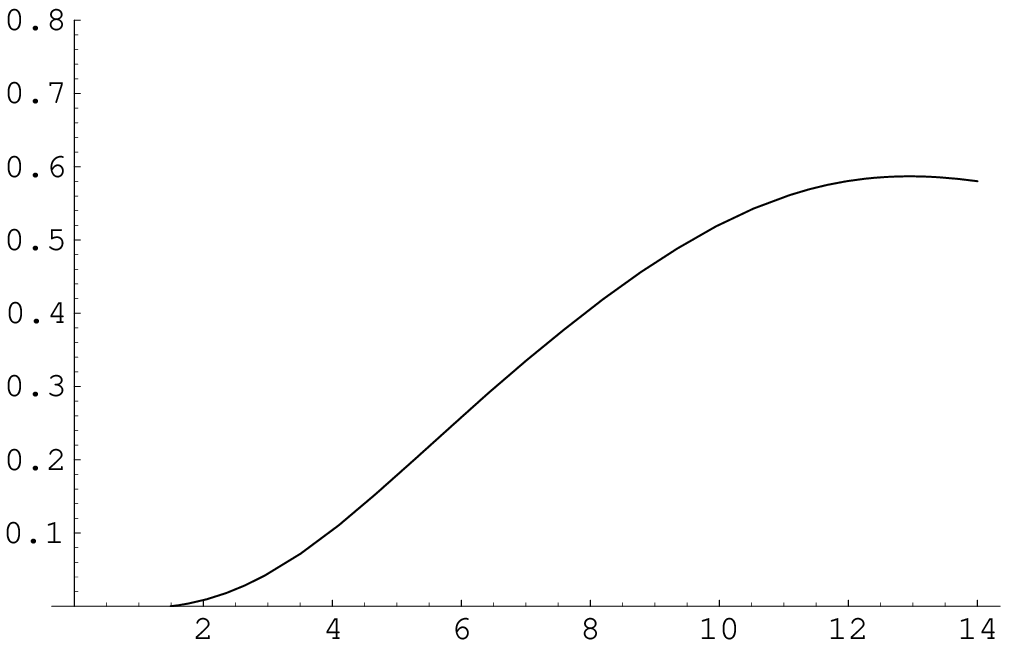
,width=20pc}
\put(-30,25){$m\qq V$}
\put(-240,160){$\Sigma(m\qq V)$}
\put(-300,110){$\re\ \xi$}
\put(-395,140){$\rho_{S\ \beta_D=4}^{(0,0)}(\xi)$}
}
\caption{The same cut as in fig. \ref{MMsig} for the 
quenched density in the $\beta_D=4$ or adjoint QCD class from MMs at rescaled 
  $\mu F_\pi\sqrt{V}=2.5$ and topology $\nu=0$. 
The corresponding resolvent is shown for
  positive arguments only.}
\label{MMsig4}
\end{figure}

To start we recall the relation between the spectral density of the
Dirac operator and spontaneous chiral symmetry breaking  
from field theory, both for zero and non-zero
$\mu$. Throughout this subsection no reference to MM results is made, 
apart from figs. \ref{MMsig} and \ref{MMsig4}. 

Let us begin by defining the spectral density of Dirac operator eigenvalues 
eq. (\ref{defDev}) by 
\beq
\rho_{\Dirac}(z)\ \equiv\ 
\left\langle \sum_k
\delta^2(z_k-z)\right\rangle_{QCD} \  ,
\label{rhoD}
\eeq
where the average is taken over the gauge fields including the 
$N_f$ Dirac determinants. In our conventions the eigenvalues $z_k$ are real at
$\mu=0$. 
In the same way higher order correlation functions can be
defined, such as density-density correlations, etc. . 
In a finite volume on the Lattice 
there is only a finite number of discrete Dirac 
eigenvalues.
When computing integrals at $\lim_{V\to\infty}$ 
with such an operator, such as sum-rules eq. (\ref{sumrule}), special care
has to be taken for UV divergencies, and we refer to
\cite{SternSmilga} for this issue at $\mu=0$. Putting QCD on the Lattice
always provides a natural regularisation here. 
Because of chiral symmetry 
\beq
\{(\Dirac+\mu\gamma_0),\gamma_5\} \ =\ 0\ , 
\eeq
eigenvalues of $(\Dirac+\mu\gamma_0)$
come in pairs $\pm iz_k$ as for each eigenfunction $\varphi_k$ of a
non-zero eigenvalue there is also an eigenfunction $\gamma_5\varphi_k$ with
eigenvalues $-iz_k$, except for $z_k=0$. 
The number of zero eigenvalues relates to the topological
charge, $\nu=|n_L-n_R|$. 

Because of this we can write for the resolvent 
\beq
\Sigma(m)
\equiv \left\langle
\sum_{k}\frac{1}{iz_k+m}\right\rangle_{QCD}
= \left\langle
\sum_{k}{}^\prime\frac{2m}{z_k^2+m^2}\right\rangle_{QCD} \ +\
\langle\nu\rangle\frac{1}{m}\ . 
\eeq
where the second sum $\Sigma^\prime$
goes only over the non-zero eigenvalues with 
$-\pi/2<$Arg$(z_k)\leq\pi/2$, the half plane $\mathbb{C}_+$. 
Here $m$ is an auxiliary mass. The partition function of course depends also 
on the $N_f$ masses $m_f$, see also the discussion after eq. (\ref{rhorep}).
The chiral condensate is then
defined as 
\beq
\qq\equiv -\lim_{m\to0}\lim_{V\to\infty}\Sigma(m)\ ,
\eeq
where the order of limits is important. 
Using the electrostatic analogy $\Sigma(m)$ is the electric field for charges
$iz_k$ on the imaginary axis for $\mu=0$, or on a stripe parallel to it
for $\mu\ne0$. The fact that chiral symmetry is broken trough the chiral
condensate $\qq$ corresponds to a jump of 
$\Sigma(m)$ along this axis, see fig. \ref{MMsig} below. 

\begin{table}[h]
\begin{tabular}{c|c|c}

 & $\mu=0 $ & $\mu\neq 0$\\[2ex]\hline\\[1ex] 

$\Sigma(m)$ 
& $\lim_{n\to0}\frac{1}{n}\partial_m\langle\det(\Dirac+m)^n\rangle_{QCD}$  
& $\lim_{n\to0}\frac{1}{n}\partial_m \langle \det|\Dirac(\mu)+m|^{2n}
\rangle_{QCD}$  
\\[2ex] \hline\\[1ex] 

$\lim_{V\to\infty} \Sigma(m)$ 
& $\int_0^\infty dx\ \rho_{\Dirac}(x)\frac{2m}{x^2+m^2} $
& $\int_{\mathbb{C}_+} d^2z\ \rho_{\Dirac}(z)\frac{2m}{z^2+m^2}$
\\[2ex]\hline\\[1ex] 

$\rho_{\Dirac}(z=x+iy)$ 
& $\lim_{\eps\to0}\left.\frac{1}{2\pi i}[\Sigma(im+\eps)-
  \Sigma(im-\eps)]\right|_{m=x}$ 
& $\left.\partial_{m^*}\Sigma(m)\right|_{m=z}$
\\[2ex]

\end{tabular}
\caption{The resolvent and spectral 
density for zero and non-zero $\mu$. Here $m$ is an auxiliary mass, the
$N_f$-dependence of $\rho_{/\!\!\!\!D}$ and $\Sigma(m)$ has been suppressed.}
\label{resolvents}
\end{table}

The relation between resolvent and spectral density however is different
for $z_k$ real or complex, as summarised in table \ref{resolvents}. 
For zero chemical potential we have 
\beq
\lim_{V\to\infty}\Sigma(m)
=\int_0^\infty dx \rho_{\Dirac}(x)\frac{2m}{x^2+m^2}\ , 
\label{SigmaR}
\eeq
with $x\in\mathbb{R}_+$, assuming that $\langle\nu^2\rangle\sim V$
is suppressed. 
This leads to the Banks-Casher \cite{BC} relation in the chiral limit
\beq
|\qq|= \frac{\pi}{V}\rho_{\Dirac}(0) \ ,
\label{bc}
\eeq
where we have used the representation 
$\lim_{m\to0}\frac{2m}{x^2+m^2}=\pi\delta^{1}(x)$
of the 1-dimensional
delta-function. 
On the right hand side of eq. (\ref{bc})
we have the macroscopic density of Dirac eigenvalues at
zero momentum. 
Its slope has also been computed in \cite{Toublan:1999hi}, 
$\rho_{\Dirac}^\prime(x=0)=
\frac{\qq^2}{16\pi^2F_\pi^4}\frac{(N_f-2)(N_f+\beta_D)}{N_f\beta_D}$,
depending on the symmetry class through the Dyson index. 
Because of eq. (\ref{bc}) the smallest eigenvalues are spaced like $1/V$ (in
contrast to free eigenvalues $\sim1/V^{\frac14}$) and we define a microscopic
density, both for $\mu=0$ and $\mu\neq0$, 
\beq
\rho_S(\xi)\equiv \rho_{\Dirac}(\xi=z/\qq V)/\qq V\ .
\label{rhomicro}
\eeq
This microscopic density can be obtained from eq. (\ref{SigmaR}) by
inverting the integral equation - if $\rho(x)$ is independent of the valence
or source quark mass $m$- 
as the discontinuity of the resolvent: 
\beq
\rho_{\Dirac}(x)= \lim_{\vareps\to0}\frac{1}{2\pi
  i}[\Sigma(ix+\vareps)- \Sigma(ix+\vareps)] \ .
\label{rhosigR}
\eeq

On the other hand for non-zero $\mu$ we have  
\beq
\lim_{V\to\infty}\Sigma(m)
=\int_{\mathbb{C}_+} d^2z\  \rho_{\Dirac}(z)\frac{2m}{z^2+m^2}\ . 
\label{SigmaC}
\eeq
Here the chiral limit $m\to0$
does {\it not} produce the 2-dimensional delta-function 
$\pi\delta^{2}(z)=\partial_{z^*}\frac{1}{z}$.
The density - resolvent relation in the complex plane instead reads:
\beq
\rho_{\Dirac}(z) = \frac{1}{\pi}\partial_{z^*}\Sigma(z) 
\label{rhosigC}
\eeq
that is outside the support the resolvent is holomorphic and inside it depends
on the complex conjugate $z^*$. For more details we refer to \cite{Zabrodin}. 
When adding a delta-function $\delta^1(y)$ to eq. (\ref{rhosigR}) we could
also write it in the form of eq. (\ref{rhosigC}).

So far we have only defined the microscopic density and the resolvent, and
related the two. The question remains of how to compute them from field
theory. We give two possibilities, replicas and supersymmetry, where again we
have to distinguish $\mu=0$ and $\mu\neq0$. Here extra fermionic and
bosonic valance or source quarks are added to generate the resolvent. 

In the supersymmetric approach at $\mu=0$ we can write 
\beq
\Sigma(m)=\left.\partial_{m_1}
\left\langle \frac{\det(\Dirac+m_1)}{\det(\Dirac+m_2)}
\right\rangle_{QCD}\right|_{m_1=m_2}\ ,
\label{susy}
\eeq
using the fact that $\det(\Dirac+m)=\exp[\Tr\log(\Dirac+m)]$. Here $m_1$ 
and $m_2$ are 1 extra fermionic and 1 bosonic quark, respectively. As long as 
$m_2$ has a small imaginary part, the generating functional containing the
ratio of determinants in addition to the $N_f$ flavours is well defined. 
In the limit of $\eps\chi$PT we obtain an integral 
over the super Riemannian manifold $Gl(N_f+1|1)$ generalising eq. 
(\ref{ZchPT}), and we refer to \cite{DOTV} for details. 

For $\mu\neq0$ the inverse determinant can no longer be regularised by an
imaginary part, as the spectrum of $\Dirac(\mu)$ now extends into the complex
plane. Here the so-called Hermitisation technique is used as follows. Defining
\beq
{\cal Z}^{(N_f+2|2,\nu)}(m_1,m_2;\kappa)\equiv
\left\langle \frac{\det[(\Dirac(\mu)+m_1)(\Dirac(\mu)+m_1)^\dag+\kappa^2]}
{\det[(\Dirac(\mu)+m_2)(\Dirac(\mu)+m_2)^\dag+\kappa^2]}
\right\rangle_{QCD},
\eeq
we can obtain the regularised resolvent by differentiation and setting
arguments equal as before, and then the density through eq. (\ref{rhosigC}):
\beq
\rho_{\Dirac}(z)
=\left.-\frac{1}{\pi}\lim_{\kappa\to0}\partial_{m_1^*}\lim_{m_2\to m_1}
\partial_{m_2}{\cal Z}^{(N_f+2|2,\nu)}(m_1,m_2;\kappa)\right|_{m_1=z}
\eeq
For a detailed calculation in a non-chiral MM framework we refer to
\cite{FKS2}. 

When using replicas the logarithm of the determinant is generated by inserting
$n$ fermionic determinants of degenerate mass (replicas) into the partition
function, leading to the resolvent as follows:
\beq
\Sigma(m)= \lim_{n\to0}\frac1n \partial_m\left\langle
\det(\Dirac+m)^n\right\rangle_{QCD}.
\label{sigrep}
\eeq
We will not comment here on the subtlety of analytically continuing $n\to0$,
as we will not use replicas in our computations. 
Eq. (\ref{sigrep}) is only valid for real eigenvalues at $\mu=0$. For 
$\mu\neq0$ the appropriate replica limit includes complex conjugated quarks
\beq
\Sigma(m)= \lim_{n\to0}\frac1n \partial_m\left\langle
\det[(\Dirac(\mu)+m)^n(\Dirac(\mu)+m)^{\dag\,n}]\right\rangle_{QCD}.
\label{sigrepmu}
\eeq
This was pointed out by Stephanov in \cite{Steph} to explain the failure of the
quenched approximation in LGT at $\mu\neq0$. Using MM eq. (\ref{ZMM1})
he showed that the critical potential is proportional to the pion mass, 
$\mu_c\sim m_\pi$, due to the presence of conjugated quarks in the correct
replica limit eq. (\ref{sigrepmu}). 
This has to be compared with a third of the lightest Baryon
mass, $\mu_c^2\sim m_B/3$, expected for unquenched QCD. 
Replicas for non-hermitian MMs and $\eps\chi$PT 
have been introduced in \cite{EK2,SplitVerb1} 
to compute correlation functions, using a relation to
integrable hierarchies to make the replica limit exact.

We close this section by giving cuts through the microscopic 
densities in the large volume limit as obtained from MMs 
in the following sections. This is to illustrate the different symmetry
classes and the effect of unquenching.
Parallel to that the
resulting resolvent is shown.  

As was pointed out in \cite{OSV} the strong oscillations in the complex valued 
unquenched spectral density are made responsible for chiral symmetry breaking,
with a discontinuous resolvent. For comparison the quenched density and
its continuous resolvent are shown as well in fig. \ref{MMsig}. 
Here the chiral condensate is vanishing in the chiral
limit and thus not responsible for symmetry breaking. It is the diquark
condensate  breaking chiral symmetry in this case.

We also give a cut through the quenched density for adjoint
QCD\footnote{Because of slow convergence for 
the double integral in density eq. (\ref{rhoQsu2}) we give smaller
volumes compared to fig. \ref{MMsig}.}. The resulting resolvent is
again smooth as in quenched QCD above. 
This implies that the chiral condensate $\qq$ is rotated to zero, 
with chiral symmetry being broken 
through a diquark condensate.


\subsection{Complex Dirac eigenvalue correlation functions from MMs}

In this section we will show how to derive complex eigenvalue correlation
functions from MMs using orthogonal polynomials in the complex plane. 
The reason we choose this technique is that it is simple and carries over
with little technical modifications to symplectic MMs or adjoint QCD in the
next section. The alternative techniques are replicas using the Toda
lattice, which has been successful only in the
$\beta_{D}=2$ class, and supersymmetry. 
They will be mentioned only very briefly.

Our starting point is 
the two-MM version of the partition function eq. (\ref{ZMM1}) 
based on eq. (\ref{D2matrix}), that 
was first introduced in \cite{James} (see also \cite{AOSV} for details)
\beq
{\cal Z}^{(N_f,\nu)}= \int d\Phi d\Psi_{N\times(N+\nu)} 
\prod_{f=1}^{N_f} \det\left(
\begin{array}{cc}
m_f& \Phi+\mu\Psi\\
-\Phi^\dag+\mu\Psi^\dag & m_f\\
\end{array}
\right)
e^{-N\Tr(\Phi\Phi^\dag+\Psi\Psi^\dag)}.
\label{ZMM2}
\eeq
In order to obtain a complex eigenvalue model a Schur decomposition for the 
following  
complex matrices is performed, $C=\Phi+\mu\Psi=U(X+R)V$ and 
$D=-\Phi^\dag+\mu\Psi=V^\dag(Y+S)U^\dag$. 
Here $X$ and $Y$ are
diagonal matrices of complex eigenvalues $x_k$ and $y_k$. $R$ and $S$ are
complex upper triangular matrices, $U$ and $V$ are unitary. 
The Jacobian is proportional to
$\Delta_N(\{z^2\})=\prod_{j>k}^N(z_j^2-z_k^2)$, the Vandermonde
determinant. Because of the Gaussian measure the
matrices $R$ and $S$ decouple and can be integrated out. A change of variables
to the complex Dirac eigenvalues $z_k=x_ky_k$ of the matrix $CD$
is made next,
and finally one of the 2 complex eigenvalue sets, $\{x_k\}$ is integrated out.
This last step is
transforming part of the Gaussian measure into a $K$-Bessel function:
\beq
{\cal Z}^{(N_f,\nu)}= \int_{\mathbb{C}} \prod_{j=1}^N d^2z_j 
\prod_{f=1}^{N_f}m_f^\nu
(z_j^2+m_f^2)\ 
|z_j|^{2\nu+2} 
K_\nu\left(\frac{N(1+\mu^2)}{2\mu^2}|z_j|^2\right)
e^{\frac{N(1-\mu^2)}{4\mu^2}(z_j^2+z_j^{*\,2})}
|\Delta_N(\{z^2\})|^2 \ .
\label{ZMM2ev}
\eeq
Here we have omitted all normalisation constants as they will drop out later
in correlation functions. 
An approximation to this complex eigenvalue model was first introduced in
\cite{A03} replacing $K_\nu(x)\to\sqrt{\pi/(2x)}\ e^{-x}$ by its asymptotic 
value. The two models coincide at $\nu=\pm\frac12$.

The concept of orthogonal polynomials (both on $\mathbb{R}$ and $\mathbb{C}$)
to compute eigenvalue correlation functions relies on the following
properties. 
Given a weight function $w(z,z^*)$ on $\mathbb{C}$ such that all moments are
finite we can define a set of monic polynomials $P_k(z)=z^k+{\cal O}(z^{k-1})$:
\beq
\int_{\mathbb{C}}  d^2z\ w(z,z^*) P_k(z)  P_l(z)^* = \delta_{kl}\ h_k
\label{OPdef}
\eeq
where $ h_k$ are the squared norms. For real positive weights the coefficients
of the polynomials can easily be seen to be real, $P_l(z)^*= P_l(z^*)$. Using 
invariance properties of determinants we can write the following identity:
\beq
\Delta_N(\{z^2\})\Delta_N(\{z^{*\,2}\})\ =\ 
\det_{1\leq k,l\leq N}[P_{k-1}(z_l)]\det_{1\leq k,l\leq N}[P_{k-1}(z_l^*)]\ =\ 
\prod_{j=0}^{N-1}h_j^{-1}\ \det_{1\leq k,l\leq N}[K_N(z_k,z_l^*)]
\label{Vanderid}
\eeq
where we have defined the kernel of orthogonal polynomials 
\beq
K_N(z_k,z_l^*)\ =\ \sum_{j=0}^{N-1} h_j^{-1}P_j(z_k)P_j(z_l^*) \ . 
\label{Kdef}
\eeq
It enjoys the following self contraction and normalisation property:
\beqn
\int_{\mathbb{C}}  d^2z\ w(z,z^*)K_N(z_k,z^*)K_N(z,z_j^*) &=& 
K_N(z_k,z_j^*)\nn\\
\int_{\mathbb{C}}  d^2z_k\ w(z_k,z_k^*)K_N(z_k,z_l^*) &=& 
\left\{ 
\begin{array}{lll}
1 & \mbox{if} &k\neq l\\
N & \mbox{if} & k=l
\end{array}
\right.
\ . 
\eeqn
For such kernels the following Theorem by Dyson (Theorem 5.2.1 in
\cite{Mehta2}) holds 
\beq 
\int_{\mathbb{C}} d^2z_n \ w(z_n,z_n^*)\det_{1 \leq i,j \leq n} 
[ K_N(z_i,z_j^*)]  \ =\ (N-n+1)
\det_{1 \leq i,j \leq n-1}[ K_{N}(z_i,z^*_j)] \ .
\label{MTh}
\eeq
Choosing the polynomials orthogonal with respect to the weight 
\beq
w^{(N_f,\nu)}(z,z^*) \equiv \prod_{f=1}^{N_f}
(z^2+m_f^2)\ 
|z|^{2\nu+2} 
K_\nu\left(\frac{N(1+\mu^2)}{2\mu^2}|z|^2\right)
\exp\left(\frac{N(1-\mu^2)}{4\mu^2}(z^2+z^{*\,2})\right),
\label{weight}
\eeq
in eq. (\ref{ZMM2ev}) we can apply the Dyson Theorem 
to compute all complex eigenvalues
$k$-point correlation functions defined as 
\beqn
R^{(N_f,\nu)}(z_1,\ldots,z_k) &\equiv& 
\frac{N!}{(N-k)!} \frac{1}{{\cal Z}^{(N_f,\nu)}}\int_{\mathbb{C}} 
\prod_{j=k+1}^N
d^2z_j \prod_{l=1}^N w^{(N_f,\nu)} (z_l,z_l^*)|\Delta_N(\{z^2\})|^2
\label{Rkdef}\\
&=& \prod_{l=1}^k w^{(N_f,\nu)}(z_l,z_l^*)
\det_{1 \leq i,j \leq k}[ K_{N}(z_i,z^*_j)] \ .
\label{Rksol}
\eeqn
It expresses the original 
$(N-k)$-fold integral as a determinant of size $k\times k$
only. Thus in the large-$N$ limit we only have to compute the asymptotic of
the kernel $ K_N(z_i,z_j^*)$. 
Note that for brevity we have suppressed 
the dependence on the complex conjugated arguments 
$z_1^*,\ldots,z_k^*$ and on the masses in the functions $R^{(N_f,\nu)}$.

Let us give an example. For the quenched weight, $w^{(0,\nu)}(z,z^*)$ with
$N_f=0$ in eq. (\ref{weight}), the orthogonal polynomials are given by
Laguerre polynomials in the complex plane \cite{James}
\beq
P_k(z)\ =\ (-)^k\frac{k!}{N^k}(1-\mu^2)^kL^\nu_k\left(
\frac{Nz^2}{1-\mu^2}\right)\ .
\label{OPNf0}
\eeq
For a proof of their orthogonality we refer to appendix A in \cite{A05}.
The corresponding kernel is given by 
\beq
K_N(z_k,z_l^*)\ =\
w(z_k,z_k^*)^\frac12 w(z_l,z_l^*)^\frac12
\frac{N^{\nu+2}}{\pi\mu^2(1+\mu^2)^{\nu}}
\sum_{j=0}^{N-1}
\frac{j!}{(j+\nu)!}
L^\nu_j\left(\frac{Nz^2}{1-\mu^2}\right)
L^\nu_j\left(\frac{Nz^{*\,2}}{1-\mu^2}\right).
\label{KNf0}
\eeq
The orthogonal polynomials for weights including $N_f>0$ flavours 
and their kernel can
be expressed in terms of the quenched quantities eqs. (\ref{OPNf0}) and
(\ref{KNf0}) using the technique of bi-orthogonal polynomials
\cite{bergere,James}, or ordinary orthogonal polynomials \cite{AV} in the case
of a positive definite weight (corresponding to $N_f>0$ phase quenched
flavours). From this all correlation functions follow for $N_f\neq0$, and 
for most explicit expressions we refer to \cite{AOSV}\footnote{Note the
  different convention there, rotating eigenvalues by $\pi/2$: $iz\to z$.}, 
see also
eq. (\ref{rhoN_f}) below. The partition functions with $N_f$
flavours themselves can
also be expressed in terms of the quenched polynomials and kernel \cite{AV},
resulting into eq. (\ref{Zmn}) in the large-$N$ limit.

As pointed out in subsection \ref{FTrel} the MM only becomes equal to the
chiral Lagrangian in the following large-$N$ limit. We rescale complex
eigenvalues, masses and chemical potential as in
eq. (\ref{scale})
\beq
\xi_j\ =\ 2Nz_j\ ,\ \ \eta_f\ =\ 2Nm_f\ ,\ \ \mbox{and}\ \ 
\alpha^2\ =\ 2N\mu^2\ ,
\label{microscale}
\eeq
which is denoted by the microscopic limit at weak non-Hermiticity. The 
concept of weak non-Hermiticity 
introduced in \cite{FKS} for non-chiral MMs implies that $\mu^2$ vanishes
at a rate $1/N$. There exists a different limit called strong non-Hermiticity
with a different scaling for the eigenvalues, 
keeping $\mu^2={\cal O}(1)$. This leads out of the
domain of the chiral Lagrangian, and we refer to \cite{AFV} for details.

\begin{figure}[-h]
\centerline{\epsfig{figure=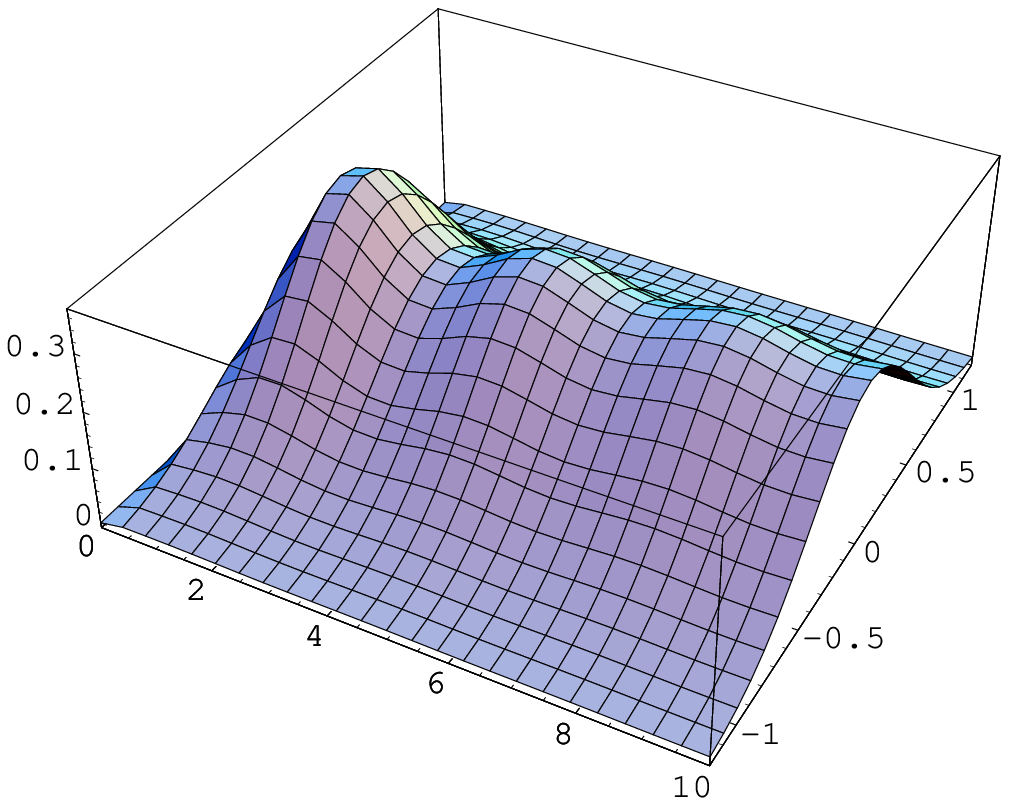,width=20pc}
\put(-160,20){$\re\ \xi$}
\put(-30,45){$\im\ \xi$}
\put(-240,140){$\rho_S^{(0,0)}(\xi)$}
\put(80,20){$\re\ \xi$}
\put(215,40){$\im\ \xi$}
\put(0,140){$\rho_S^{(0,2)}(\xi)$}
\epsfig{figure=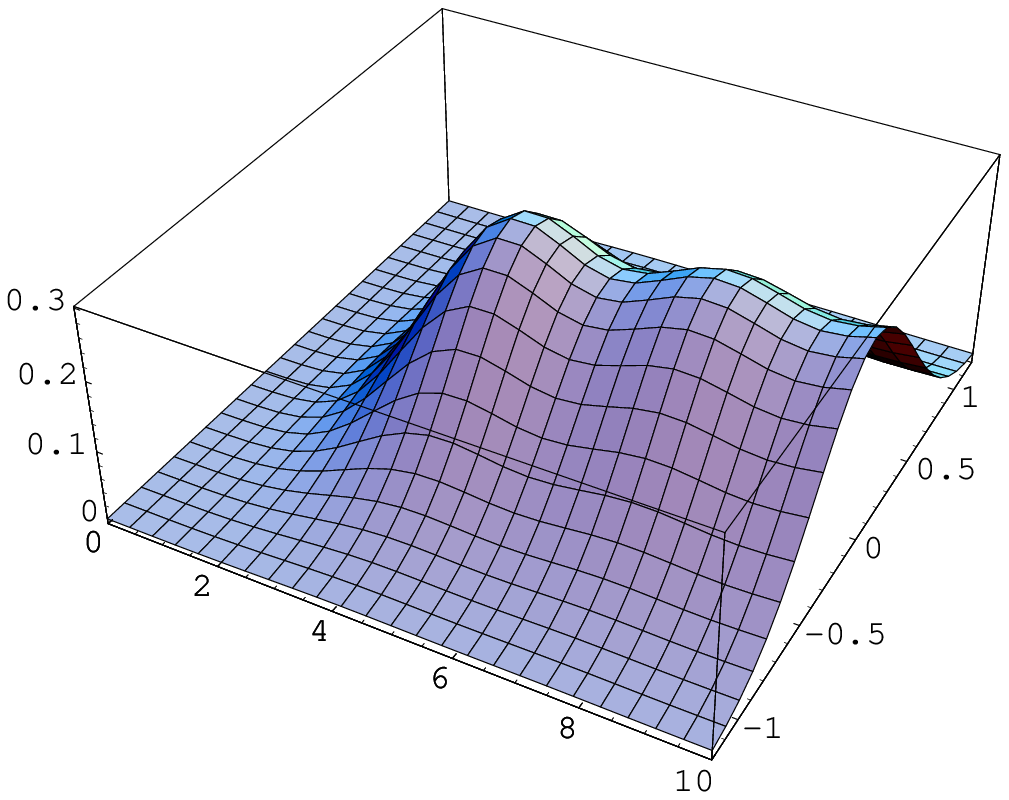,width=20pc}}
\centerline{\epsfig{figure=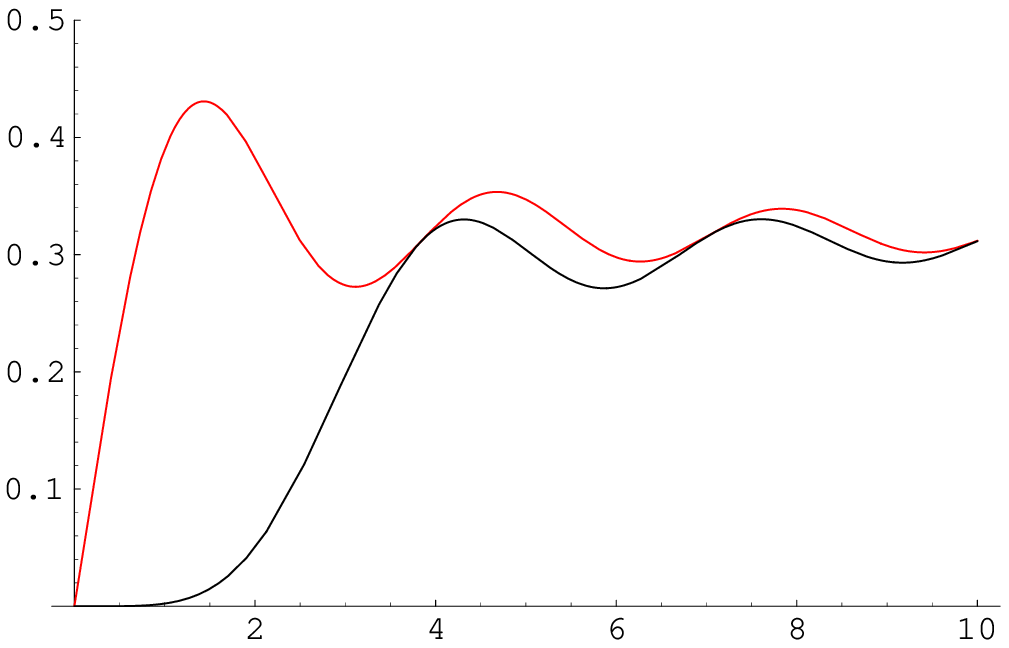,width=20pc}
\put(10,0){$\xi_x$}
\put(-240,160){$\rho_S^{(0,\nu)}(\xi_x)$}
\put(80,20){$\re\ \xi$}
\put(215,40){$\im\ \xi$}
\put(0,140){$\rho_S^{(0,0)}(\xi)$}
\epsfig{figure=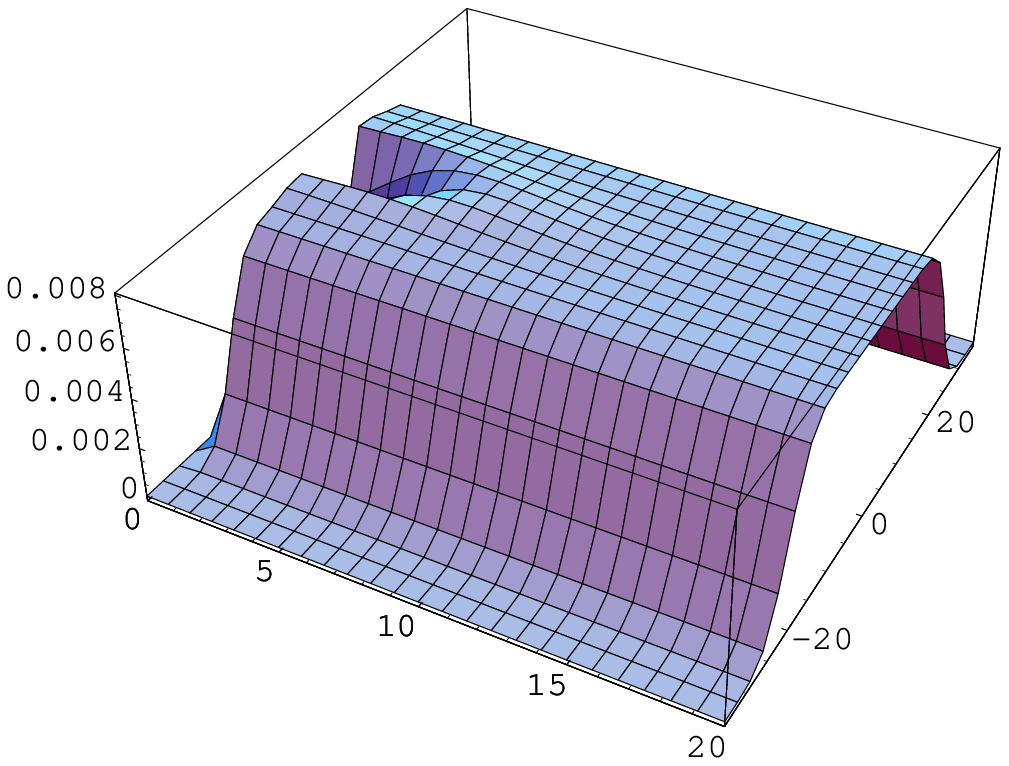,width=20pc}}
\caption{
The quenched microscopic density eq. (\ref{rhoQ})
at $\al=0.4$ for $\nu=0$ (top left) and $\nu=2$ 
(top right), and at $\al=3.28$ for $\nu=0$ (bottom right).
For comparison we also show the density of real eigenvalues
eq. (\ref{rhoreal}) (bottom left) 
for $\nu=0$ (left curve) and $\nu=2$ (right curve), respectively.
Note that in the complex plane the normalisation becomes a 
delta-function in the Hermitian limit $\al\to0$.
}
\label{weakplot}
\end{figure}

The microscopic spectral density in the large-$N$ limit eq. (\ref{microscale})
is defined as follows
\beq
\rho_S^{(N_f,\nu)}(\xi)\ =\ 
\lim_{N\to\infty}\frac{1}{2N}R^{(N_f,\nu)}\left(z=\frac{\xi}{2N}\right)\ .
\label{rhomicdef}
\eeq
All higher order correlations are rescaled accordingly. The quenched
microscopic density obtained from 
eq. (\ref{KNf0}) thus reads
\beq
\rho_S^{(0,\nu)}(\xi)= 
\frac{1}{2\pi\alpha^2}|\xi|^2 K_\nu\left(\frac{|\xi|^2}{4\alpha^2}\right)
e^{\frac{\xi^2+\xi^{*\,2}}{8\alpha^2}} 
\int_0^1dt\ t\ e^{-2\alpha^2t^2}J_\nu(t\xi)J_\nu(t\xi^*) \ .
\label{rhoQ}
\eeq
In the Hermitian limit $\alpha\to0$ it reduces to the density of 
real eigenvalues $\rho_S^{(0,\nu)}(\xi_x)$
\cite{SV93} times $\delta^1(\Im m(\xi))$ 
\beq
\rho_S^{(0,\nu)}(\xi_x)\ \sim \
|\xi_x|
\int_0^1dt\ t\ J_\nu(t\xi_x)^2 \ =\ \frac{\xi_x}{2}(J_\nu(\xi_x)^2-
J_{\nu-1}(\xi_x)J_{\nu+1}(\xi_x))\ ,
\label{rhoreal}
\eeq
where $\Re e(\xi)=\xi_x$ denotes the real part.
Fig. \ref{weakplot} compares the two functions 
for different values of $\nu$ and $\alpha$. 
For small $\alpha$ the density in the complex plane eq. (\ref{rhoQ}) 
is very similar to the real one eq. (\ref{rhoreal}), times a Gaussian decay
in imaginary direction. Introducing exact zero eigenvalues by increasing $\nu$
pushes the density away from the origin through level repulsion. For
increasing $\alpha$ the density quickly becomes a constant along a strip as
from mean field, washing out the oscillations. 

For the microscopic partition function with $N_f$ flavours we obtain
eq. (\ref{ZLS}) which is $\mu$-independent for $B=\id$ as it should. 
We can now check 
that the same sum rule eq. (\ref{sumrule}) indeed follows
from the quenched density eq. (\ref{rhoQ}): 
\beq
\frac{1}{4\nu}\ =\ \int_\mathbb{C} d^2\xi\
\frac{1}{\xi^2}\ \rho_S^{(0,\nu)}(\xi)  
\ \ \mbox{for} \ \ \nu\geq1 \ ,
\eeq
which we have confirmed 
numerically\footnote{I am indebted to Leonid Shifrin for
  his help with Mathematica.}  up to $\nu=5$ for two different values of
$\alpha<1$.  
Thus despite that for $\mu\neq 0$ the eigenvalues spread  
into the complex plane the sum rule remains unchanged. 
An analytic check, in particular of 
the unquenched, massive sum rules \cite{poulmass} using the
complex valued densities for $N_f\geq1$ 
given below would be highly desirable as well. 
For $\mu=0$ the Leutwyler-Smilga sum rule eq. (\ref{sumrule}) was shown
analytically to follow from the real density eq. (\ref{rhoreal}) \cite{VZ}.

Before turning to more flavours let us comment on other methods of deriving
complex eigenvalue correlations and their relation to field theory. 
As was pointed out in subsect. \ref{XSBD} 
the spectral density can also be derived using replicas \cite{SplitVerb1}
\beq
\rho_S^{(N_f,\nu)}(\xi) \ =\ \lim_{n\to0}\frac{1}{n\pi}
\partial_{\xi^*}\partial_{\xi}\log[ {\cal Z}^{(N_f+2n,\nu)}(\{\eta\},
\{i\xi,i\xi^*\})]\ ,
\label{rhorep}
\eeq 
where the first derivative generates the resolvent. 
Using replicas 
for complex eigenvalues (see table \ref{resolvents})
we have introduced $n$ pairs\footnote{Usually a different notation is used,
  counting only the number of complex conjugated pairs $n$ in the
  superscript.}  of complex conjugate quarks of masses $i\xi$ and
$i\xi^*$ respectively, having opposite
$\mu$. Thus the corresponding partition functions eq. (\ref{Zmn}) now depend
on $\mu$. It was shown in \cite{SplitVerb1} that such partition functions obey 
the Toda Lattice equation 
\beq
\frac{1}{n\pi}\xi^*\partial_{\xi^*}\xi
\partial_{\xi}{\cal Z}^{(N_f+2n,\nu)}(\{\eta\},\{\xi,\xi^*\})\ =\ 
\frac12 (\xi\xi^*)^2
\frac{{\cal Z}^{(N_f+2n+2,\nu)}(\{\eta\},\{\xi,\xi^*\})
{\cal Z}^{(N_f+2n-2,\nu)}(\{\eta\},\{\xi,\xi^*\})}
{({\cal    Z}^{(N_f+2n,\nu)}(\{\eta\})(\{\eta\}))^2} \ ,
\label{toda}
\eeq
leading to 
\beq
\rho_S^{(N_f,\nu)}(\xi) \ =\ 
\frac12 \xi\xi^*
\frac{{\cal Z}^{(N_f+2,\nu)}(\{\eta\},\{i\xi,i\xi^*\})
{\cal    Z}^{(N_f-2,\nu)}(\{\eta\},\{i\xi,i\xi^*\})}
{({\cal Z}^{(N_f,\nu)}(\{\eta\}))^2}\ .
\label{rhotoda}
\eeq
Here the minus 2 means that this partition function contains 2 conjugate
bosons instead of fermions. In \cite{Splittorff:2006uu} 
the corresponding bosonic group
integral given by eq. (\ref{ZchPTboson}) 
was computed for $N_f=0$, thus deriving the quenched density from
eq. (\ref{rhotoda}) entirely from $\eps\chi$PT, without using knowledge
from MMs. For $N_f\geq1$ this result is still lacking, and using MMs we know
that the corresponding group integral will factorise:
${\cal Z}^{(N_f-2,\nu)}(\{\eta\},\{\xi,\xi^*\})\sim 
{\cal Z}^{(N_f,\nu)}(\{\eta\}){\cal Z}^{(-2,\nu)}(\{\xi,\xi^*\})$.
The bosonic partition function has to be regularised and is proportional to
the weight function eq. (\ref{weight}) to leading order.
Using this as an input the equivalence of the Toda lattice and 
the orthogonal polynomials approach in MMs has been shown in \cite{AOSV}, see
eq. (\ref{rhoN_f}).  

Another method to compute the spectral density is supersymmetry. Here the
additional inverse determinants are introduced as
Gaussian integrals over bosonic Grassmann variables, see the discussion after
eq. (\ref{susy}), 
and then treated along
the lines similar to subsect. \ref{FTrel}. We refer to \cite{SplitVerb3} 
where the equivalence of the supersymmetric and the Toda lattice
approach was shown. 

After comparing different techniques of solving MMs we would like to comment
on the concept of universality. 
Sometimes the fact that instead of using a field theory such as $\eps\chi$PT to
compute correlation functions one can use a simpler, chiral random MM, is
called universality. As we have seen in sect. \ref{FTrel}
these two are equivalent on the level of partition functions. 

Here, we would like speak about MM universality, meaning that in the large-$N$
limit different MMs, Gaussian or non-Gaussian, containing 1 or 2 random
matrices, lead to the very same eigenvalue correlation functions. 
What is know about universality for MM with complex eigenvalues? So far we
know that the Gaussian one-MM  eq. (\ref{ZMM1}) and the Gaussian 
two-MM eq. (\ref{ZMM2}) have the same partition functions for any number of
$N_f$ fermions with chemical potentials $\pm\mu$, as well as the same
quenched spectral density. Higher order and unquenched 
correlation functions have been obtained only in the second model. 

At present little is known about the large-$N$ limit of
non-Gaussian chiral MMs with complex eigenvalues. The determinantal
\cite{A03,James,AOSV}
or Pfaffian structure \cite{A05,ABa} of the correlations functions 
remain the same for non-Gaussian weights at finite-$N$, and from
heuristic universality results \cite{A02} 
for the non-chiral complex ensemble at $\beta_D=2$ we expect that they are also
universal. 
\begin{figure}[-h]
\centerline{\epsfig{figure=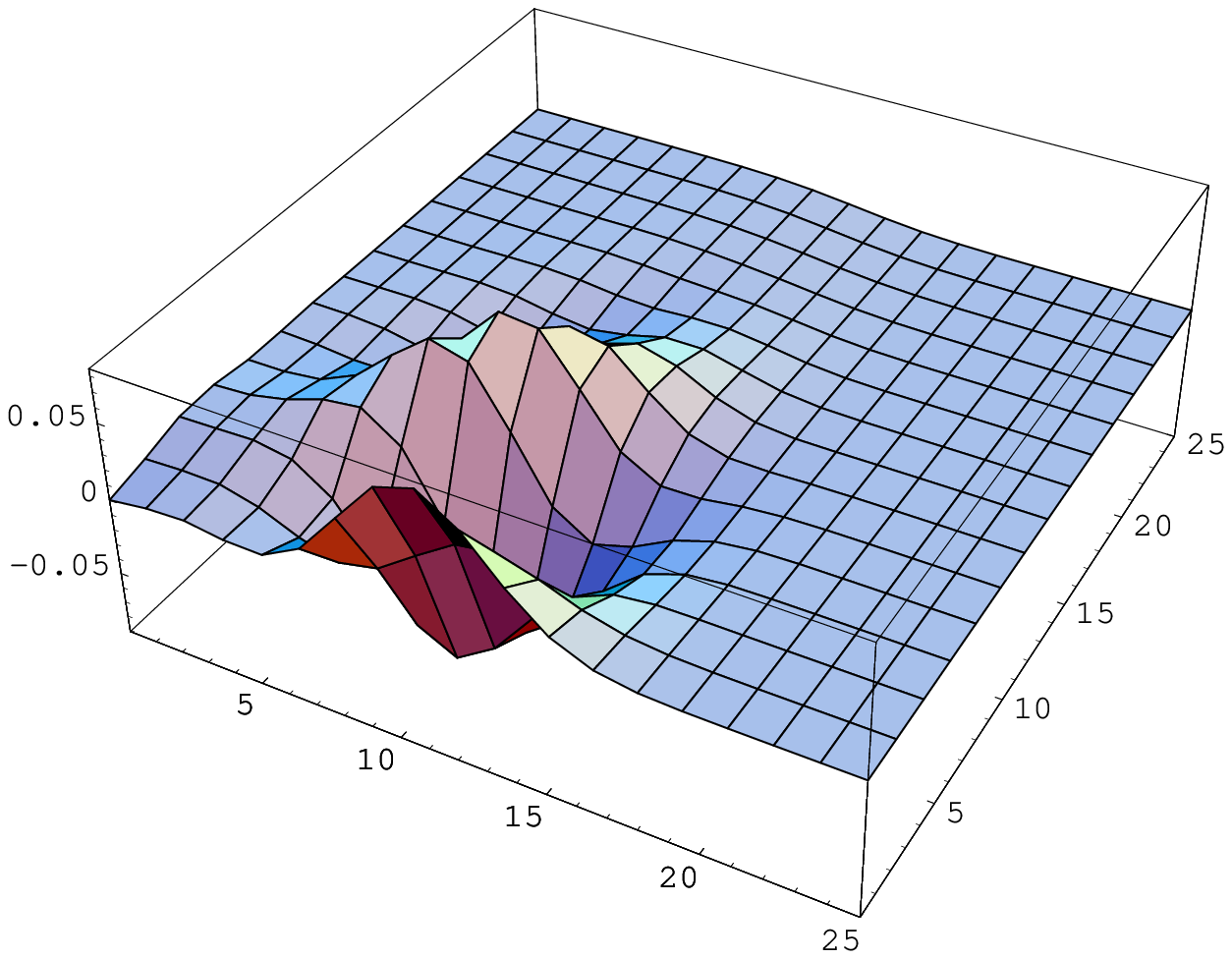,width=18pc}
\epsfig{figure=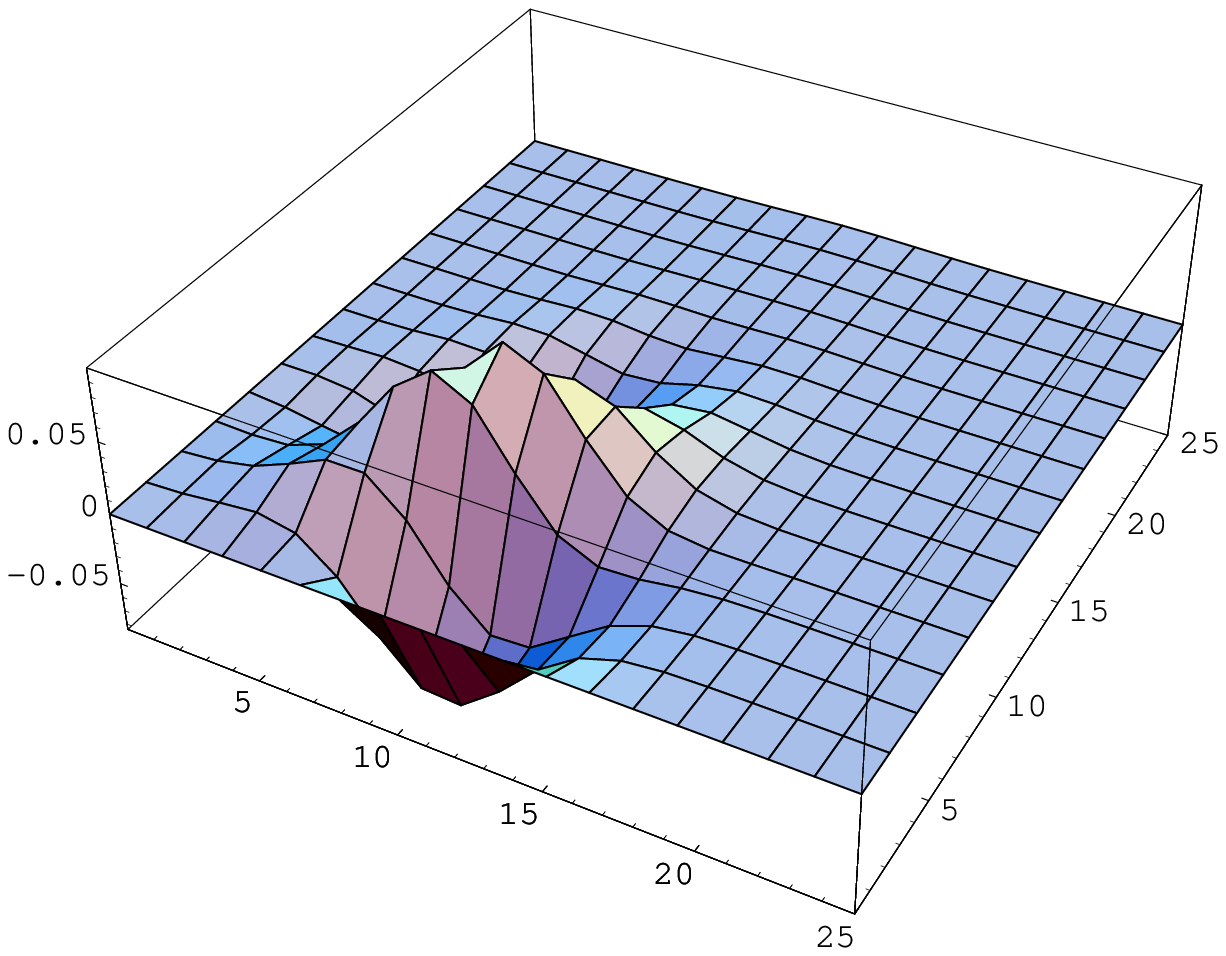,width=18pc}
\put(-450,180){$\re(\rho_{S\ \beta_D=2}^{(N_f=2,0)}(\xi))$}
\put(-200,180){$\im(\rho_{S\ \beta_D=2}^{(N_f=2,0)}(\xi))$}
\put(-160,20){$\im\ \xi$}
\put(-30,45){$\re\ \xi$}
\put(-380,20){$\im\ \xi$}
\put(-250,45){$\re\ \xi$}
}
\centerline{\epsfig{figure=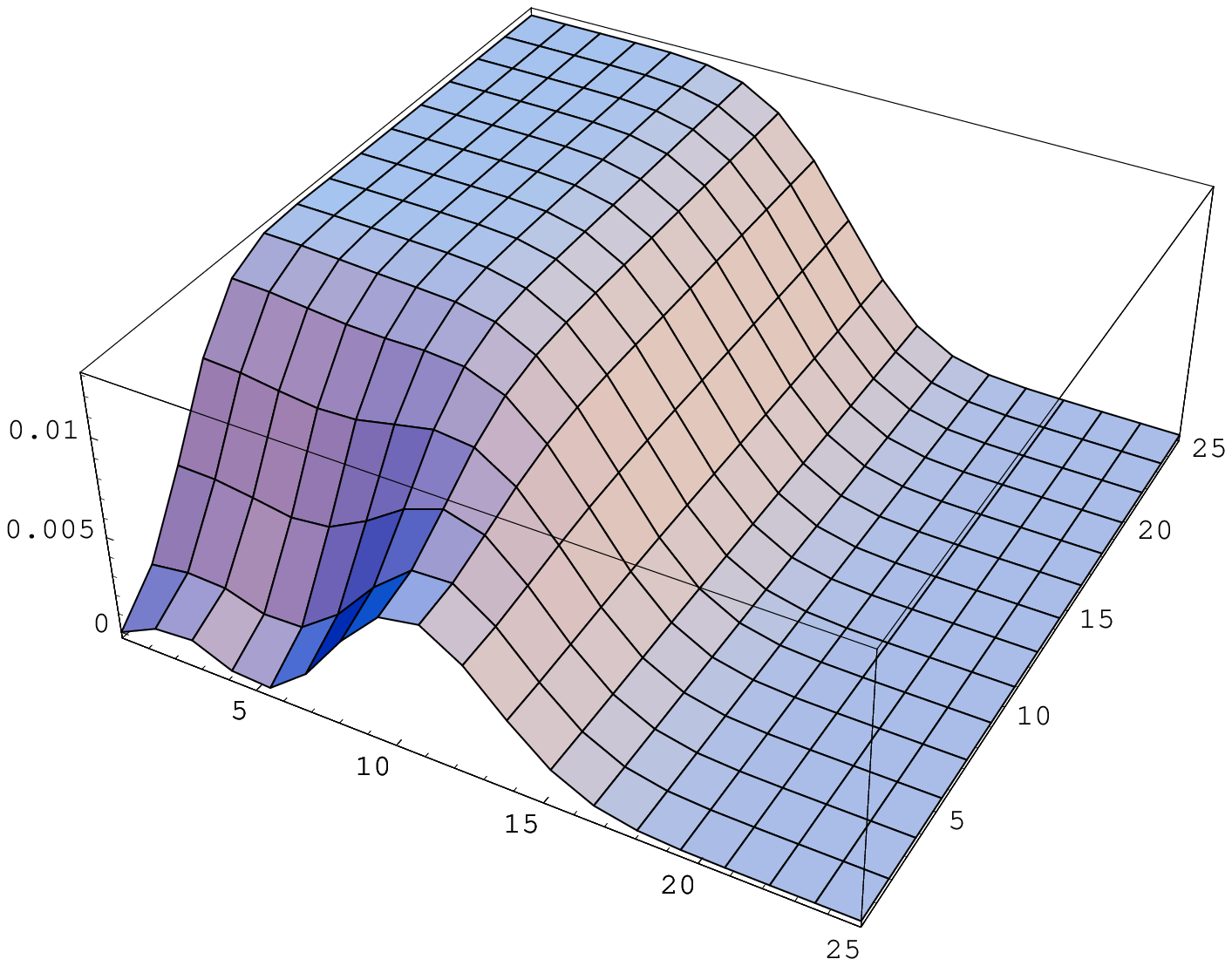,width=18pc}
\epsfig{figure=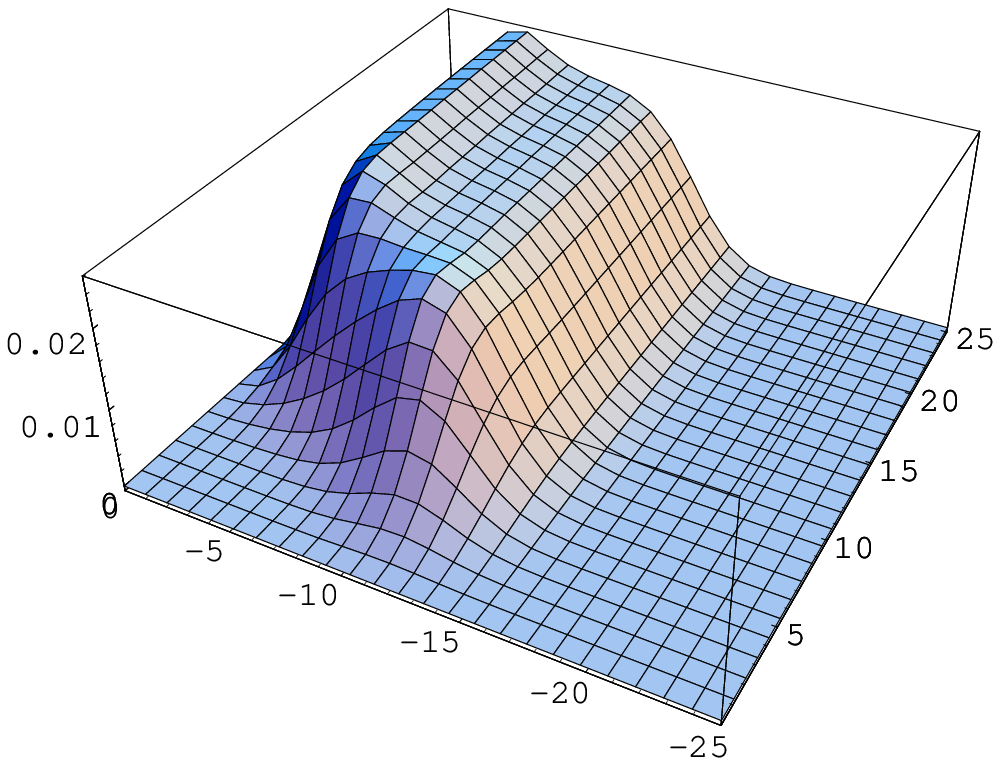,width=18pc}
\put(-450,160){$\rho_{S\ \beta_D=2}^{(|N_f|=2,0)}(\xi)$}
\put(-200,160){$\rho_{S\ \beta_D=4}^{(N_f=2,0)}(\xi)$}
\put(-160,20){$\im\ \xi$}
\put(-25,45){$\re\ \xi$}
\put(-380,20){$\im\ \xi$}
\put(-250,45){$\re\ \xi$}
}
\caption{Comparison of densities for different theories 
in the first quadrant of $\mathbb{C}$ (rotated by $\pi/2$). 
Upper plots: the complex spectral density for 
unquenched QCD with $N_f=2$, 
$\Re e(\rho(z))$ (left) and $\Im m(\rho(z))$ (right), from \cite{AOSV}.
Lower plots:
phase quenched QCD (left) and adjoint QCD (right), both with $N_f=2$. All plots
are at the same values $\alpha=2$, mass $\eta=5$.
}
\label{rhoUQ}
\end{figure}

We finish this subsection by giving the results for more flavours. 
With the definition 
\beq
{\cal I}_\nu(\xi_1,\xi_2)\ \equiv\ \int_0^1dt\ t\ e^{-2\alpha^2t^2}
J_\nu(t\xi_1)J_\nu(t\xi_2) \ ,
\label{Ks}
\eeq
the unquenched density for $N_f=1$ reads \cite{AOSV,James}
\beq
\rho_S^{(N_f=1,\nu)}(\xi)= 
\rho_S^{(0,\nu)}(\xi)\left( 1- 
\frac{J_\nu(\xi)
{\cal I}_\nu(i\eta,\xi^*)
}{J_\nu(i\eta)
{\cal I}_\nu(\xi,\xi^*)
}\right) ,
\label{rho1}
\eeq
where we have factored out the quenched density. Note that the support of the 
quenched and unquenched density is the same. It can easily be seen  
that the density 
$\rho_S^{(N_f=1,\nu)}(\xi)$ is no longer real and positive. 
For general $N_f$ the structure remains the same : the factor multiplying 
the quenched density becomes a determinant of size $N_f+1$  \cite{AOSV,James}:
\beq
\rho_S^{(N_f,\nu)}(\xi)= 
\frac{1}{2\pi\alpha^2}|\xi|^2 K_\nu\left(\frac{|\xi|^2}{4\alpha^2}\right)
e^{\frac{\xi^2+\xi^{*\,2}}{8\alpha^2}} 
\prod_{f=1}^{N_f}(\eta_f^2+\xi^2) 
\frac{{\cal Z}^{(N_f+2,\nu)}(\{\eta\},i\xi,i\xi^*)}{{\cal Z}^{(N_f,\nu)}
(\{\eta\})} \ ,
\label{rhoN_f}
\eeq
where the additional pair ``$+2$'' has masses $i\xi$ and $i\xi^*$. 
The weight function
originates from the bosonic partition function in the Toda lattice picture, 
and we refer to \cite{AOSV} for more details. 
It is not the explicit mass factor in front 
that makes the densities complex as it gets
cancelled by the Vandermonde from eq. (\ref{Zmn}). We recall that 
the partition functions themselves become complex valued for 
complex mass arguments $\xi$.

For illustration we give the spectral density for $N_f=2$ in different
theories. In the upper plots 
fig.\ref{rhoUQ} real and imaginary part start displaying
oscillation for moderate values of rescaled chemical potential. The spectral 
densities for phase quenched can also be computed using orthogonal polynomials,
and we refer to \cite{AOSV} for details. For two phase quenched flavours with 
$|\det(\Dirac(\mu)+m)|^2$ the density with the very same parameter values is
given in fig. \ref{rhoUQ} by the lower left plot. 
The oscillations have disappeared, with the real
density simply vanishing at the locations of the mass. Note the difference in
scale, the height of the supporting strip is hardly visible in the upper left
plot. We
also show the same unquenched density in the $\beta_D=4$ or adjoint class,
from section \ref{aQCD}. 
Apart from the additional repulsion from the axes it looks very similar to
phase quenched QCD.


\subsection{Comparison to quenched Lattice simulations} 

In this section we compare the predictions for complex eigenvalue correlation
functions from the previous section to Lattice simulation. Because of the sign 
problem
only quenched or phase quenched simulations can be compared with so far, and
we will focus on the former. Phase quenched simulations with two flavours have
been performed (see e.g. \cite{Sasai:2003py}) but the Dirac spectrum has 
not yet been compared to MMs.

The first comparison with the spectral density of quenched QCD on the lattice
was performed in \cite{AW} for 3 different volumes $4^4,\ 6^4$ and $10^4$
using staggered fermions at gauge coupling $\beta=6/g^2=5.0$ for chemical
potentials varying from $\mu=10^{-3}-0.2$. 
The reason for choosing a rather strong coupling was to make the window in
which MMs apply large enough for the small lattices studied. This assumed
that the concept of a Thouless energy \cite{OV}, 
the maximal eigenvalue up to which MMs
and $\eps\chi$PT agree, can be generalised to $\mu\neq0$. Evidence for this
picture has been provided later from Lattice data in \cite{OW}. 
\newpage

\begin{figure}[-h]
\centerline{
{\epsfig{figure=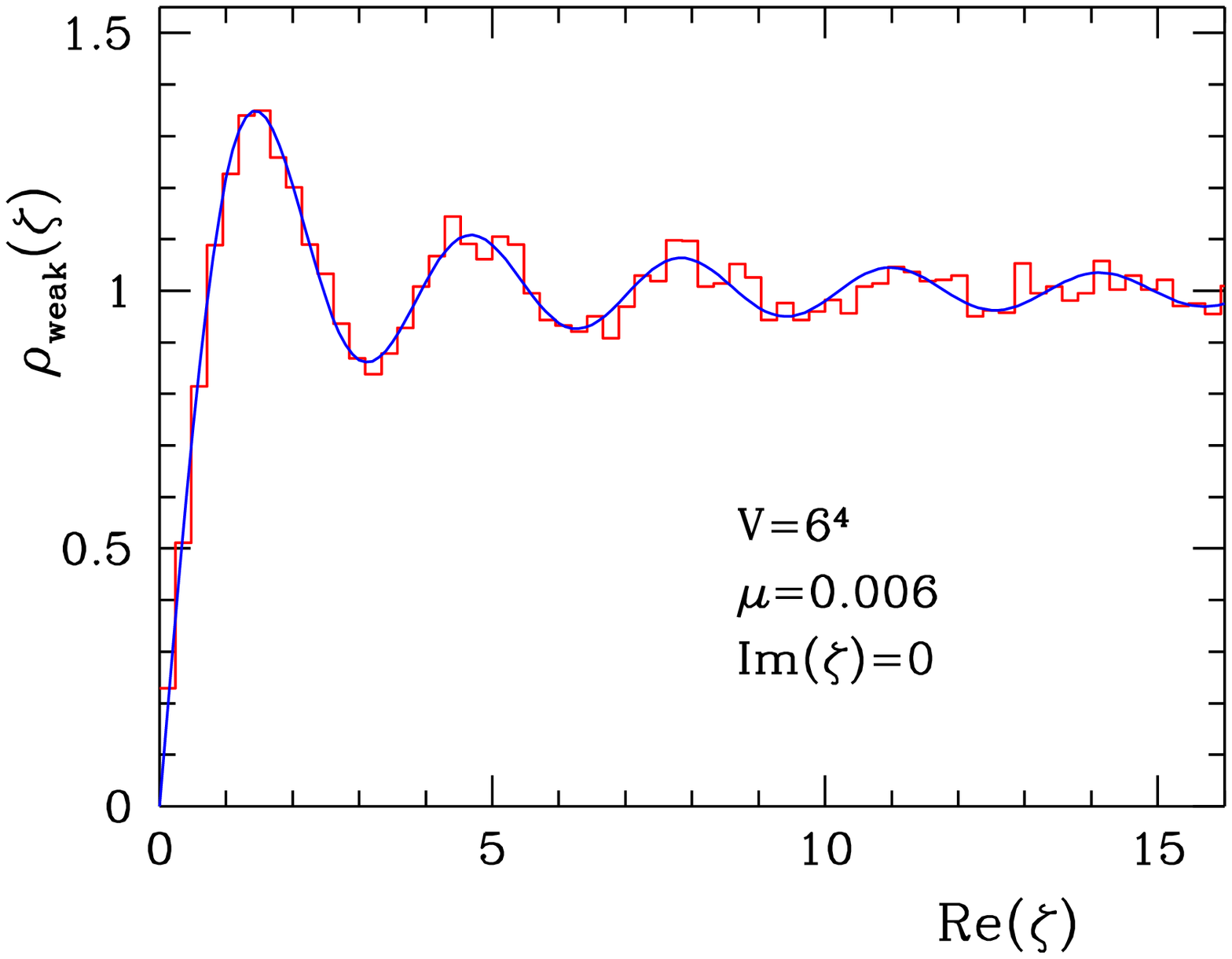,clip=,width=8cm}}
{\epsfig{figure=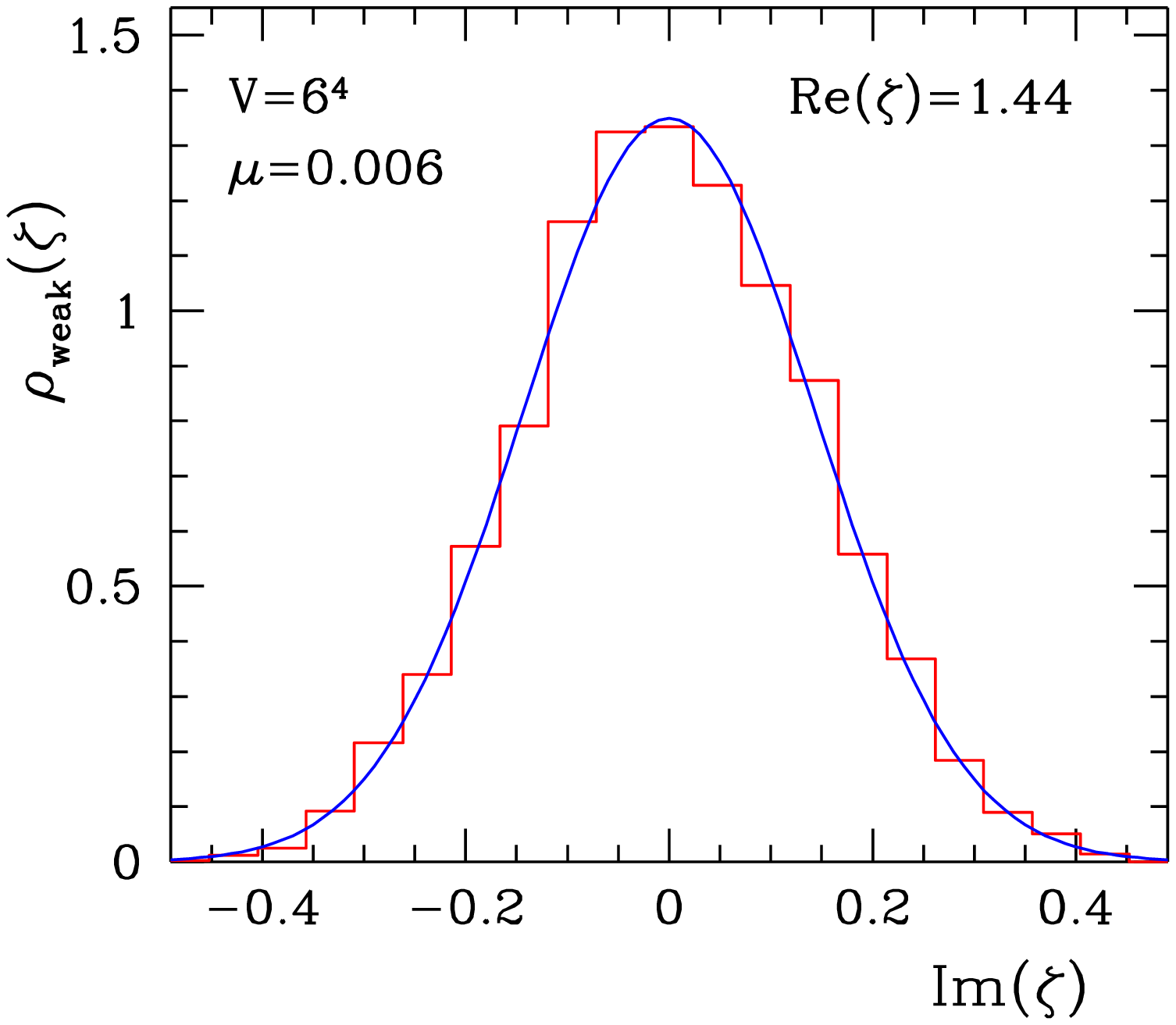,clip=,width=7.1cm}}
}
\centerline{
{\epsfig{figure=dr5065.eps,clip=,width=8cm}}
{\epsfig{figure=dr5066.eps,clip=,width=8cm}}
}
\caption{Comparison of the quenched density eq. (\ref{rhoQ}) at $\al=0.2$
(full line) and
  Lattice data (histograms) for $V=6^4$ and $\mu=0.006$ (upper plots), from
  \cite{AW}. The left curve is cut along the real-axis, and the right curve is 
on the
  first maximum of the left curve, cut perpendicular to the real-axis.
\newline
Fitted integrated density (lower plots) 
eq. (\ref{rhotheta}) vs. small Dirac eigenvalues 
from Lattice data \cite{OW} for $V=6^4$ and $\mu=0.1$ (left)
  and  $\mu=0.2$ (right). The two curves are indistinguishable.
}
\label{rhosmallal}
\end{figure}

Instead of comparing to the 3-dimensional plots of fig. \ref{weakplot} 
we show cuts compared to data. For small $\alpha$ shown in fig. 
\ref{rhosmallal} we cut along the real axis and parallel to the imaginary axis
on the first maximum. The agreement between the data and eq. (\ref{rhoQ}) is
excellent. In the plot the asymptotic form of the $K$-Bessel function in
eq. (\ref{rhoQ}) was used, as predicted from \cite{A03}. The exact expression 
eq. (\ref{rhoQ}) was only found later in \cite{SplitVerb1}. However, at the
fitted value of $\al=0.200(2)$ 
the two curves are not distinguishable from the data
(see fig. 1 in the second of ref.  \cite{SplitVerb1}). 

\begin{figure}[-h]
\centerline{
\epsfig{figure=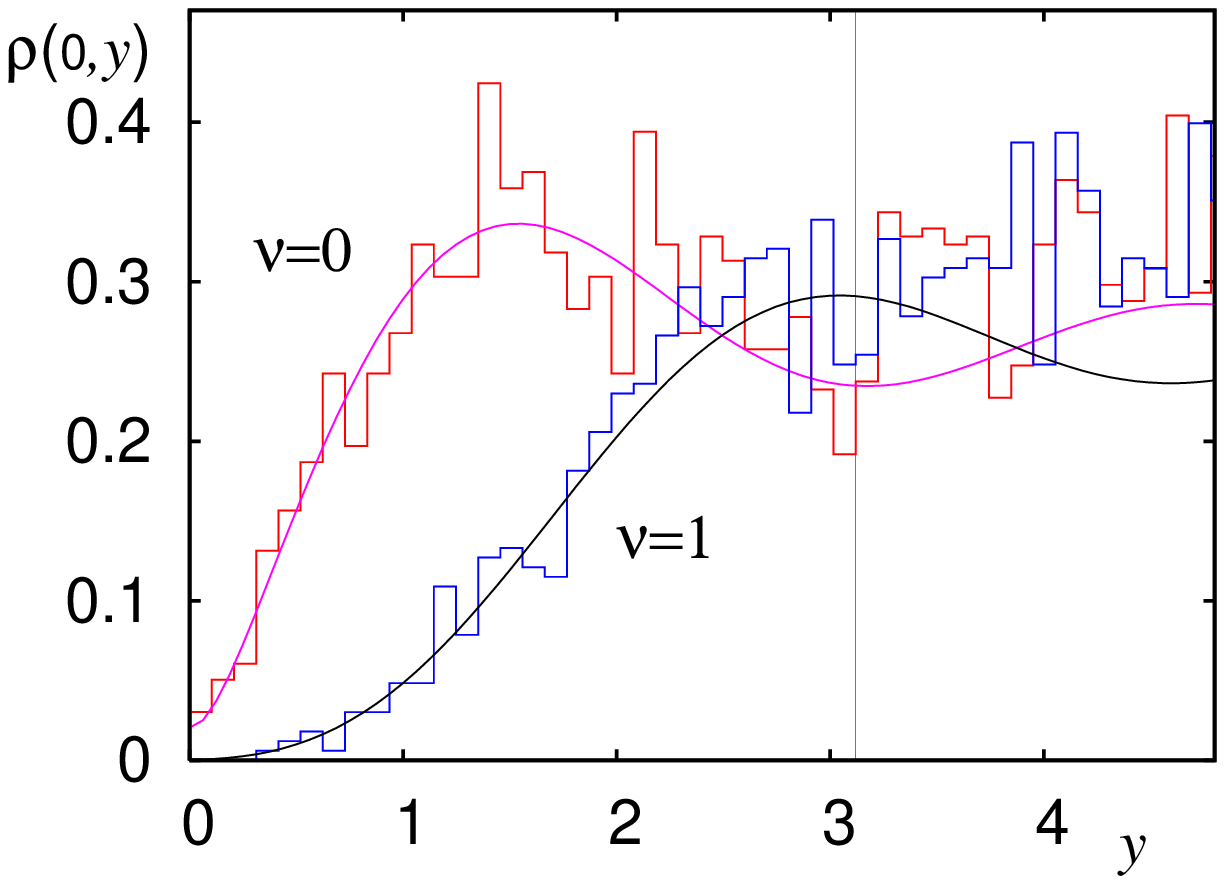,width=18pc}
\epsfig{figure=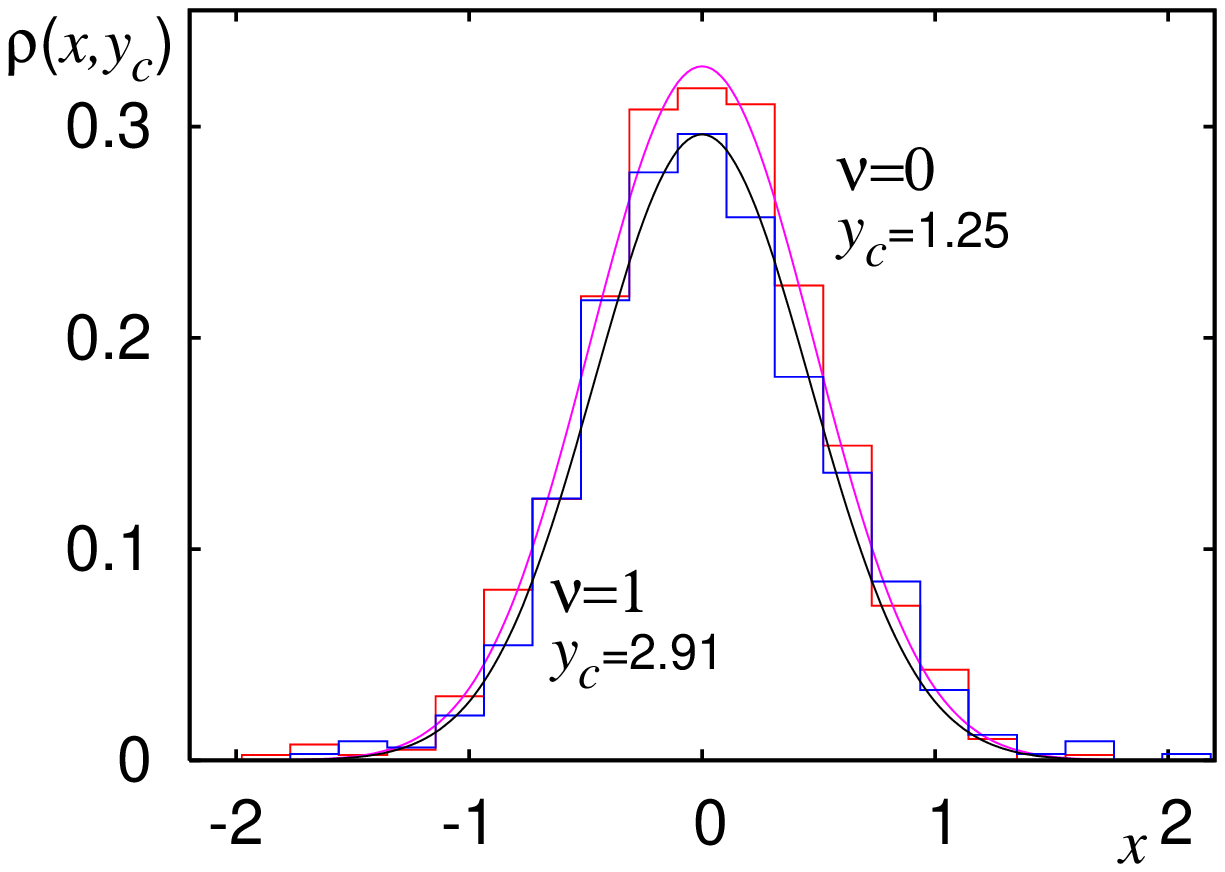,width=18pc}
\put(-460,80){$\mu=0.1$}}
\centerline{
\epsfig{figure=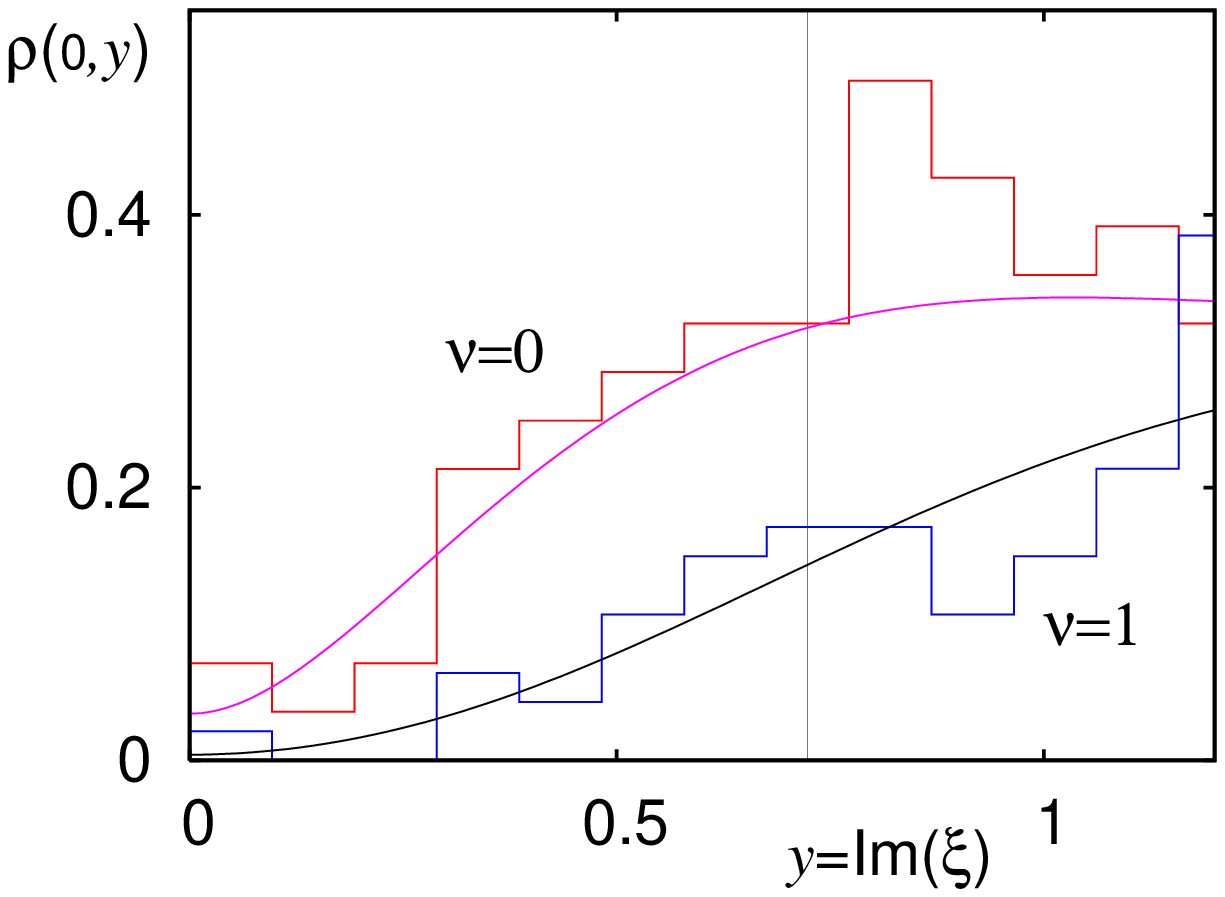,width=18pc}
\epsfig{figure=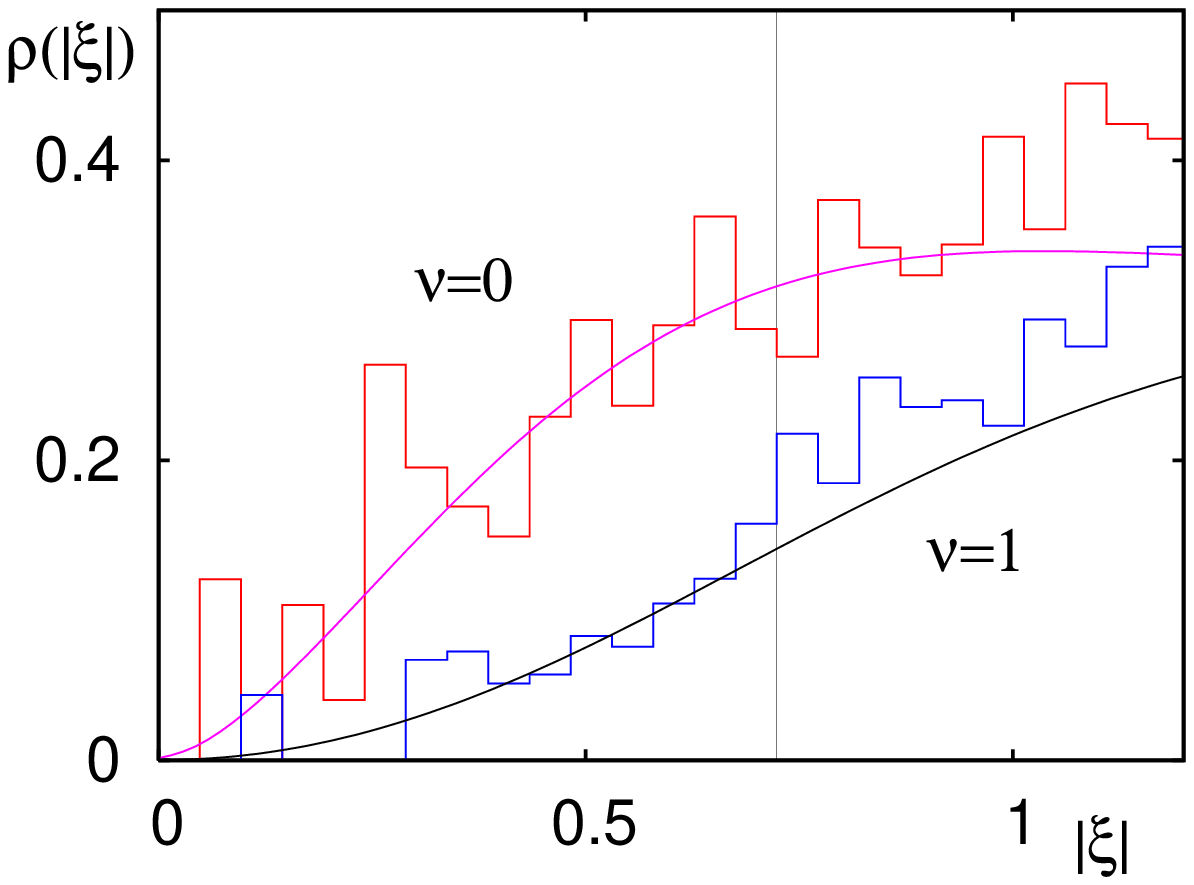,width=18pc}
\put(-460,80){$\mu=1.0$}}
\caption{Cuts through the quenched density eq. (\ref{rhoQ}) for topological
  sectors $\nu=0$ and 1, vs. Lattice data from \cite{BW} for $V=4^4$ and 
$\mu=0.1$ (upper plots) and $\mu=1.0$ (lower plots).
($x=\re\xi$ and  $y=\im\xi$ are interchanged w.r.t. fig. 
\ref{rhosmallal}). In the left plots the vertical bar indicates the fit range.
For the largest value $\mu=1.0$ the density is almost rotational invariant, as 
indicated by the lower right plot fitted to the asymptotic density 
$\rho_S(|\xi|)=|\xi|^2/2\pi\alpha^2K_\nu(|\xi|^2/4\alpha^2)
I_\nu(|\xi|^2/4\alpha^2)$. In this 
case only the ratio $\qq/F_\pi$ can be fitted.
}
\label{rhoover}
\end{figure}

For all other plots we use the exact density,
eq. (\ref{rhoQ}). The lower plots of fig. \ref{rhosmallal} from \cite{OW} 
show reanalysed and extended
data from \cite{AW}. At larger values of $\al$ the density at the origin
becomes asymptotically rotational invariant, see also fig. \ref{rhoover}. 
Therefore 
the integrated density eq. (\ref{rhoQ}) 
\beq
{\cal I}(r,\theta) = \int_{0}^{r} s\;ds \int_{0}^{\theta} d\phi
 ~ \rho_S(\xi=s\,{e}^{i\phi}) ~.
\label{rhotheta}
\eeq
is compared to the data, finding again excellent agreement. 
Because of the rotational invariance only the ratio $\Sigma/F_\pi$ can be
determined, and we refer to \cite{OW} for details. 

Recently progress has been made concerning chiral fermions with
chemical potential. In \cite{BW} it has been shown how to introduce $\mu\neq0$ 
into the overlap operator \cite{overlapp} using a generalisation of the sign
function in the complex plane. 
This opens up the possibility to compare to different
topological sectors, and successful comparisons to the density 
eq. (\ref{rhoQ}) have been made for $\nu=0$ and $\nu=1$ shown in fig. 
\ref{rhoover}. Here the same cuts as in fig.  \ref{rhosmallal} have been made 
(real and imaginary axis have been interchanged rotating by $\pi/2$), 
finding very good agreement. 
For the simulation and details of the complex overlap operator, in particular
concerning the projection from the complex Ginsparg-Wilson circle to the 
complex plane we refer to \cite{BW}.

The results presented here are not the first comparisons between non-Hermitian
MMs and Lattice data. In \cite{Markum:1999yr} quenched QCD lattice data were 
compared for a large range of $\mu$
to predictions of the non-chiral 
$\beta_D=2$ Ginibre ensemble for the spacing between 
eigenvalues in the complex plane. These predictions are only valid away from
the origin where chiral symmetry is irrelevant, in the so-called bulk region. 
Here no relation to an effective theory as the chiral Lagrangian is known.
A transition between the Wigner surmise for $\mu=0$, 
the Ginibre, and the Poisson distribution in the complex plane was found 
\cite{Markum:1999yr}.
Unfortunately to date no prediction is known for the spacing or the
distribution of individual eigenvalues in the origin region relevant for
chiral symmetry breaking.

\sect{Adjoint QCD and Symplectic Matrix Model}
\label{aQCD}

In this section we summarise the known results for the $\beta_D=4$ symmetry
class of matrices with quaternion real elements, being relevant for $SU(N_C)$
gauge theories in the adjoint representation. 
In the first subsection \ref{MM4} explicit results for the complex eigenvalue
correlation functions are given. Since the methods are similar to the previous
section we will be brief here, and only highlight the main differences: 
the fact that for $N_f\neq0$ the densities remain real and positive as well
as the different repulsion pattern of complex eigenvalues 
(see figs. \ref{HOVfig} and \ref{MMsig4}). 
The positivity  
makes both quenched and unquenched lattice simulations possible, and a
comparison to our MM predictions is done
in the next subsection \ref{latt4}.

\subsection{Complex eigenvalue correlations from a symplectic MM} 
\label{MM4}

\begin{figure}[-h]
\centerline{\epsfig{figure=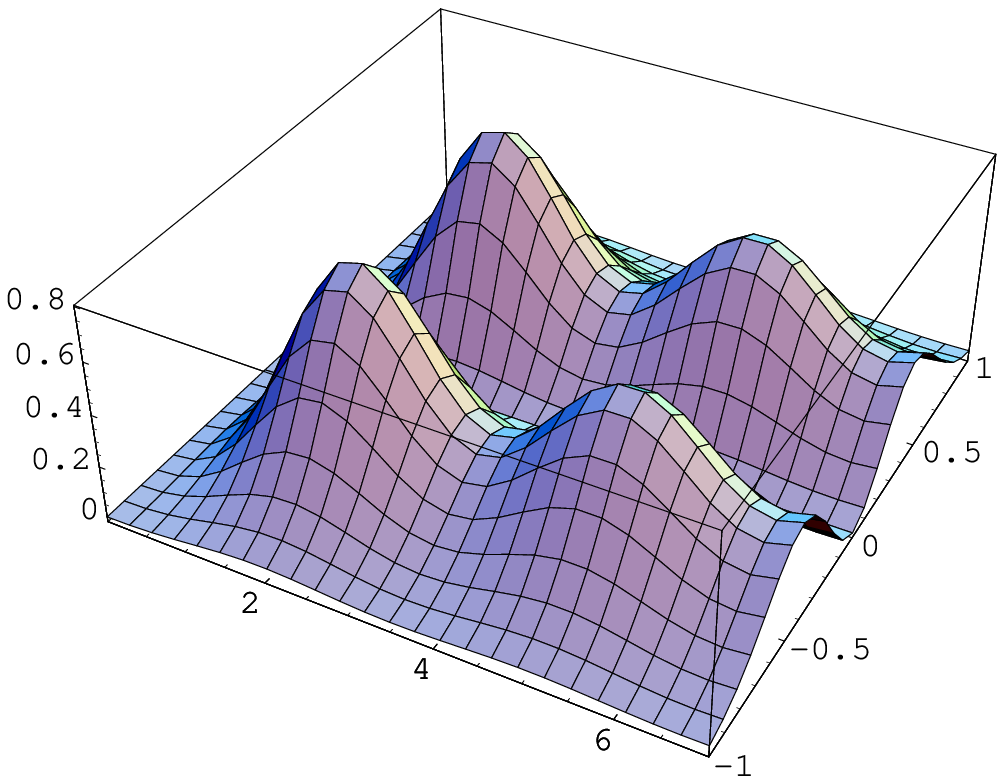,width=18pc}
\put(-160,20){$\re\ \xi$}
\put(-25,45){$\im\ \xi$}
\put(-240,140){$\rho_S^{(0,0)}(\xi)$}
\put(80,15){$\re\ \xi$}
\put(215,40){$\im\ \xi$}
\put(0,140){$\rho_S^{(N_f=2,0)}(\xi)$}
\epsfig{figure=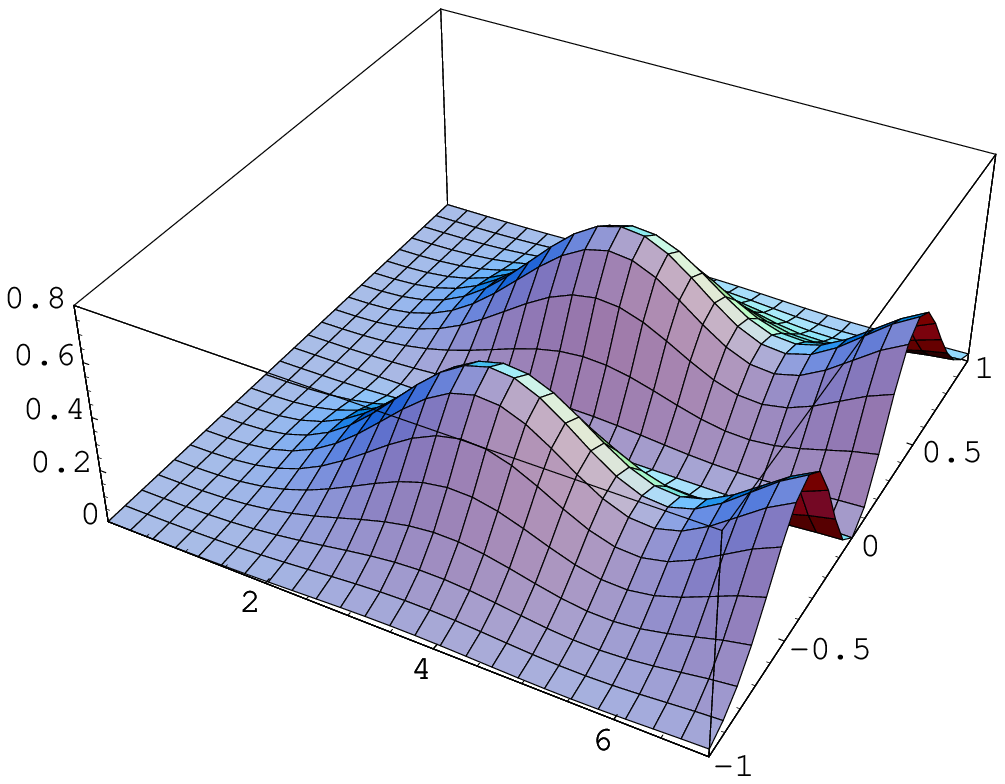,width=18pc}
}
\centerline{\epsfig{figure=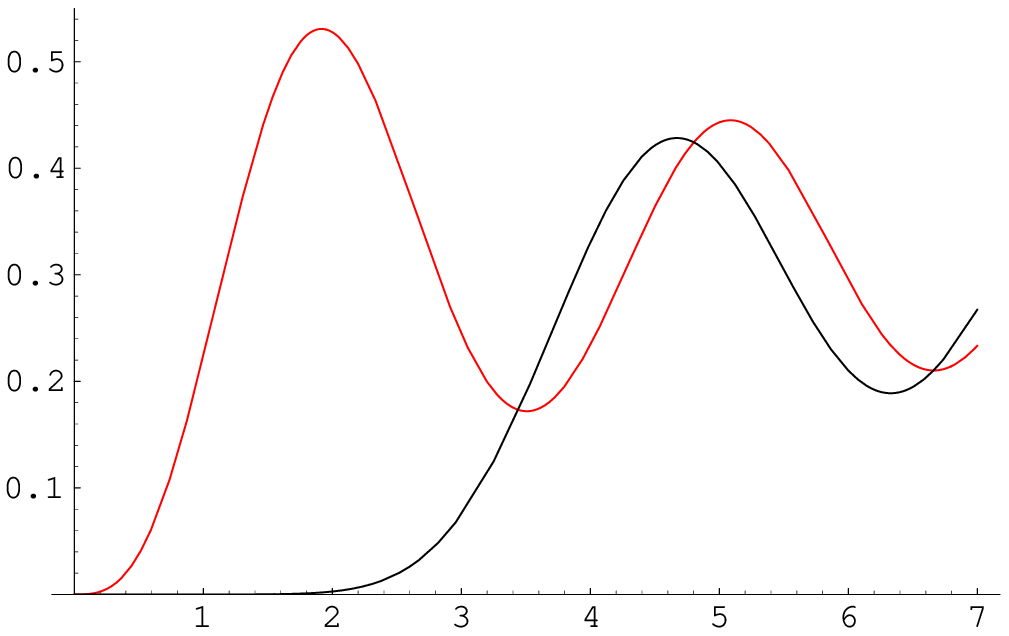,width=18pc}
\put(10,0){$\xi_x$}
\put(-220,140){$\rho_S^{(0,\nu)}(\xi_x)$}
\put(80,20){$\re\ \xi$}
\put(215,40){$\im\ \xi$}
\put(0,140){$\rho_S^{(N_f=2,0)}(\xi)$}
{\epsfig{figure=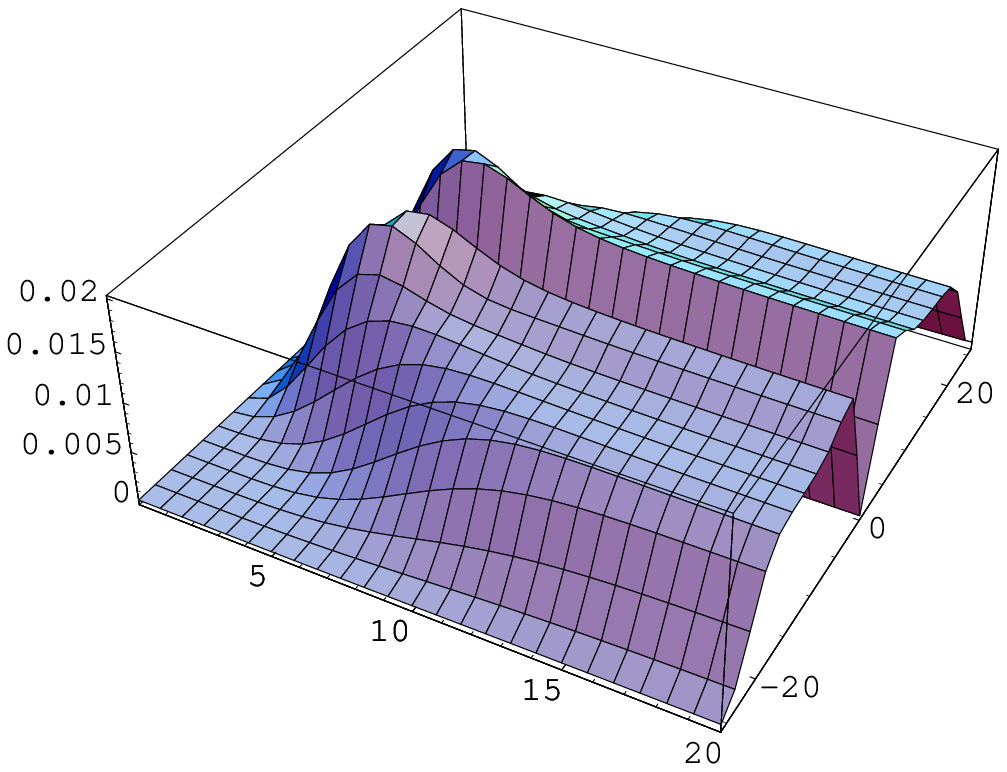,clip=,width=18pc}
}}
\caption{The quenched density eq. (\ref{rhoQsu2}) 
  (top left) vs. the unquenched density eq. (\ref{rhoNf2b4}) for $N_f=2$
  flavours of equal rescaled mass $\eta=4.26$ (top right), both at $\nu=0$ for
  $\alpha=0.4$, and the unquenched density eq. (\ref{rhoNf2b4}) for $N_f=2$
  at $\alpha=3.28$
  and rescaled mass $\eta=19$ (bottom right). For comparison we
  also give the quenched density of real eigenvalues (bottom
  left) for $\nu=0$ (left curve) and $\nu=2$ (right curve), respectively. 
}
\label{rhobeta4}
\end{figure}

Our starting point is the MM eq. (\ref{ZMM2}) with the matrices having
quaternion real elements instead. The very same procedure described in the
previous section can be employed to achieve a complex eigenvalue model. We
refer to \cite{A05} for details of the derivation where this ensemble was
first defined. The MM partition function for $N_f$ flavours with equal
chemical potential is reading in terms of complex eigenvalues 
\beqn
{\cal Z}^{(N_f,\nu)}(\{m_f\}) &\sim&  
\int  \prod_{j=1}^N d^2z_j \prod_{f=1}^{N_f} m_f^{2\nu}|z_j^2 + m_f^2|^2 \ 
|z_j|^{4\nu+2} K_{2\nu}\!\left(\frac{N(1+\mu^2)}{2\mu^2}|z_j|^2\right)
e^{\frac{N(1-\mu^2)}{4\mu^2}(z_j^2+z_j^{*\,2})}
\nn
\\
&&\times 
\prod_{k>l}^N |z_k^2-z_l^2|^2\ |z_k^2-z_l^{\ast\,2}|^2
\prod_{h=1}^N |z_h^2-z_h^{\ast\,2}|^2 \ .
\label{ZMM4ev}
\eeqn
There are two main differences to eq. (\ref{ZMM2ev}) for $\beta_D=2$. 
First, the mass term is automatically positive definite here because the
eigenvalues of quaternion real matrices always come in complex conjugated
pairs \cite{Mehta2}. 
It thus resembles QCD with pairs of $N_f$ conjugated flavours also
called phase quenched (see fig. \ref{rhoUQ}).
The same pairing of eigenvalues is also true for the zero modes, leading to a
shift $\nu\to2\nu$ compared to QCD. 
Second, the Jacobian is different from the squared  absolute value of the 
Vandermonde determinant for $\beta_D=2$\footnote{There exists a different 
symplectic MM of normal matrices with Jacobian $|\Delta\{z\}|^4$, see
e.g. \cite{Zabrodin}.}.   
The last factor in eq. (\ref{ZMM4ev})
leads to vanishing correlations along the $x$- and $y$-axis here, as 
$|z_h^2-z_h^{\ast\,2}|^2=16x_h^2y_h^2$ vanishes there.
We will find this feature explicitly in our results for the correlation
functions, see fig. \ref{rhobeta4}.
In the limit $\mu\to0$ for both $\beta_D=2,4$ 
the Jacobians reduce to $\Delta_N(x^2)^{\beta_D}$,
counting the number of independent degrees of freedom per matrix
element.

The complex eigenvalue correlation functions are defined in complete analogy
to eq. (\ref{Rkdef}) with the corresponding weight and Jacobian. 
The result can be written as a Pfaffian \cite{A05,ABa}
\beq
R^{(N_f,\nu)}(z_1,\ldots,z_k)\ =\ 
\prod_{j=1}^k(z_j^{\ast\,2}-z_j^2) w(z_j,z_j^\ast)\ 
\mbox{Pf}_{i,j=1,\ldots,k}
\left[\begin{array}{ll}
\kappa_N(z_i,z_j) &\kappa_N(z_i,z_j^\ast)\\
\kappa_N(z_i^\ast,z_j)      &\kappa_N(z_i^\ast,z_j^\ast)
\end{array}\right] \ .
\label{detformula}
\eeq
For antisymmetric matrices $A$ of even dimension the Pfaffian is the square
root of the determinant, Pf$(A)=\sqrt{\det(A)}$.
A similar form uses a quaternion determinant, 
see \cite{ABa}. 
Here we have introduced the pre-kernel $\kappa_N(z_1,z_2^\ast)$,
\beq
\kappa_N(z_1,z_2^\ast)\ \equiv\  
\sum_{k=0}^{N-1} \frac{1}{r_k} \left(q_{2k+1}(z_1)q_{2k}(z_2^\ast)-
q_{2k+1}(z_2^\ast)q_{2k}(z_1)\right)  \ ,
\label{prekernel}
\eeq
containing skew orthogonal polynomials $q_k(z)$. These are defined by the
following skew-orthogonality relation 
\beq
\int_{\mathbb{C}} d^2z\ w(z,z^\ast)
(z^{\ast\,2}-z^2)\left[q_{2k+1}(z)q_{2l}(z^\ast)-q_{2k+1}(z^\ast)
q_{2l}(z)\right] \ =\ r_k\ \delta_{kl}\ ,
\label{skewOP}
\eeq
where the remaining integrals with even/even and odd/odd indices vanish. 
For the quenched weight with $N_f=0$ above 
these polynomials are again given in terms of Laguerre polynomials
in the complex plane \cite{A05},
\beq
q_{2k+1}(z) \ \sim\ L_{2k+1}^{2\nu}\left( \frac{Nz^2}{1-\mu^2}\right)\ , \ \ \
\ 
q_{2k}(z) \ \sim\ \sum_{j=0}^k c_j(\mu)  
L_{2j}^{2\nu}\left( \frac{Nz^2}{1-\mu^2}\right),
\label{qQ}
\eeq
with some known $\mu$-dependent coefficients $c_j(\mu)$ 
\cite{A05}. The unquenched correlation functions
can again be expressed in terms of the quenched pre-kernel \cite{A05,ABa},
where for an odd number of $N_f$ flavours also the quenched 
skew orthogonal polynomials of even order $q_{2k}(z)$ explicitly 
appear \cite{ABa}.

The large-$N$ limit is defined by the same rescaling of 
the chemical potential, eq. (\ref{microscale}), 
of the complex eigenvalues and masses\footnote{The reason for this 
difference is that
  both ensembles in eqs. (\ref{ZMM4ev}) and (\ref{ZMM2ev}) have the same
  variance at $\mu=0$, instead of $\exp[-\beta_Dx^2/2]$.}, 
$\sqrt{2}\,z=\xi$ and $\sqrt{2}\,m=\eta$,
and correspondingly of the microscopic density
eq. (\ref{rhomicdef}). The simplest examples for the resulting correlation
functions in the large-$N$ limit is the 
quenched microscopic spectral density 
\beq
\rho_S^{(0,\nu)}(\xi)\ =\ \frac{1}{32\al^4}
(\xi^{\ast\,2}-\xi^2)\ |\xi|^2\ 
K_{2\nu}\left(\frac{|\xi|^2}{2\al^2}\right)
\mbox{e}^{+\frac{1}{4\al^2}(\xi^2+\xi^{*\,2})}{\cal I}_\nu(\xi,\xi^*) \ , 
\label{rhoQsu2}
\eeq
where the limiting pre-kernel eq. (\ref{prekernel}) 
leads to a double integral here 
due to the extra sum in eq. (\ref{qQ}) (compared to eq. (\ref{rhoQ}) for
$\beta_D=2$) 
\beq
{\cal I}_\nu(\xi,\xi^*)\equiv
\int_0^1 ds \int_0^1 \frac{dt}{\sqrt{t}}\ \mbox{e}^{-2s(1+t)\al^2}
\left(J_{2\nu}(2\sqrt{st}\ \xi)J_{2\nu}(2\sqrt{s}\ \xi^\ast)
\ -\ J_{2\nu}(2\sqrt{s}\ \xi)J_{2\nu}(2\sqrt{st}\ \xi^\ast)
\right)\ .
\label{I4}
\eeq
Its antisymmetry together with the pre-factor $(\xi^{\ast\,2}-\xi^2)$ leads to
a real positive density eq. (\ref{rhoQsu2}). 

Our second example is the 
unquenched microscopic spectral density for $N_f=2$ flavours of degenerate 
masses: 
\beq
\rho_S^{(N_f=2,\nu)}(\xi)\ =\ \rho_S^{(0,\nu)}(\xi)\left(1- 
\frac{{\cal I}_\nu(\xi^*,i\eta){\cal I}_\nu^\prime(\xi,i\eta)
-{\cal I}_\nu(\xi,i\eta){\cal I}_\nu^\prime(\xi^*,i\eta)
}{{\cal I}_\nu(\xi,\xi^*){\cal I}_\nu^\prime(i\eta,i\eta)}\right) \ , 
\label{rhoNf2b4}
\eeq
where 
\beq
{\cal I}_\nu^\prime(\xi,\xi^*)\equiv
\int_0^1\!ds\int_0^1dt \sqrt{\frac{s}{t}}\ \mbox{e}^{-2s(1+t)\al^2}
\left( \sqrt{t} J_{2\nu+1}(2\sqrt{st}\ \xi)J_{2\nu}(2\sqrt{s}\ \xi^*)
- J_{2\nu}(2\sqrt{st}\ \xi^*)J_{2\nu+1}(2\sqrt{s}\ \xi)\right)
\eeq
is the derivative of the kernel eq. (\ref{I4}). It appears because 
of the degenerate masses in our example. 
Eq. (\ref{rhoNf2b4}) now contains 3 terms being a Pfaffian, and it is
again explicitly real. 

The quenched and unquenched densities eqs. (\ref{rhoQsu2}) and
(\ref{rhoNf2b4}) are compared in fig. \ref{rhobeta4}
for different values of rescaled chemical potential $\alpha$ and mass $\eta$.
The vanishing correlations along $x$- and $y$-axis due to the Jacobian is
clearly observed. The introduction of masses has a mild effect, pushing the
density further away from the origin when lowering the mass, without drastic
change of its shape. At zero mass the
resulting density becomes indistinguishable from the quenched density at
$\nu=2$. However, this flavour-topology duality holds only approximately for
$\mu\neq0$  due to the explicit $\nu$-dependence of the $K$-Bessel weight. 

For large values of  $\alpha$ the density becomes almost constant along a
strip, apart from the repulsion from the axis. In fig. \ref{rhobeta4} bottom
right the mass at $\eta=19$ deforms this strip, as the location of the mass
is a zero of the partition function, and of the density as well. 
The described effect of the masses on the density will be used in the next
subsection to test the effect of unquenching in Lattice result. 

Finally let us give the microscopic density for an arbritrary number of
flavours: 
\beq
\rho_S^{(N_f,\nu)}(\xi)= 
\frac{1}{2\pi\alpha^2}(\xi^{*\,2}-\xi^2)
|\xi|^2 K_{2\nu}\left(\frac{|\xi|^2}{4\alpha^2}\right)
e^{\frac{z^2+z^{*\,2}}{8\alpha^2}} 
\prod_{f=1}^{N_f}|\eta_f^2+\xi^2|^2 
\frac{{\cal Z}^{(N_f+2,\nu)}(\{\eta\},i\xi,i\xi^*)}
{{\cal Z}^{(N_f,\nu)}(\{\eta\})} \ ,
\label{rhoN_f4}
\eeq
where the additional two flavours have masses $i\xi$ and $i\xi^*$. 
The structure is exactly the same as for the unquenched densities for QCD, 
eq. (\ref{rhoN_f}). The difference here is that due to the quaternion
structure  
the extra complex mass pair $\xi$ and $\xi^*$ does not spoil the reality and
positivity of the
corresponding partition functions. 
For more details and higher order correlation functions we refer to 
\cite{A05,ABa}. 

Let us close this subsection with a few remarks. For $\beta_D=4$ the complex
eigenvalue correlation functions are simpler than the corresponding ones for
$\mu=0$. This seemingly paradoxical statement originates in the simpler skew
product and pre-kernel compared to the MM of real eigenvalues, explaining
features of the latter \cite{ABa}.

In the previous section on the $\beta_D=2$ class several alternative and
equivalent derivations of the same results have been stated. In our 
$\beta_D=4$ class to date no such alternative derivation is 
known so far using replicas, even for $\mu=0$.
Also the supersymmetric method has only been applied to non-chiral MMs
with $\beta_D=4$ (and $\beta_D=1$) 
\cite{Efetov}, where the microscopic density was derived. For
a complete solution from skew-orthogonal polynomials of this non-chiral MM 
we refer to \cite{EK}. 
As in the $\beta_D=2$ case a supersymmetric treatment would be simpler 
to do first in the 
one-MM eq. (\ref{ZMM1}). 

Consequently little is known about the universality of the microscopic
correlation functions and their relation to field theory in this symmetry
class (the same statement is true for $\mu=0$, except for all masses
degenerate). 

The only exception is for one pair of complex conjugate flavours $N_f=2$, 
where the same result for the partition function 
\beq
{\cal Z}^{(N_f=2,\nu)}_{\eps\chi PT}(\eta)\ \sim\ \int_0^1dt \
e^{-2t^2\alpha^2}I_{2\nu}(2t\eta)  
\eeq
can also be derived from the corresponding group integral in the $\eps\chi$PT
\cite{B}. For more general MM partition functions with $N_f>2$ we refer to
\cite{A05,ABa}.

\begin{figure}[-h]
\centerline{
{\epsfig{figure=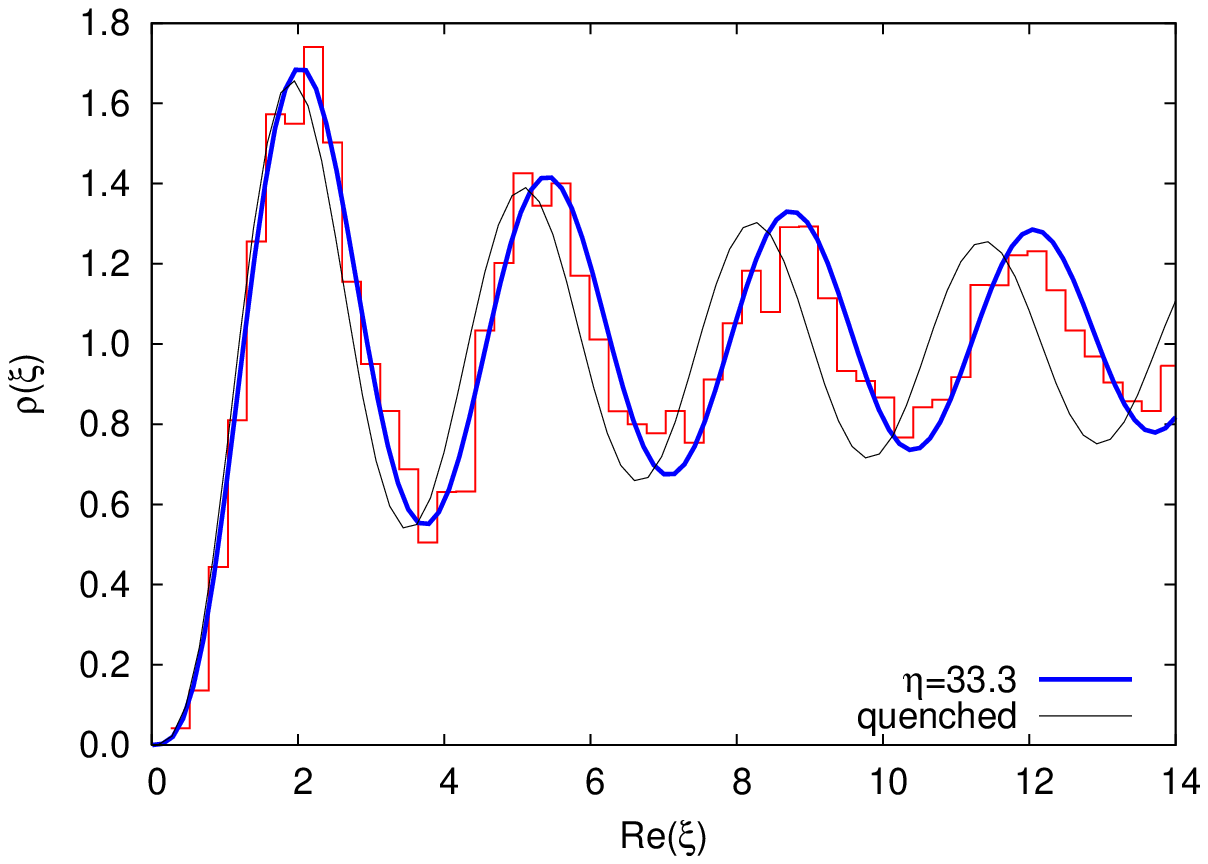,clip=,width=8cm}}
{\epsfig{figure=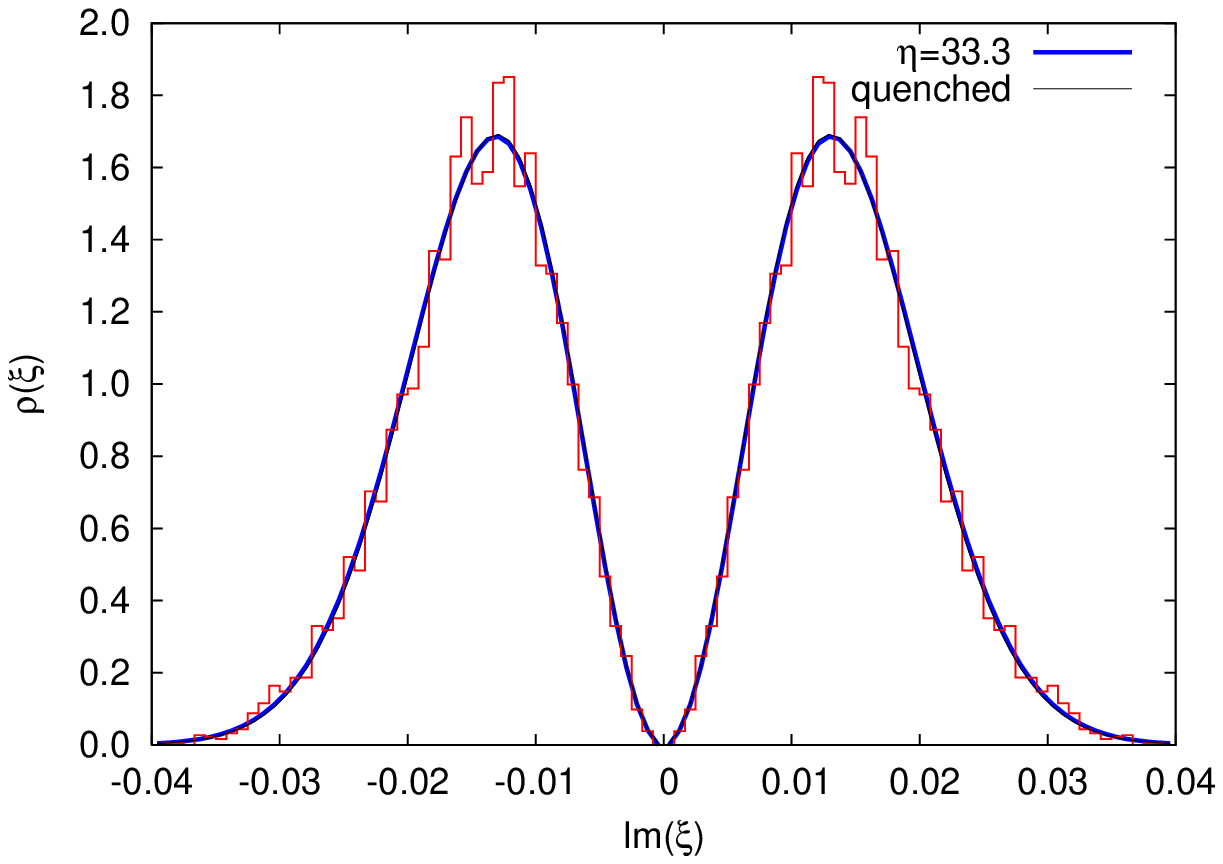,clip=,width=8cm}}
}
\caption{Lattice data (histograms) 
at mass $ma=0.07$ in lattice unit $a$, volume 
$V=6^4$ and $\mu=10^{-3}$ from \cite{AB}, 
vs. MM prediction eq. (\ref{rhoNf2b4}) at $\al=0.0127$ 
(full curve). The cuts are made parallel to the $x$-axis
along the maxima (left), and perpendicular to it on the first pair of maxima. 
For comparison the quenched curve eq. (\ref{rhoQsu2}) is included as well,
which is clearly shifted away from the data (left plot).
} 
\label{beta4asmall}
\end{figure}


\subsection{Comparison to unquenched Lattice data using 
staggered fermions for $SU(2)$} 
\label{latt4}

In this subsection we compare the complex eigenvalue correlation functions
from the previous subsection to lattice simulations \cite{ABLMP,AB}. Because
for eigenvalues in the complex plane much statistics was needed
staggered fermions were chosen in these simulations, using the code of
\cite{HKLM}.  
In order to remain in the  $\beta_D=4$
symmetry class while using staggered fermions this comparison is made between
our symplectic MM results and an $SU(2)$ gauge theory in the fundamental
representation realised as a staggered operator (see table \ref{Dsym}). 
We recall that in the continuum $SU(2)$ in the fundamental
representation belongs to the class $\beta_D=1$.

The lattice simulations \cite{AB} were done using two different volumes,
$V=6^4$ and $8^4$ at fixed gauge coupling $\beta=1.3$ for chemical potentials
ranging from $\mu=10^{-6}-0.2$. Starting at large masses where the smallest
eigenvalues effectively quench \cite{ABLMP} 
the masses were varied down to values where an
effect of unquenching was observed. 
For more details of the simulations we refer to \cite{AB}.
 
For a quantitative comparison we show cuts trough the 3 dimensional density
plots of fig. \ref{rhobeta4}: the curves are cut along the maxima parallel to
the $x$-axis and on the (first) maximum perpendicular to that parallel to the
$y$-axis. 

For small $\alpha<1$ the effect of unquenching is only seen in the cut in
$x$-direction through a shift of the curve, see fig. \ref{beta4asmall} left. 
The shift compared to the quenched curve from eq. (\ref{rhoQsu2}) 
is clearly visible. Note that it is very difficult to perform simulations at
sufficiently small mass. In the perpendicular direction which is used to fit 
$\alpha$ no difference to the quenched curve is seen (when appropriately
normalised) as we expect. The data follow very nicely the MM prediction. A
similar comparison was made for large quark masses in \cite{ABLMP}, nicely
matching with the quenched result.

\begin{figure}[-h]
\centerline{
{\epsfig{figure=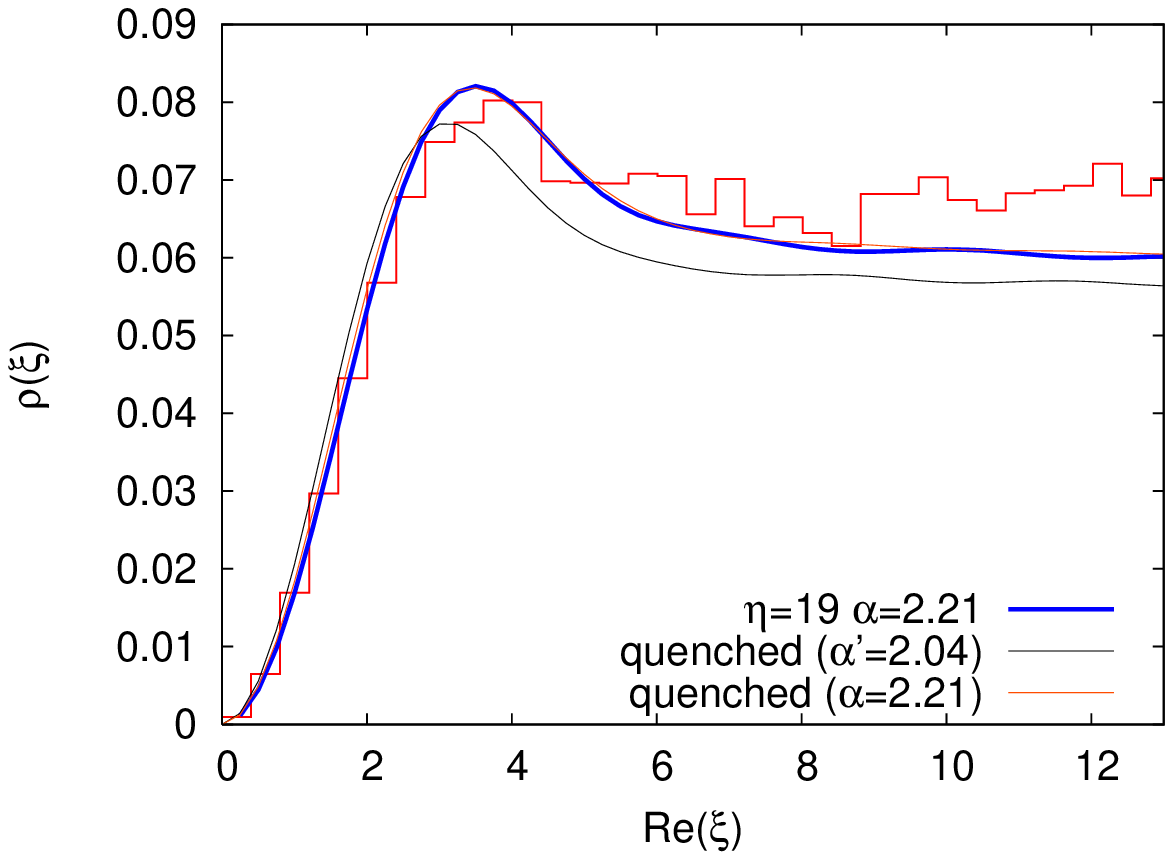,clip=,width=8cm}}
{\epsfig{figure=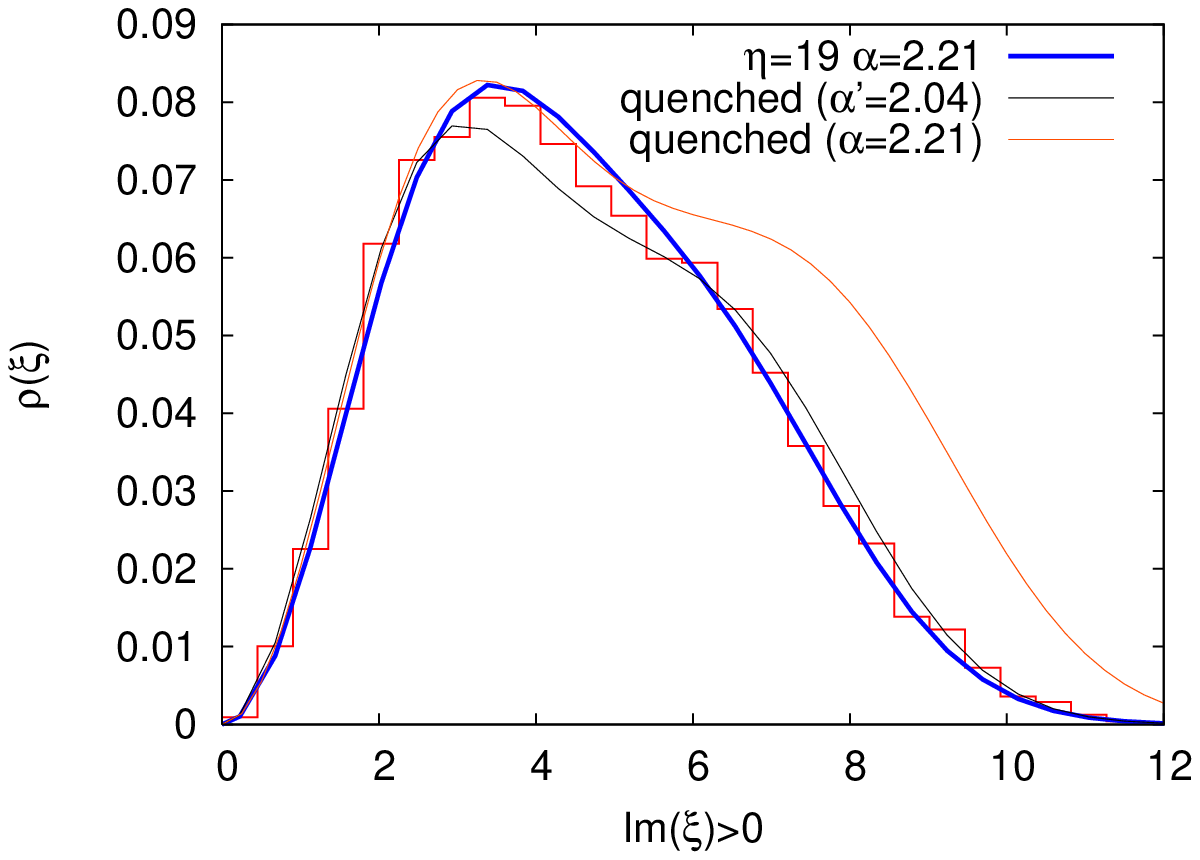,clip=,width=8cm}}
}
\caption{The same cuts as in fig. \ref{beta4asmall} but for larger $\al$:
Lattice data (histograms) at mass  $ma=0.06$, volume  $V=6^4$ and $\mu=0.2$, 
from \cite{AB}, 
versus MM prediction eq. (\ref{rhoNf2b4}) (full curve). 
For a better resolution we only
display half of the cut in $y$-direction compared to fig. \ref{beta4asmall} 
(right). Here two different quenched densities from eq. (\ref{rhoQsu2}) are
inserted: the with the same $\al$ as unquenched, and a fit to the right cut. 
Both deviate from the data.}
\label{beta4alarge}
\end{figure}

Turning to large values of $\alpha$ the situation is reversed. At the smallest
mass values available the cut in $x$-direction is not distinguishable from
quenched. However, the perpendicular cut in $y$-direction shows the
deformation of the density through the mass, as is clearly seen in fig.  
\ref{beta4alarge}. Since in this case the  cut in $y$-direction is also used
to fit $\alpha$, two different fits are compared: 
the $\alpha$ determined from the unquenched curve is compared to the quenched
curve at the same value of  $\alpha$, see the
right curve in fig. \ref{beta4alarge}. 
Second, a different value of $\alpha$ is obtained by directly
fitting the quenched curve, leading however to a bigger mismatch in fig.
\ref{beta4alarge} left, underscoring the data. Clearly the unquenched curve
describes the data best.

The same procedure has been repeated for different values of the mass and
chemical potential in \cite{AB}. 
Comparing the two volumes and keeping the product $V\mu^2$
fixed confirms the scaling predicted 
by $\eps\chi$PT and MMs from the data in the $\beta_D=4$ symmetry class.

In previous Lattice simulations using staggered fermions for both classes
$\beta_D=4$ and 1 \cite{BLMP} no comparison to MMs was available, and still is
not for the latter class.

In summary we have seen that our symplectic MM is able to describe unquenched
Lattice data in the case where the density remains real and positive.

\sect{Matrix Model of QCD with Imaginary Isospin Chemical Potential}
\label{Imu}

In this section we discuss a MM for imaginary isospin chemical potential. The
main difference to the previous MM is that is has real Dirac operator
eigenvalues. This model was first introduced in \cite{DHSS} in terms of the
corresponding $\eps\chi$PT. The motivation was to have a model that allows to
determine the pion decay constant $F_\pi$ on the Lattice while keeping the 
Dirac eigenvalues real, in order to allow for unquenched simulations. The 
quenched results in \cite{DHSS}
were then extended to the unquenched two-point function with $N_f=2$ and
confirmed on the Lattice \cite{DHSST}. 
We note that $F_\pi$ can also be determined on the lattice for real $\mu$ 
using quenched QCD with complex Dirac eigenvalues \cite{OW,AW}.

Here we follow the presentation of \cite{ADOS} (see also \cite{James06})
where the corresponding MM was
introduced and solved for all unquenched correlations functions. The cases 
known from field theory 
\cite{DHSS,DHSST} were confirmed.

\subsection{Real eigenvalue correlations}

We start by writing down the Dirac operators for  
imaginary isospin chemical potential $\pm i\mu$: 
\beqn
\Dirac_\pm(\mu) = \left( \begin{array}{cc}
0 &  \Phi \pm  \mu \Psi \\
-( \Phi^{\dagger} \pm  \mu \Psi^{\dagger}) & 0
\end{array} \right) ~.
\eeqn
For both signs it is anti-hermitian thus having real eigenvalues. 
The symmetry classification in table \ref{Dsym} is not affected by this
rotation and thus we can immediately write down the MM for $m$ copies with
$+i\mu$ and $n$ copies with $-i\mu$
\beqn
{\cal Z}^{(N_f=m+n,\nu)} \sim
 \int\! d\Phi  d\Psi 
\prod_{f+=1}^{m}\! \det[\Dirac_+ + m_{f+}]\!
\prod_{f-=1}^{n} \!
\det[\Dirac_- + m_{f-}]\ 
e^{-N{\rm Tr}\left(\Phi^{\dagger}\Phi+\Psi^{\dagger}\Psi\right)}\ .
\label{ZNfiso}
\eeqn
However, for $\mu\neq0$ the operators $\Dirac_+(\mu)$ and 
$\Dirac_-(\mu)$ now have different real eigenvalues (or more precisely real
positive singular values): $\{x_k\}$ and $\{y_k\}\in\mathbb{R}_+$
respectively.  
After the change of variables 
\beq
\Phi_\pm \ \equiv\  \Phi \pm \mu \Psi 
\label{Phipm}
\eeq
the Gaussian weight becomes replaced by 
\beq
V(\Phi_+,\Phi_-) =
  c_+(\Phi_{+}^{\dagger}\Phi_+ +\Phi_{-}^{\dagger}\Phi_-)
- c_- \left(\Phi_{+}^{\dagger}\Phi_-
+\Phi_{-}^{\dagger}\Phi_+\right)
\label{V}
\eeq
where $c_\pm=(1\pm\mu^2)/(4\mu^2)$. A transformation to eigenvalues of this
two-MM can thus only be performed for the $\beta_D=2$ symmetry class including
QCD, to which we restrict ourselves in all the following. Only 
here the following matrix integral is available \cite{GW}
\beq
\int dUdV \exp\left[c_- N {\rm Tr}(VXUY) ~~{\rm + ~h.c.}~ \right] 
~=~ \frac{\det[I_{\nu}(2 c_- N x_iy_j)]}
{\prod(x_iy_i)^{\nu}\Delta_N(\{x^2\})\Delta_N(\{y^2\})} ~.
\eeq
If we parametrise $\Phi_+=U_+XV_+$ and $\Phi_-=U_-YV_-$ and use the invariance
of the Haar measure we obtain the following eigenvalue model:
\beqn
{\cal Z}^{(N_f,\nu)}(\{m_1\},\{m_2\}) &\sim&
\int_0^{\infty} \prod_{i=1}^N\left(dx_idy_i (x_iy_i)^{\nu+1} 
\prod_{f+=1}^{m} (x_i^2+m_{f+}^2)
\prod_{f-=1}^{n} (y_i^2+m_{f-}^2) \right) \cr
&&\times\Delta_N(\{x^2\})\Delta_N(\{y^2\})\det_{k,l}\left[I_{\nu}(2 c_- N x_k
  y_l)\right]  
e^{-N \sum_i^N c_+(x_i^2 +y_i^2) }. 
\label{evrep}
\eeqn
While for the partition function the determinant can be replaced by its
diagonal part due to the antisymmetry of the Vandermonde determinants, this
is not the case for the eigenvalue correlations. We note that also in this
model the initial Gaussian weight has become non-Gaussian containing an
$I$-Bessel function, due to the group integral performed.

The MM eq. (\ref{evrep}) 
can be completely solved by introducing bi-orthogonal polynomials 
\beq
\int_0^{\infty} dx dy \ w^{(N_f)}(x,y)\ 
P_n^{(N_f)}(x^2)\ 
Q_k^{(N_f)}(y^2)
= h_n^{(N_f)} \delta_{nk} ~, 
\label{bio}
\eeq
with respect to the weight function
\beq
w^{(N_f)}(x,y)\ \equiv\  
(xy)^{\nu+1}\prod_{f+=1}^{m}(x^2+m_{f+}^2)
\prod_{f-=1}^{n}(y^2+{m}_{f-}^2)\ 
I_{\nu}\left(2c_-Nxy\right) e^{-Nc_+( x^2+  y^2)}\ .
\label{weightimu}
\eeq
The monic polynomials are in general different. Together with their 
integral transform  
\beqn
\chi_k^{(N_f)}(y) &\equiv& 
\int_0^{\infty}dx\ w^{(N_f)}(x,y) P_k^{(N_f)}(x^2) \ ,\nn\\
\hchi_k^{(N_f)}(x) &\equiv& 
\int_0^{\infty}dx\ w^{(N_f)}(x,y) Q_k^{(N_f)}(y^2) 
\ .\label{chidef}
\eeqn
four different kernels can be constructed 
\beqn
K_N^{(N_f)}(y,x) 
&=& 
\sum_{k=0}^{N-1}\frac{Q_k^{(N_f)}(y^2)P_k^{(N_f)}(x^2)}{h_k}\ ,\ \ 
H_N^{(N_f)}(x_1,x_2) \ =\ 
\sum_{k=0}^{N-1}\frac{\hchi_k^{(N_f)}(x_1)P_k^{(N_f)}(x_2^2)}{h_k} \ ,
\nn
\\
M_N^{(N_f)}(x,y) &=& 
\sum_{k=0}^{N-1}\frac{\hchi_k^{(N_f)}(x)\chi_k^{(N_f)}(y)}{h_k}
\ ,\ \ \ \ \ \ \hH_N^{(N_f)}(y_1,y_2) \ =\ 
\sum_{k=0}^{N-1}\frac{Q_k^{(N_f)}(y_1^2)\chi_k^{(N_f)}(y_2)}{h_k} .
\label{Hh}
\eeqn
These are needed to express all correlation functions of mixed $x_k$ and $y_l$
eigenvalues defined as
\beqn
R_{(n,k)}^{(N_f,\nu)}(\{x\}_n;\{y\}_k) &\equiv&
\frac{N!^2}{(N-n)!(N-k)!}\frac{1}{{\cal Z}_{\nu}^{(N_f)}(\{m_1\},\{m_2\})} 
\nn\\
&\times&
\int_0^{\infty}  \prod_{i=n+1}^N dx_i \prod_{j=k+1}^N dy_j 
\det_{p,q}\left[w^{(N_f)}(x_p,y_q)\right]
\Delta_N(\{x^2\})\Delta_N(\{y^2\})
\label{Revrep}\\
&=&\!\! 
\!\!\!\det_{1\leq i_1,i_2\leq n;\ 1\leq  j_1,j_2\leq k}\!\left[\!\!
\begin{array}{cc}
H_N^{(N_f)}(x_{i_1},x_{i_2}) &
M_N^{(N_f)}(x_{i_1},y_{j_2})-w^{(N_f)}(x_{i_1},y_{j_2})\\ 
K_N^{(N_f)}(y_{j_1},x_{i_2}) & \hH_N^{(N_f)}(y_{j_1},y_{j_2})\\
\end{array}
\!\!\!\right]
\label{allRnk}
\eeqn
In the definition in the first line we have written the integrand of
eq. (\ref{evrep}) in a more compact form. The solution in terms of a
determinant of 4 blocks follows from \cite{EM}, adopted to our chiral two-MM 
\cite{ADOS}. Note that in contrast to eqs. (\ref{Rksol}) 
and (\ref{detformula}) 
the kernels built from the integral transforms of the polynomials
contain part of the weight inside the
determinant. 

For $N_f\neq0$ all polynomials, their transforms and the corresponding kernels
can be expressed through the quenched quantities as detailed in \cite{ADOS}. 
Below we give the quenched quantities as an explicit simple example. 

For the quenched weight $N_f=0$ in eq. (\ref{weightimu}) the bi-orthogonal
polynomials are equal and both given by Laguerre polynomials of real variables:
\beqn
P_n(x^2) \ ,\ Q_n(x^2) 
&\sim& L_n^{\nu} \left(\frac{N}{1+\mu^2} x^2\right) .
\label{polP}
\eeqn
In addition the integral transforms also become proportional to Laguerre
polynomials, times the Laguerre weight:
\beqn
\chi_k(y)\ ,\ \hat{\chi}_k(y)
&\sim&
y^{2\nu+1}e^{-\frac{N}{1+\mu^2}y^2}L_k^{\nu} 
\left(\frac{N}{1+\mu^2} y^2\right).
\eeqn
The reason is that for this special case the bi-orthogonal polynomials and
their weight can be constructed starting from two ordinary orthogonal Laguerre
polynomials of a factorised weight, 
where for details we refer to appendix B of ref. \cite{ADOS}.

We finish this subsection by giving two examples in the microscopic large-$N$
limit given by eq. (\ref{microscale}). 
First, the partition function eq. (\ref{evrep}) in this limit becomes
identical to eq. (\ref{Zmn}), with the replacement $\mu^2\to-\mu^2$ in
eq. (\ref{hIdef}) \cite{ADOS}. It agrees with the partition functions computed
from $\eps\chi$PT field theory, simply because the effective chiral Lagrangian
eq. (\ref{ZchPT}) is obtained by this rotation to $B=i\mu$
diag$(\id_m,-\id_n)$. The group integral is convergent for both signs of
$\mu^2$ and the rotation is trivial. 

On the level of correlation functions however, this rotation is highly
nontrivial. Leading to complex eigenvalues on the one side and two sets of
real eigenvalues on the other side their determinantal structure is very
different, comparing  eqs. (\ref{Rksol}) and (\ref{allRnk}) above.

The simplest nontrivial correlation function is the quenched 
probability to find one
eigenvalues $\xi$ of $\Dirac_+$ and one eigenvalues $\zeta$ 
of  $\Dirac_-$. In the
microscopic limit eq. (\ref{scale}) we obtain from eq. (\ref{allRnk})
\beq
\rho_{S\ (1,1)}^{(N_f,\nu)}(\xi;\zeta) 
~\equiv~ \rho_S^{(N_f,\nu)}(\xi)\ \rho_S^{(N_f,\nu)}(\zeta)\ +\
\rho_{S\ (1,1)}^{(N_f,\nu)\ conn}(\xi;\zeta) \ . 
\label{rhoQ11}
\eeq
Here the first product contains the quenched 
microscopic density eq. (\ref{rhoreal}) at
zero $\mu$,
\beqn
\rho_S^{(N_f,\nu)}(\xi) &=& \frac{\xi}{2}\left[J_{\nu}^2(\xi)
- J_{\nu+1}(\xi)J_{\nu-1}(\xi)\right] ~, 
\nn
\eeqn 
whereas the connected part is given by 
\beqn
\rho_{S\ (1,1)}^{(N_f,\nu)\ conn}(\xi;\zeta) &=&
\mathcal{I}^+_\nu(\xi,\zeta) ~
\left( \frac{1}{\al^2} \exp\left(-\frac{\xi^2+\zeta^2}{2\al^2}\right)
 I_{\nu}\left(\frac{\xi\zeta}{\al^2}\right) 
-  \mathcal{I}^-_\nu(\xi,\zeta)\right)
\ .
\label{rhoQ11int}
\eeqn
Here the same building block as in eq. (\ref{hIdef}) appears (see also
eq. (\ref{rhoQ})), containing the real eigenvalues 
\beqn
{\cal I}^\pm_\nu(\xi,\zeta) &\equiv& 
 \int_0^1 dt\ t \, e^{\pm 2\mu^2t^2}J_{\nu}(\xi t)J_{\nu}(\zeta t) ~. 
\label{shortint}
\eeqn
The connected density is plotted in the next subsection, fig. \ref{rhoreal2mm}.


\subsection{Comparison to Lattice data}

Because the introduction of an imaginary isospin chemical potentials 
does not change the anti-Hermiticity of the Dirac operator, standard
Monte-Carlo simulations are possible in the unquenched case as well. 
In this section we give a very brief
account of the lattice results of \cite{DHSS,DHSST} to where we refer for
details. There, staggered unimproved fermions were used, which does not lead
out of the $\beta_D=2$ symmetry class. The simulations covered several 
volumes $V=6^4,\ 8^4$ and $10^4$ at various small values of isospin 
chemical potentials $\mu_{iso}=0.002-0.01$.

In order to better understand the pictures below let us rewrite the
correlation function eq. (\ref{rhoQ11}) used there:
\beq
\rho_{S\ (1,1)}^{(N_f,\nu)}
(\xi=2Nx;\zeta=2Ny) \ =\ \lim_{N\to\infty}\frac{1}{(2N)^2}
\left\langle \sum_k\delta^1(x-x_k)  
\sum_j\delta^1(y-y_j)\right\rangle
\eeq
The connected part that will be compared to data is obtained by subtracting
the factorised correlation function in eq. (\ref{rhoQ11}). 
For $\mu\neq0$ the eigenvalues of the two
operators $\Dirac_\pm(\mu)$ are clearly distinct. In the limit $\mu\to 0$
however, they will coincide, leading to a delta-function in the connected
correlator $\sim\delta^1(y-x) $. The effect $\mu\neq0$ has is to smooth
out this delta-function, with the width of the resulting curve being sensitive
to $F_\pi$ contained in the rescaling factor of $\mu$. 
\begin{figure}[-h]
\centerline{
{\epsfig{figure=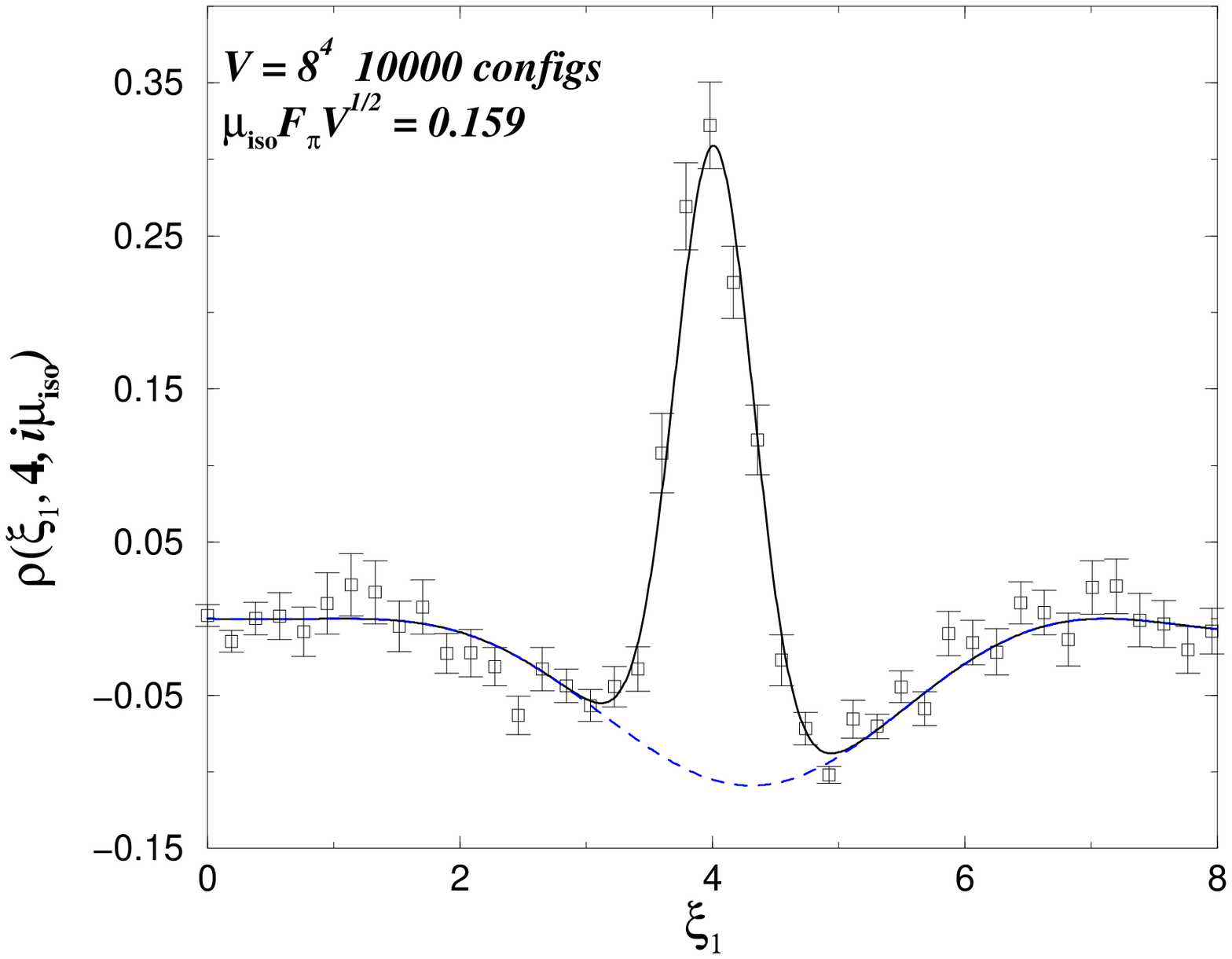
,clip=,width=8cm}}
{\epsfig{figure=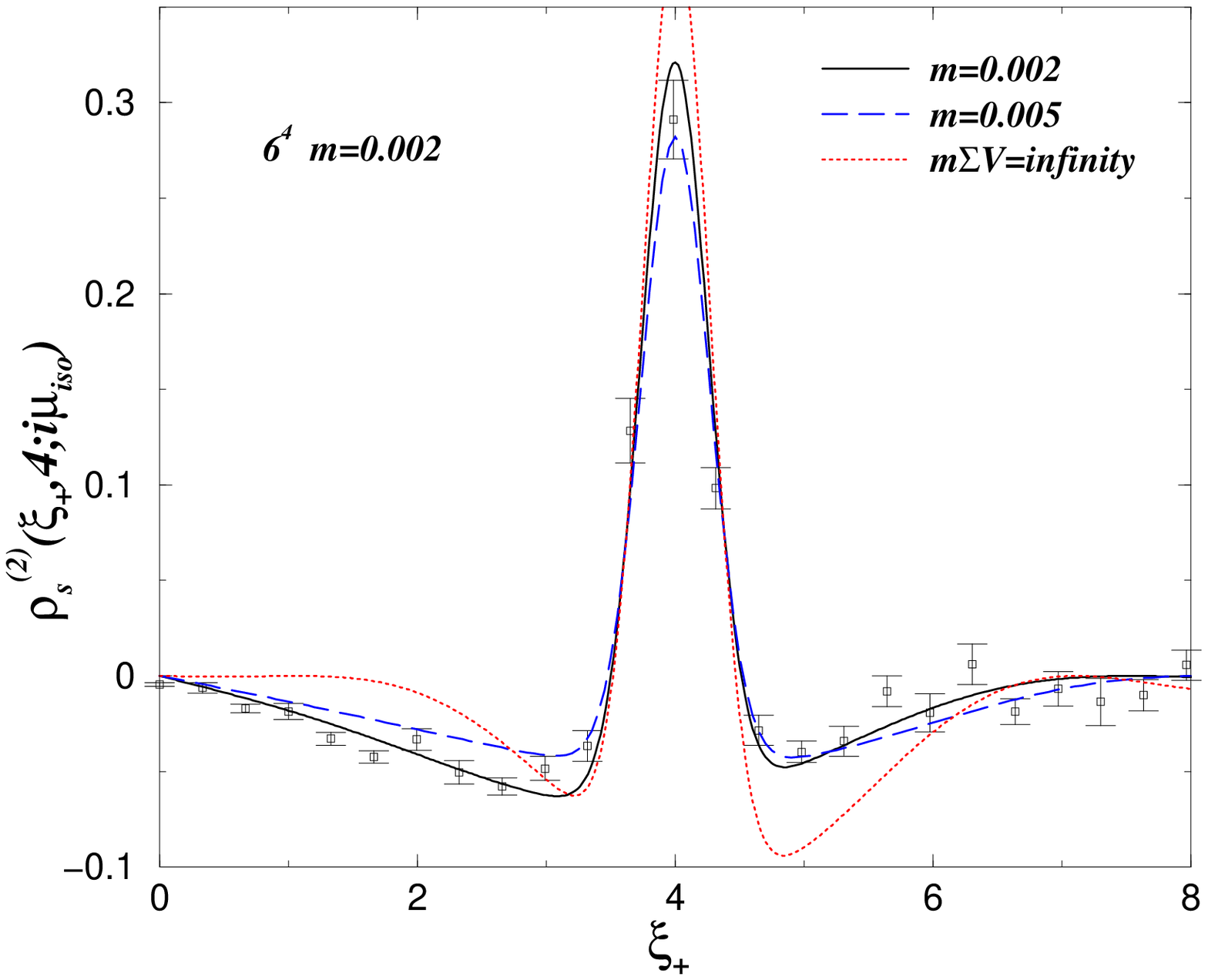,clip=,width=7.6cm}}
}
\caption{Quenched \cite{DHSS} (left) and unquenched \cite{DHSST} (right) 
Lattice data at $V=8^4$ (left) and 
  $\alpha=0.152$ at $V=6^4$ (right), vs. the connected correlation 
$\rho_{S\ (1,1)}^{(N_f,\nu)\ conn}(\xi,\zeta)$ 
(see eq. (\ref{rhoQ11int}) for quenched).}
\label{rhoreal2mm}
\end{figure}
In fig. \ref{rhoreal2mm} left the quenched connected
$\rho_{S\ (1,1)}^{(N_f,\nu)\ conn}(\xi,\zeta)$ 
density eq. (\ref{rhoQ11int}) is plotted versus lattice data \cite{DHSS}: 
here the second argument is kept fixed at $\zeta=4$ while the first argument
$\xi$ is varied. For comparison 
the same correlation function in the limit $\mu\to0$ is
inserted in dashed line, omitting the delta-spike at $\xi=\zeta=4$. 

In the second plot  fig. \ref{rhoreal2mm} right the unquenched density
generalising  eq. (\ref{rhoQ11int}) (see \cite{DHSST,ADOS} for details) is
compared to the data. 
The data nicely fit the curve corresponding to bare mass $m=0.002$ (lowest
curve to the left), distinguishing well from the parameter set inserted
for comparison $m=0.005$ (see \cite{DHSST} for the corresponding simulations).
The quenched limit obtained by setting $2Nm=\infty$ is inserted for comparison
as well (lowest curve on the right).
This example underlines once more 
the fruitful interplay between MMs, effective field
theory and Lattice simulations.

\sect{Conclusions and Open Problems}
\label{COP}

We have seen that Matrix Models are a useful tool to investigate QCD with both
real and imaginary chemical potential, extending their previous success for
zero $T$ and zero $\mu$. 
MMs provide both qualitative predictions such as for the phase
diagram, and quantitative results for the Dirac operator spectrum that can be
confronted with Lattice simulations. 

Following the symmetry classification into 3 classes we
have explicitly shown how to construct and solve MMs with complex eigenvalues 
for two of these 3 classes, the class with three or more colours $N_C\geq3$ 
containing QCD, and the class of adjoint QCD. 
All complex eigenvalue correlation functions could be computed for these two
classes as a function of
the number of quark flavours $N_f$, their masses $m_f$ and 
chemical potential of equal modulus $\pm\mu$. 
In particular this included unquenched
correlation functions in QCD 
where the spectral densities become complex valued quantities. 

The same construction was then repeated to build a MM for the 
QCD class with imaginary isospin chemical potential. Here all real eigenvalues
correlation functions were computed for the two sets of Dirac operators
$\Dirac(\pm i\mu)$.

Several issues have been understood on the way. The equivalence with the
underlying effective field theory, the epsilon regime of chiral perturbation
theory was shown for the QCD class both for real and imaginary $\mu$, on the
level of the partition function for an arbitrary set of chemical potentials
$\mu_f$. 
We have seen for QCD why the partition function with ordinary quarks and the
corresponding sum rules remain $\mu$-independent, while the generating
functional for the density is $\mu$-dependent. 
The strong oscillations and the complex nature of the unquenched eigenvalue
density in QCD have been made responsible for chiral symmetry breaking,
in contrast to quenched and phase quenched QCD. 

The detailed MM predictions for the microscopic density explained the
different signatures of complex Dirac eigenvalues, including their attraction
or repulsion from the real and imaginary axis. 
Excellent agreement was found when comparing these quantitatively to Lattice
simulations. We made comparisons to quenched QCD using staggered
and overlap fermions, allowing to compare to different topological
sectors of the gauge fields. For the adjoint QCD class we could compare to  
unquenched simulations. Here staggered fermions with two colours and fermions
in the fundamental representation were used, which are in the same symmetry
class. Again excellent agreement was found when lowering the mass to
effectively unquenched the eigenvalues close to the origin. 

For imaginary isospin chemical potential in the QCD class we could make the
same comparison between analytic predictions for two Dirac operators with
different real eigenvalues and Lattice data using staggered fermions. 
Both for quenching and unquenching an excellent agreement was found. 

Several interesting problems remain open within the MM approach. Fortunately
the symmetry class including unquenched QCD is technically the simplest within
MMs. Nevertheless partially quenched correlation functions with several
different $\mu_f$ have so far only been computed
for imaginary chemical potential
for 2 flavours, $\mu_1\neq\mu_2$. It would be very interesting to extend this
to more flavours, and in particular to real $\mu$, allowing for instance 
to compute
correlations with $\Dirac(\mu_f)$ in a background of zero chemical potential. 

To match with the partially quenched $\chi$PT we would have to compute the
corresponding group integral for a general baryon charge matrix $B$. 
Although we have shown that the expression for MMs and $\chi$PT are equal, 
we have not yet been able to compute the integral in either way. 

Turning to the 2 other symmetry classes only the class for adjoint QCD
corresponding to symplectic MMs has bee completely solved. Here the link to 
$\eps\chi$PT is an open question for $N_f>2$. 
The $\beta_D=1$ class for two-colour QCD remains the most challenging problem 
for MMs. Even the easier corresponding non-chiral model of real non-symmetric
matrices has resisted a solution for 40 years. This class would be very
interesting also from a Lattice perspective as it has a sign problem:
although the determinant $\det[\Dirac(\mu)+m]$ 
remains real its sign fluctuates. 

Lattice simulations of these 2 latter classes may now be feasible with chiral
fermions at non-vanishing $\mu$ due to the existence of the corresponding
overlap Dirac operator at $\beta_D=2$, without having to use the staggered
formulation. Such simulations would be very interesting to perform.
The most exciting challenge of course remains to see the MM for unquenched QCD
realised on the Lattice.

As a final remark it is well known that a much larger class of MMs with complex
eigenvalues exists, compared to the 3 chiral classes for $\mu=0$. It would be
very interesting to identify and solve some of the 
other classes that may be relevant in wider applications as well.

\indent

\underline{Acknowledgements}: 
I would like to thank
E. Bittner,
Y. Fyodorov, E. Kanzieper,  M. Lombardo, H. Markum, 
J. Osborn, A. Pottier, 
R. Pullirsch, L. Shifrin, G. Vernizzi 
and T. Wettig 
for discussions and stimulating collaborations over the past few years, 
without which this would not have
been possible.
I am particularly indebted to  F. Basile, P. Damgaard, 
K. Splittorff and J. Verbaarschot for valuable comments on the manuscript.
This work is supported by 
EU network ENRAGE MRTN-CT-2004-005616 
and by EPSRC grant EP/D031613/1.


\end{document}